\documentclass[12pt]{article}
\usepackage{lineno}
\usepackage{graphicx}
\usepackage{amsmath}
\usepackage{hhline}
\usepackage{amssymb}
\usepackage{times}
\usepackage{cite}
\usepackage{array}
\usepackage{rotating}
\usepackage{multirow}
\usepackage{bigstrut}
\usepackage{color}
\definecolor{rltred}{rgb}{0.75,0,0}
\definecolor{rltgreen}{rgb}{0,0.5,0}
\definecolor{rltblue}{rgb}{0,0,0.75}
\usepackage{dcolumn}
\newcolumntype{d}{D{.}{.}{1}}

\usepackage{ifpdf}
\ifpdf
\usepackage{thumbpdf}
\usepackage[
        colorlinks=true,
        urlcolor=rltblue,       
        filecolor=rltgreen,     
        linkcolor=rltred,       
        pdftitle={Example File for pdflatex},
        pdfauthor={H1},
        pdfsubject={Template file},
        pdfkeywords={High-Energy Physics, Particle Physics},
        pdfpagemode=None,
        bookmarksopen=true]{hyperref}
 
\fi

\newlength{\dinwidth}
\newlength{\dinmargin}
\begin{document}  

\newcommand{\pom}{{I\!\!P}}
\newcommand{\reg}{{I\!\!R}}
\newcommand{\slowpi}{\pi_{\mathit{slow}}}
\newcommand{\fiidiii}{F_2^{D(3)}}
\newcommand{\fiidiiiarg}{\fiidiii\,(\beta,\,Q^2,\,x)}
\newcommand{\n}{1.19\pm 0.06 (stat.) \pm0.07 (syst.)}
\newcommand{\nz}{1.30\pm 0.08 (stat.)^{+0.08}_{-0.14} (syst.)}
\newcommand{\fiidiiiful}{F_2^{D(4)}\,(\beta,\,Q^2,\,x,\,t)}
\newcommand{\fiipom}{\tilde F_2^D}
\newcommand{\ALPHA}{1.10\pm0.03 (stat.) \pm0.04 (syst.)}
\newcommand{\ALPHAZ}{1.15\pm0.04 (stat.)^{+0.04}_{-0.07} (syst.)}
\newcommand{\fiipomarg}{\fiipom\,(\beta,\,Q^2)}
\newcommand{\pomflux}{f_{\pom / p}}
\newcommand{\nxpom}{1.19\pm 0.06 (stat.) \pm0.07 (syst.)}
\newcommand {\gapprox}
   {\raisebox{-0.7ex}{$\stackrel {\textstyle>}{\sim}$}}
\newcommand {\lapprox}
   {\raisebox{-0.7ex}{$\stackrel {\textstyle<}{\sim}$}}
\def\gsim{\,\lower.25ex\hbox{$\scriptstyle\sim$}\kern-1.30ex%
\raise 0.55ex\hbox{$\scriptstyle >$}\,}
\def\lsim{\,\lower.25ex\hbox{$\scriptstyle\sim$}\kern-1.30ex%
\raise 0.55ex\hbox{$\scriptstyle <$}\,}
\newcommand{\pomfluxarg}{f_{\pom / p}\,(x_\pom)}
\newcommand{\dsf}{\mbox{$F_2^{D(3)}$}}
\newcommand{\dsfva}{\mbox{$F_2^{D(3)}(\beta,Q^2,x_{I\!\!P})$}}
\newcommand{\dsfvb}{\mbox{$F_2^{D(3)}(\beta,Q^2,x)$}}
\newcommand{\dsfpom}{$F_2^{I\!\!P}$}
\newcommand{\gap}{\stackrel{>}{\sim}}
\newcommand{\lap}{\stackrel{<}{\sim}}
\newcommand{\fem}{$F_2^{em}$}
\newcommand{\tsnmp}{$\tilde{\sigma}_{NC}(e^{\mp})$}
\newcommand{\tsnm}{$\tilde{\sigma}_{NC}(e^-)$}
\newcommand{\tsnp}{$\tilde{\sigma}_{NC}(e^+)$}
\newcommand{\st}{$\star$}
\newcommand{\sst}{$\star \star$}
\newcommand{\ssst}{$\star \star \star$}
\newcommand{\sssst}{$\star \star \star \star$}
\newcommand{\tw}{\theta_W}
\newcommand{\sw}{\sin{\theta_W}}
\newcommand{\cw}{\cos{\theta_W}}
\newcommand{\sww}{\sin^2{\theta_W}}
\newcommand{\cww}{\cos^2{\theta_W}}
\newcommand{\trm}{m_{\perp}}
\newcommand{\trp}{p_{\perp}}
\newcommand{\trmm}{m_{\perp}^2}
\newcommand{\trpp}{p_{\perp}^2}
\newcommand{\alp}{\alpha_s}

\newcommand{\alps}{\alpha_s}
\newcommand{\sqrts}{$\sqrt{s}$}
\newcommand{\LO}{$O(\alpha_s^0)$}
\newcommand{\Oa}{$O(\alpha_s)$}
\newcommand{\Oaa}{$O(\alpha_s^2)$}
\newcommand{\PT}{p_{\perp}}
\newcommand{\JPSI}{J/\psi}
\newcommand{\sh}{\hat{s}}
\newcommand{\uh}{\hat{u}}
\newcommand{\MP}{m_{J/\psi}}
\newcommand{\PO}{I\!\!P}
\newcommand{\xbj}{x}
\newcommand{\xpom}{x_{\PO}}
\newcommand{\ttbs}{\char'134}
\newcommand{\xpomlo}{3\times10^{-4}}  
\newcommand{\xpomup}{0.05}  
\newcommand{\dgr}{^\circ}
\newcommand{\pbarnt}{\,\mbox{{\rm pb$^{-1}$}}}
\newcommand{\gev}{\,\mbox{GeV}}
\newcommand{\WBoson}{\mbox{$W$}}
\newcommand{\fbarn}{\,\mbox{{\rm fb}}}
\newcommand{\fbarnt}{\,\mbox{{\rm fb$^{-1}$}}}
%
%
\newcommand{\qsq}{\ensuremath{Q^2} }
\newcommand{\gevsq}{\ensuremath{\mathrm{GeV}^2} }
\newcommand{\et}{\ensuremath{E_t^*} }
\newcommand{\rap}{\ensuremath{\eta^*} }
\newcommand{\gp}{\ensuremath{\gamma^*}p }
\newcommand{\dsiget}{\ensuremath{{\rm d}\sigma_{ep}/{\rm d}E_t^*} }
\newcommand{\dsigrap}{\ensuremath{{\rm d}\sigma_{ep}/{\rm d}\eta^*} }

\def\Journal#1#2#3#4{{#1} {\bf #2} (#3) #4}
\def\NCA{\em Nuovo Cimento}
\def\NIM{\em Nucl. Instrum. Methods}
\def\NIMA{{\em Nucl. Instrum. Methods} {\bf A}}
\def\NPB{{\em Nucl. Phys.}   {\bf B}}
\def\PLB{{\em Phys. Lett.}   {\bf B}}
\def\PRL{\em Phys. Rev. Lett.}
\def\PRD{{\em Phys. Rev.}    {\bf D}}
\def\ZPC{{\em Z. Phys.}      {\bf C}}
\def\EJC{{\em Eur. Phys. J.} {\bf C}}
\def\CPC{\em Comp. Phys. Commun.}

\begin{titlepage}
\begin{flushleft}
{\tt DESY 09-096  \hfill ISSN 0418-9833} \\
{\tt October 2009} \\
\end{flushleft}

\vspace{2cm}

\begin{center}
\begin{Large}
  
   {\bf {\boldmath {Measurement of the Charm and Beauty \\
Structure Functions  using \\
the H1 Vertex Detector at HERA}}}

\vspace{2cm}

H1 Collaboration

\end{Large}
\end{center}

\vspace{2cm}

\begin{abstract}
  \noindent Inclusive charm and beauty cross sections are
  measured in $e^-p$ and $e^+p$ neutral current collisions at HERA in
  the kinematic region of photon virtuality $5 \le Q^2 \le 2000~{\rm
  GeV}^2$ and Bjorken scaling variable $0.0002 \le x \le 0.05$. The
  data were collected with the H1 detector in the years 2006 and 2007
  corresponding to an integrated luminosity of 189~${\rm pb^{-1}}$.
  The numbers of charm and beauty events are determined using variables
  reconstructed by the H1 vertex detector including the impact
  parameter of tracks to the primary vertex and the position of the
  secondary vertex. The measurements are combined with previous data
  and compared to QCD predictions.
\end{abstract}

\vspace{1.5cm}

\begin{center}
Accepted by  Eur. Phys. J.\ C.
\end{center}

\end{titlepage}

\begin{flushleft}

F.D.~Aaron$^{5,49}$,           
M.~Aldaya~Martin$^{11}$,       
C.~Alexa$^{5}$,                
K.~Alimujiang$^{11}$,          
V.~Andreev$^{25}$,             
B.~Antunovic$^{11}$,           
A.~Asmone$^{33}$,              
S.~Backovic$^{30}$,            
A.~Baghdasaryan$^{38}$,        
E.~Barrelet$^{29}$
% ...

\bigskip{\it
% ...
$ ^{5}$ National Institute for Physics and Nuclear Engineering (NIPNE) ,
          Bucharest, Romania \\
% ...
$ ^{11}$ DESY, Hamburg, Germany \\
% ...
$ ^{25}$ Lebedev Physical Institute, Moscow, Russia$^{ e}$ \\
% ...
$ ^{29}$ LPNHE, Universit\'{e}s Paris VI and VII, IN2P3-CNRS,
          Paris, France \\
 $ ^{30}$ Faculty of Science, University of Montenegro,
          Podgorica, Montenegro$^{ e}$ \\
% ...
$ ^{33}$ Dipartimento di Fisica Universit\`a di Roma Tre
          and INFN Roma~3, Roma, Italy \\
% ...
$ ^{38}$ Yerevan Physics Institute, Yerevan, Armenia \\
% ...
$ ^{49}$ Also at Faculty of Physics, University of Bucharest,
          Bucharest, Romania \\
% ...
$ ^e$ Supported by the Deutsche Forschungsgemeinschaft \\
% ...
}
\end{flushleft}

\newpage

\section{Introduction}

The measurement of the inclusive charm ($c$) and beauty ($b$) quark cross sections
and the derived structure functions $F_2^{c\bar{c}}$ and
$F_2^{b\bar{b}}$ in DIS at HERA is an important test of the theory of the
strong interaction, quantum chromodynamics (QCD), within the Standard
Model.  These measurements uniquely constrain the parton density
functions (PDFs) of the proton, in particular its $b$ and $c$ content.
Precise knowledge of
the PDFs is for example essential at the Large Hadron
Collider (LHC).  The predictions of the `standard candle' QCD
processes at the LHC, such as the inclusive production of $W$ and $Z$
bosons, are sensitive to the theoretical treatment of heavy quarks~\cite{collins,VFNS1,VFNS2,VFNS3,VFNS4,cteqvfns,mstw}.  
The $b$ quark density is important in Higgs production at
the LHC in both the Standard Model and in extensions to the Standard Model
such as supersymmetric models at high values of the mixing parameter 
$\tan\beta$\cite{bbh}.

This paper reports on measurements made in neutral current deep
inelastic scattering (DIS) at HERA of the charm and beauty 
contributions to the inclusive proton structure function $F_2$ in the
range of virtuality of the exchanged photon $5 \le Q^2
\le 2000$~${\rm GeV}^2$ and Bjorken $x$  $0.0002 \le x \le 0.05$.
The analysis uses the precise spatial information from the H1 vertex
detector to separate events containing $c$ and $b$ flavoured hadrons
 from light quark events. The analysis extends to
lower and higher $Q^2$ than previous H1 measurements~\cite{Aktas:2004az,Aktas:2005iw} which used a similar technique to the one used in this paper.

The analysis is based on a dataset with an integrated luminosity of
$189$~${\rm pb}^{-1}$, which is about three times greater than in the
previous measurements.  The data was recorded in the years 2006 and 2007 with
$54$~${\rm pb}^{-1}$ taken in $e^-p$ mode and $135$~${\rm pb}^{-1}$ in
$e^+p$ mode. The $ep$ centre of mass energy is $\sqrt{s} = 319~{\rm
GeV}$, with a proton beam energy of $E_p=920~{\rm GeV}$ and electron beam
energy of $E_e=27.6~{\rm GeV}$.  This dataset is referred to here as HERA~II.
Many details of the analysis are similar to the
previous measurements~\cite{Aktas:2004az,Aktas:2005iw}, referred to here as 
HERA~I.
The HERA~I and HERA~II measurements are combined to produce a complete
 HERA dataset.  Measurements of the
charm contribution to the proton structure function have also been
made at HERA using  $D$ or $D^*$ meson
production~\cite{H1ZEUSDstar,H1Dstar}.
There are also measurements of charm and beauty in DIS using
semi-leptonic decays~\cite{H1ZEUSmuons}.

Events containing heavy quarks are distinguished from those containing
only light quarks using variables that are sensitive to the longer
lifetimes of heavy flavour hadrons. The most important of these
variables are the transverse displacement of tracks from the primary
vertex and the reconstructed position of a secondary vertex in the
transverse plane.  For events with three or more tracks in the vertex
detector the reconstructed variables are used as input to an artificial neural
network.  This method has better discrimination between $c$ and
$b$ compared to previous methods~\cite{Aktas:2004az,Aktas:2005iw},
which used only the transverse displacement of tracks from the primary
vertex. Lifetime based methods have the advantage over more
exclusive methods, such as $D^*$ or muon tagging, in that a higher
fraction of heavy flavour events may be used, although the background
from light quark events is larger. The charm structure function
$F_2^{c\bar{c}}$ and the beauty structure function $F_2^{b\bar{b}}$
are obtained from the measured $c$ and $b$ cross sections after
applying small corrections for the longitudinal structure functions
$F_L^{c\bar{c}}$ and $F_L^{b\bar{b}}$.

\section{Monte Carlo Simulation}

Monte Carlo simulations are used to correct for the effects of the
finite detector resolution, acceptance and efficiency.  The Monte
Carlo program RAPGAP\cite{Jung:1993gf} is
used to generate DIS events for the processes 
$ep \rightarrow eb\bar{b}X$, $ep \rightarrow ec\bar{c}X$ and 
$ep \rightarrow eq X$ where $q$ is a light
quark of flavour $u$, $d$ or $s$.  
RAPGAP combines $\cal{O}$($\alpha_s$) matrix elements with higher
order QCD effects modelled by parton showers. The
heavy flavour event samples are generated according to the `massive'
photon gluon fusion (PGF) matrix element~\cite{massive} with the mass of the $c$ and
$b$ quarks set to $m_c=1.5~{\rm GeV}$ and $m_b=4.75~{\rm GeV}$,
respectively.  The DIS cross section is calculated using the leading
order (LO) 3-flavour PDFs from MRST (MRST2004F3LO~\cite{Martin:2006qz}).  
The partonic system for the generated events is fragmented according
to the Lund string model~\cite{Andersson:1983ia} implemented within the PYTHIA
program~\cite{Sjostrand:2001yu}.  
The $c$ and $b$ quarks are hadronised according to the Bowler fragmentation
function~\cite{bowler}.
The HERACLES
program\cite{Kwiatkowski:1990es} interfaced to RAPGAP
calculates single photon radiative
 emissions off the lepton line, virtual and electroweak corrections.
 The Monte Carlo program PHOJET\cite{Engel:1995yd} is used to simulate
 the background contribution from photoproduction 
$\gamma p \rightarrow X$.

The samples of generated events are passed through a detailed
simulation of the detector response based on the GEANT3
program\cite{Brun:1978fy}, and through the same reconstruction
software as is used for the data.

\section{H1 Detector}
\label{sec:h1detector}
Only a short description of the H1 detector is given here; a more complete
description may be found in\cite{Abt:1997xv}. 
A right handed coordinate system is employed
with the origin at the position of the nominal interaction point
that has its $Z$-axis pointing in the proton beam, or forward,
direction and $X$ ($Y$) pointing in the horizontal (vertical)
direction. The pseudorapidity is related to the polar angle $\theta$
by $\eta = - \ln \tan(\theta/2)$.

Charged particles are measured in the central tracking detector (CTD).
This device consists of two cylindrical drift chambers interspersed
with $Z$-chambers to improve the $Z$-coordinate reconstruction and
multi-wire proportional chambers mainly used for triggering. The CTD
is operated in a uniform solenoidal $1.16\,{\rm T}$ magnetic field, enabling the
momentum measurement of charged particles over the polar angular
range $20^\circ< \theta<160^\circ$.

The CTD tracks are linked to hits in the vertex detector, the central
silicon tracker CST~\cite{cst}, to provide precise spatial track
reconstruction. The CST consists of two layers of double-sided silicon
strip detectors surrounding the beam pipe, covering an angular range
of $30^\circ< \theta<150^\circ$ for tracks passing through both
layers.   The information on the $Z$-coordinate of the CST tracks 
is not used in the analysis presented in this paper. For CTD tracks
with CST hits in both layers the transverse distance of closest
approach (DCA) to the nominal vertex in $X$--$Y$,
averaged over the azimuthal angle, is measured to have
a resolution of $43 \ \mu{\rm m} \oplus 51 \ \mu{\rm m} /(P_T
[{\rm GeV}])$ where $P_T$ is the transverse momentum of the particle.
The first term represents the intrinsic
resolution (including alignment uncertainty) and the second term is
the contribution from multiple scattering in the beam pipe and the
CST.

The track detectors are surrounded in the forward and central
directions ($4^\circ<\theta<155^\circ$) by a fine-grained liquid argon
calorimeter (LAr) and in the backward region
($153^\circ<\theta<178^\circ$) by a lead-scintillating fibre
calorimeter (SPACAL)~\cite{Nicholls:1996di} 
with electromagnetic and
hadronic sections. These calorimeters are used in this analysis to measure 
and identify the scattered
electron\footnote{In this paper we use `electron' to also denote `positron'
unless explicitly stated.}  and also provide energy and angular
reconstruction for final state particles from the hadronic system.  

Electromagnetic calorimeters situated downstream in the electron beam
direction allow detection of photons and electrons scattered at very
low $Q^2$. The luminosity is measured with these calorimeters
from the rate of photons
produced in the Bethe-Heitler process $ep\rightarrow ep\gamma$.

\section{Experimental Method}

\subsection{DIS Event Selection}
\label{disselect}
The events are triggered by requiring a compact, isolated
electromagnetic cluster in either the LAr or SPACAL calorimeters with
an overall trigger efficiency of almost $100$\%.  The electromagnetic
cluster with the highest transverse energy, which also passes stricter
offline criteria is taken as the scattered electron.  The $Z$-position
of the interaction vertex, reconstructed by one or more charged tracks
in the tracking detectors, must be within $\pm 20~{\rm cm}$
 to match the acceptance of the CST. The interaction vertex  approximately
follows a Gaussian distribution with a standard deviation of $13~{\rm cm}$.

Photoproduction events and DIS events with a hard photon radiated from
the initial state electron are suppressed by requiring $\sum_{i} (E_i
- p_{z,i}) >35~{\rm GeV}$.  Here, $E_i$ and $p_{z,i}$ denote the
energy and longitudinal momentum components of a particle and the sum
is over all final state particles including the scattered electron and
the hadronic final state (HFS). The HFS particles are reconstructed
using a combination of tracks and calorimeter deposits in an energy
flow algorithm that avoids double counting\cite{peezportheault}.

The event kinematics, including the photon virtuality $Q^2$, the
Bjorken scaling variable $x$ and the inelasticity variable $y$, are
reconstructed with the `$e\Sigma$' method\cite{Bassler:1994uq}, which
uses the scattered electron and the HFS.  The variables obey the
relation $x = Q^2/sy$. In order to have good acceptance in the SPACAL
and to ensure that the HFS has a significant transverse momentum,
events are selected for $Q^2 >4.5 \ {\rm GeV^2}$. The analysis is
restricted to $y>0.07$ in order to ensure that the direction of the
quark which is struck by the photon is mostly in the CST angular
range.  An upper $y$ cut is applied that varies from $0.5$ at low
$Q^2$ to $0.85$ at high $Q^2$ in order to suppress photoproduction
background.  The measurement is made differentially by dividing the
data into discrete $y$--$Q^2$ intervals. This binning scheme is
preferable to one using $x$--$Q^2$ boundaries as it avoids event
losses near the cuts on $y$. The RAPGAP Monte Carlo program is used to
estimate the acceptance of $c$ ($b$) events that have a $c$ ($b$)
quark with tranverse momentum greater than $0.3$~GeV and within the
angular acceptance of the CST. The overall $c$ ($b$) quark acceptance is
$89\%$ ($90\%$) with a minimum of $75\%$ ($75\%$) in any $y$--$Q^2$
bin.

The position of the beam interaction region in $X$--$Y$ (beam spot) is
calculated from information of tracks with CST hits and updated
regularly to account for drifts during beam storage. The beam
interaction region has an elliptical shape with a size of around
$90 \ \mu {\rm m}$ in $x$ and around $22 \ \mu {\rm m}$ in $Y$.

\subsection{Track Selection}
\label{tracksel}
\begin{figure}[h!]
  \begin{center} \includegraphics[width=0.75\textwidth]{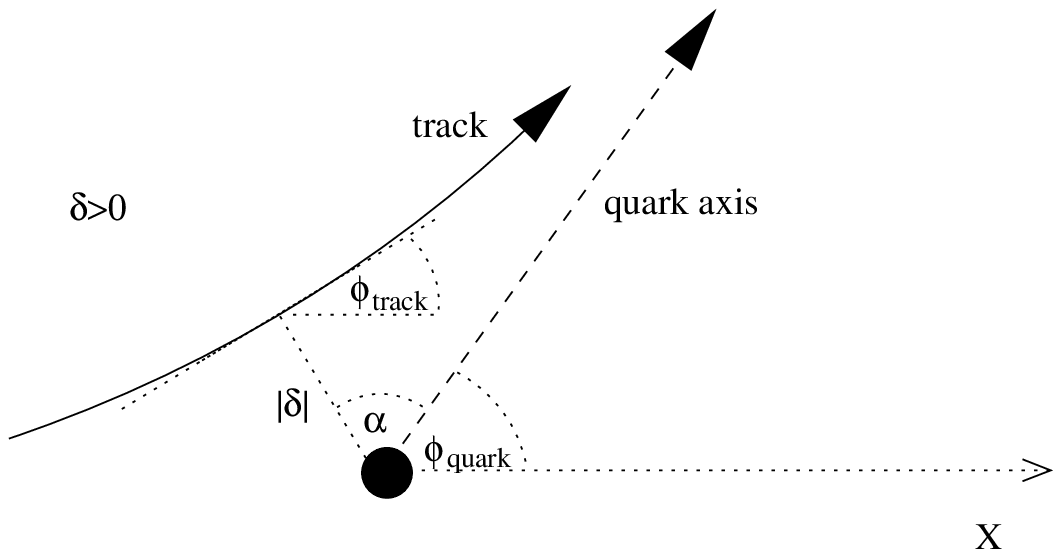}
	\vspace{0.6cm}
	 \includegraphics[width=0.72\textwidth]{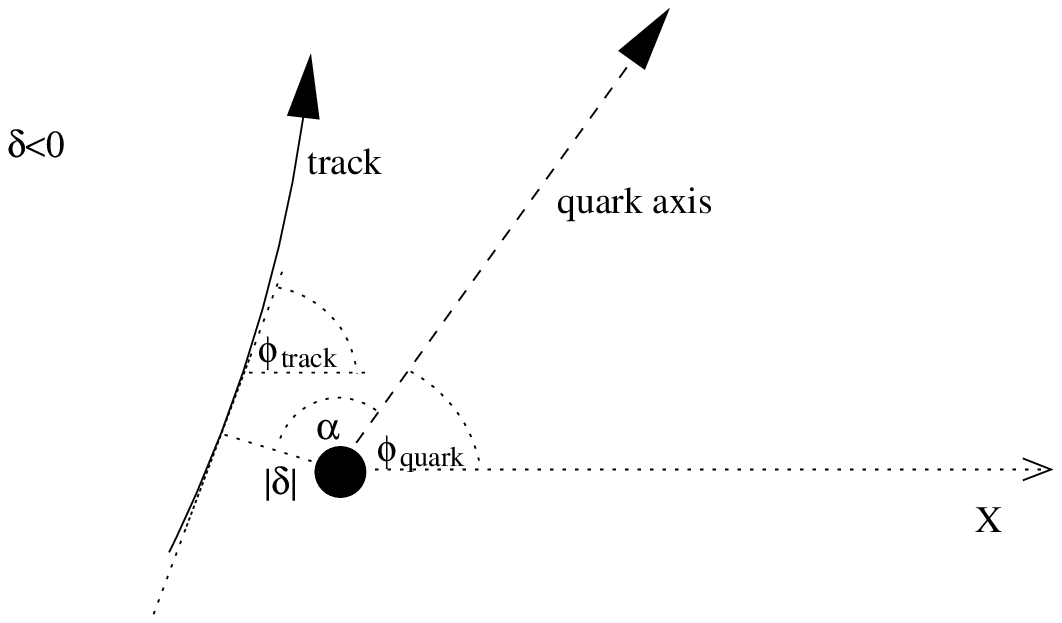}
  \caption{Diagrams of  a track in the $X$--$Y$ plane.
If the angle $\alpha$ is less than $90^\circ$,
$\delta$ is defined as positive  otherwise $\delta$ is defined as negative. 
}  \label{fig:alpha}
  \end{center}
\end{figure}

The impact parameter $\delta$ of a track, which is the transverse DCA
of the track to the primary vertex point (see figure~\ref{fig:alpha}),
is only determined for those tracks which are measured in the CTD and
have at least two CST hits linked (referred to as CST tracks). The
 beam spot is used as the position of the primary vertex.
CST tracks are required to have a transverse momentum 
$P_T^{\rm track}> 0.3~{\rm GeV}$.

The direction of the struck quark is used in the determination of the
sign of $\delta$.  The vector of the struck quark ($P_T^q$,
$\theta^q$, $\phi^q$) is reconstructed as the azimuthal angle of the
highest transverse momentum jet in the event. If there is no jet
reconstructed in the event the vector is reconstructed from the
electron~\cite{eflow} so that $P_T^q=P_T^{\rm elec}$, $\cos \theta^q
=1-8E^2_ey^2/(4E^2_ey^2+Q^2(1-y))$, $\phi^q=180^\circ-\phi^{\rm
elec}$, where $P_T^{\rm elec}$ and $\phi^{\rm elec}$ are
the transverse momentum and the azimuthal angle in degrees of the
scattered electron, respectively. Jets are reconstructed using the
inclusive longitudinally invariant $k_T$ algorithm with massless $P_T$
recombination scheme and distance parameter $R_0=1$ in the $\eta-\phi$
plane\cite{KTJET}. The algorithm is run in the laboratory frame using
all reconstructed HFS particles and the resultant jets are required to
have transverse momentum greater than $1.5$ \ ${\rm GeV}$ and to be in
the angular range $15^\circ < \theta < 155^{\rm o}$.  Approximately
$95\%$ ($99\%$) of $c$ ($b$) events have $\phi^q$ reconstructed from a
jet, as determined from the Monte Carlo simulation. Tracks with
azimuthal angle $\phi_{\rm track}$ outside $\pm 90^\circ$ of $\phi^q$
are assumed not to be associated to the struck quark and rejected.

\begin{figure}[h!]
  \begin{center} \includegraphics[width=0.9\textwidth]{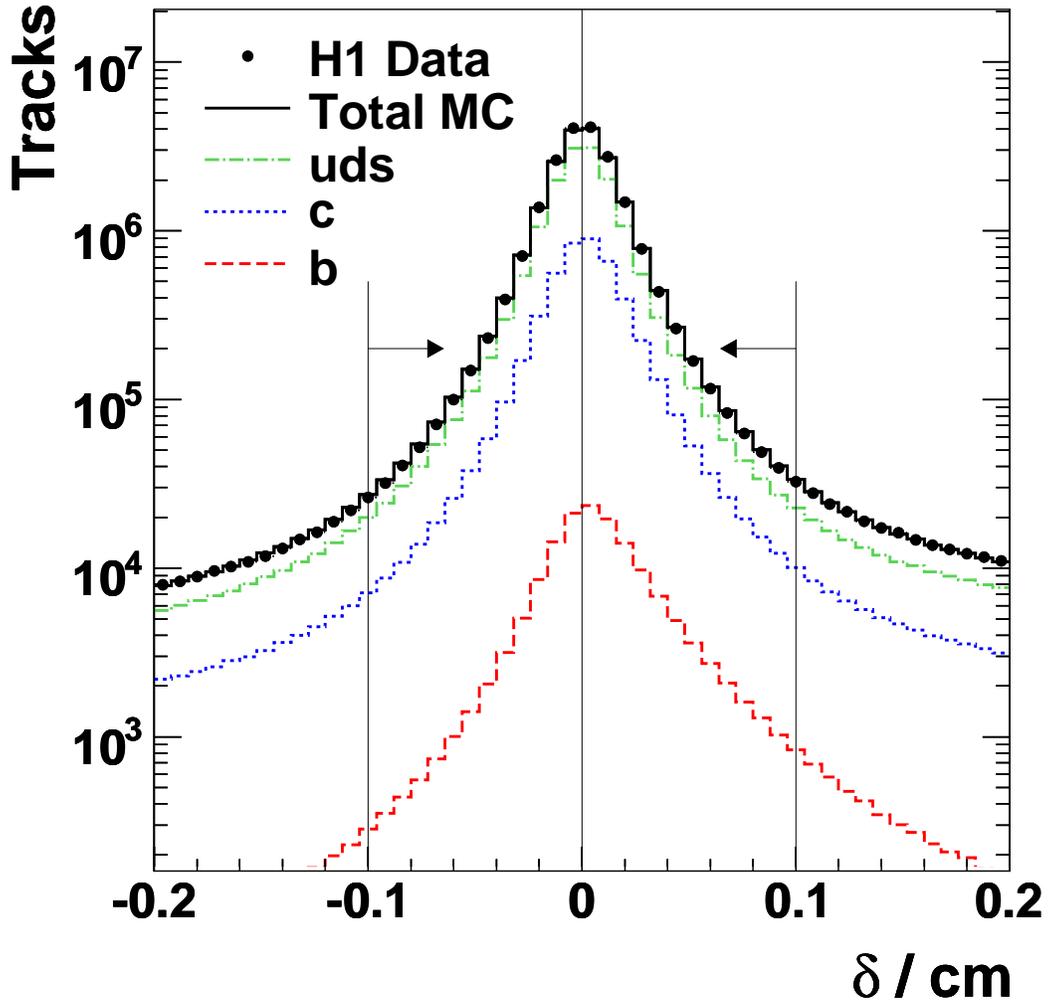}
  \caption{The distribution of the signed impact parameter $\delta$ of
  a track to the primary vertex in the $X$--$Y$ plane. Included in the
  figure is the expectation from the Monte Carlo simulation for
  light, $c$
  and $b$ quarks. The contributions from the various quark flavours
  in the Monte Carlo are shown after applying the scale factors $\rho_l$, $\rho_c$ and $\rho_b$ obtained from the fit to the complete data sample (see
  section~\ref{quarkflavourseparation}).
The arrows indicate the range over which tracks are selected for analysis.
}  \label{fig:dca}
  \end{center}
\end{figure}

If the angle $\alpha$ between $\phi^q$ and the line joining the
primary vertex to the point of DCA is less than $90^\circ$, $\delta$
is defined as positive. It is defined as negative otherwise.
Figure~\ref{fig:alpha} shows an example of a track with positive
$\delta$ and one with negative $\delta$. The $\delta$ distribution,
shown in figure~\ref{fig:dca}, is seen to be asymmetric with positive
values in excess of negative values indicating the presence of
long-lived particles. It is found to be well described by the Monte
Carlo simulation.  CST tracks with $|\delta|>0.1~{\rm cm}$ are
rejected to suppress light quark events containing long-lived strange
particles.

The number of reconstructed CST tracks $N_{\rm track}$ associated to
the struck quark is an important quantity since for higher track
multiplicities more information can be used. In the kinematic range of
this measurement $22\%$ of the events have no selected track, $26\%$ of
the events have $N_{\rm track}=1$, $23\%$ have $N_{\rm track}=2$ and
$29\%$ have $N_{\rm track} \ge 3$.

\subsection{Secondary Vertex Reconstruction}
\label{secvertex}
The complete set of reconstructed tracks in the event is used to
simultaneously reconstruct a secondary and primary vertex in the
transverse plane. Two vertices are reconstructed in each event even if
they are not statistically separable. Each track is assigned a weight
for each vertex, which is a measure of the probability that the track
originated at that vertex\cite{svref1}. In this approach tracks are
not assigned unambiguously to one vertex or the other.  A simultaneous
fit to the primary and secondary vertex is made, with the weights of
all tracks of the event considered for the primary vertex, but only
the weights of CST tracks considered for the secondary vertex.  The
beam spot together with its spread is used as an additional constraint
to the primary vertex. The vertex configuration that minimises the
global $\chi^2$ is found iteratively using deterministic
annealing\cite{svref2}.

The transverse distance between the secondary and primary vertex is
defined as $L_{xy}$.  The secondary vertex significance $S_L$ is
$L_{xy} / \sigma (L_{xy})$, where $\sigma (L_{xy})$ is the uncertainty
on $L_{xy}$.  If the absolute difference between the azimuthal angle
of the secondary vertex (taking the primary vertex as the origin) and
$\phi^q$ is less than $90^\circ$, $S_L$ is signed positive
and negative otherwise. A measure of the decay multiplicity $N^{\rm
SV}_{\rm track}$ is made by counting the number of tracks with weight
greater than $0.8$ to the secondary vertex. This method was shown
in~\cite{Aktas:2004az} to yield consistent results with the default
method that used track significances.

\subsection{Quark Flavour Separation}
\label{quarkflavourseparation}

The track significance is defined as $\delta/\sigma(\delta)$, where
$\sigma(\delta)$ is the uncertainty on $\delta$. The significances
$S_1$, $S_2$ and $S_3$, are defined as the significance of the CST
track with the highest, second highest and third highest absolute
significance, respectively.  The significances take the sign of
$\delta$ (see section~\ref{tracksel}). Tracks with a negative sign for
significance are likely not to arise from particles with a large
lifetime and are used in this analysis to estimate the light quark
contribution.

\begin{figure}[htb]
\includegraphics[width=0.5\textwidth]{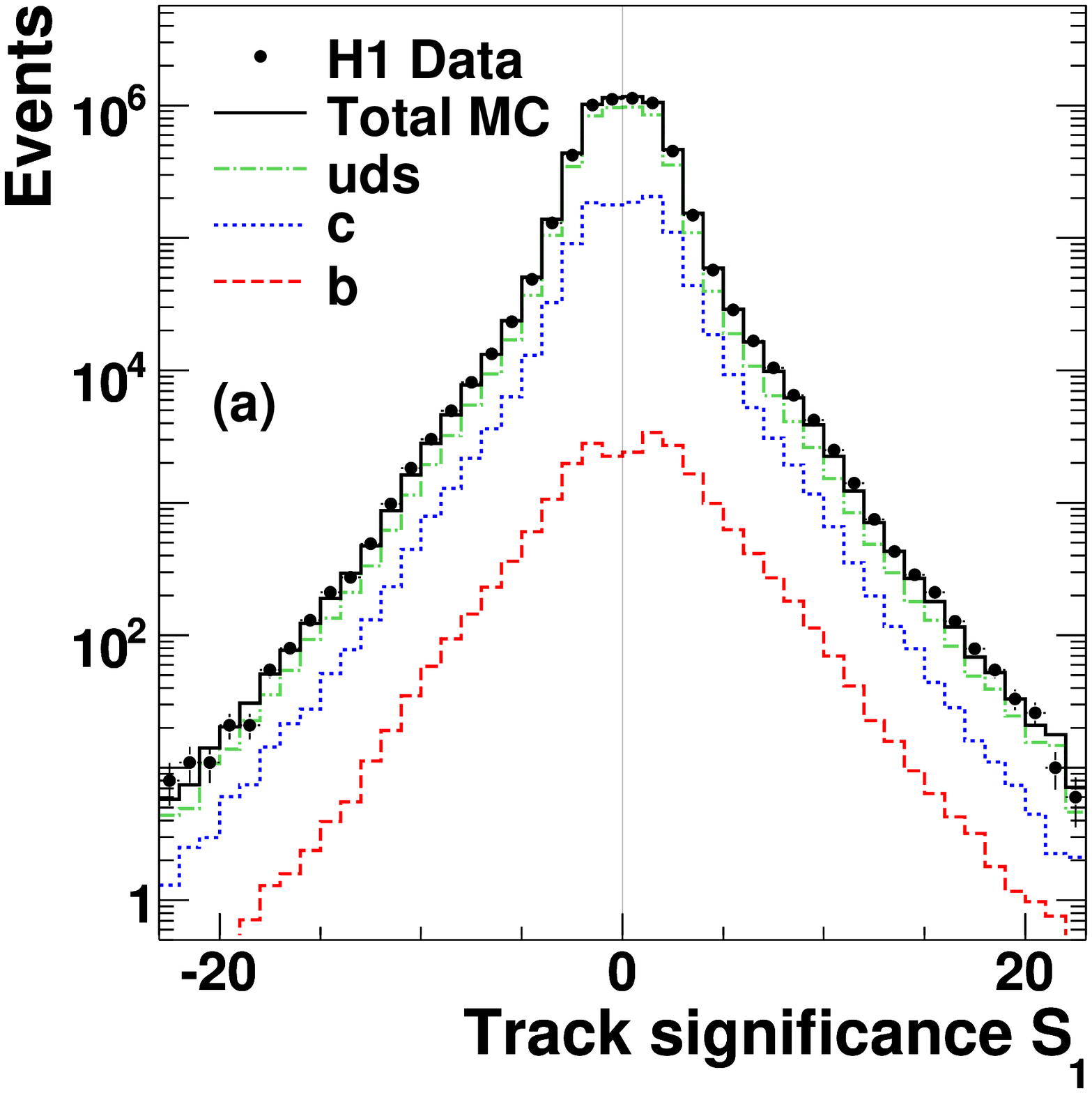}
 \includegraphics[width=0.5\textwidth]{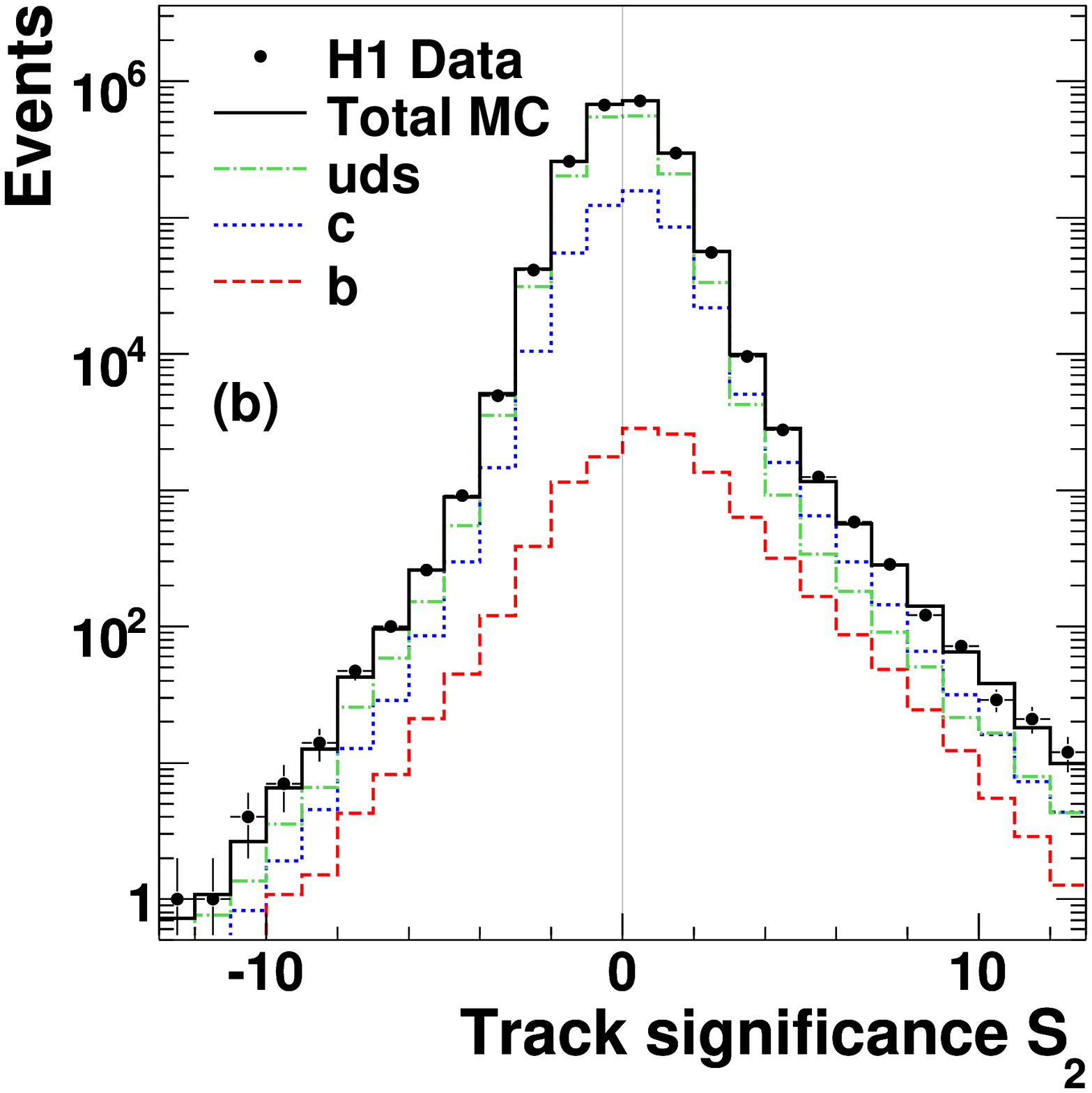}
 \caption{The
  significance $\delta /\sigma(\delta)$ distribution (a) of the
  highest absolute significance track ($S_1$) 
	 and (b) of the track with
  the second highest absolute significance ($S_2$).
  Included in the figure is the expectation from the Monte
  Carlo simulation for light, $c$ and $b$ quarks. The contributions from the various
  quark flavours in the Monte Carlo are shown after applying the scale factors $\rho_l$, $\rho_c$ and $\rho_b$ obtained from the fit to the complete data sample.}
  \label{fig:s1s2}
\end{figure}

Tracks that do not have the same significance sign as $S_1$ are
ignored. The $S_1$ distribution is used for events with one remaining
CST track after this selection and the $S_2$ distribution is used if
there are two remaining CST tracks. For events with three or more
remaining CST tracks, where more information is available, information 
is combined from the significance distributions and the reconstructed
secondary vertex using an artificial neural network (NN) that takes 
into account
correlations of the input variables~\cite{jetnet}. In this way each
event with at least $1$ CST track appears in exactly one distribution
and the distributions are statistically independent.  The $S_1$ and
$S_2$ distributions are shown in figure~\ref{fig:s1s2}.

\begin{figure}[htb]
\begin{center}
\includegraphics[width=0.494\textwidth]{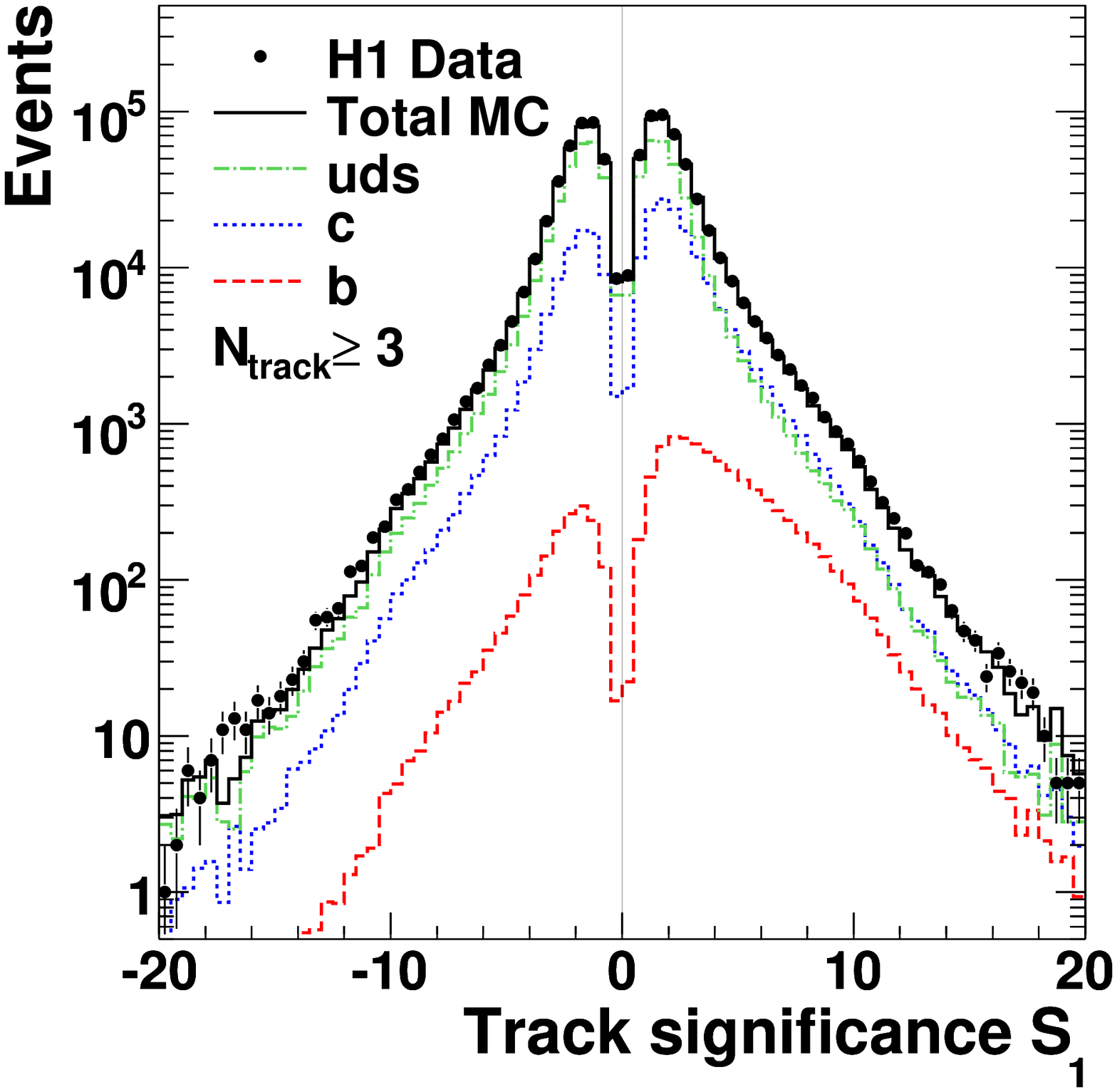}
\includegraphics[width=0.494\textwidth]{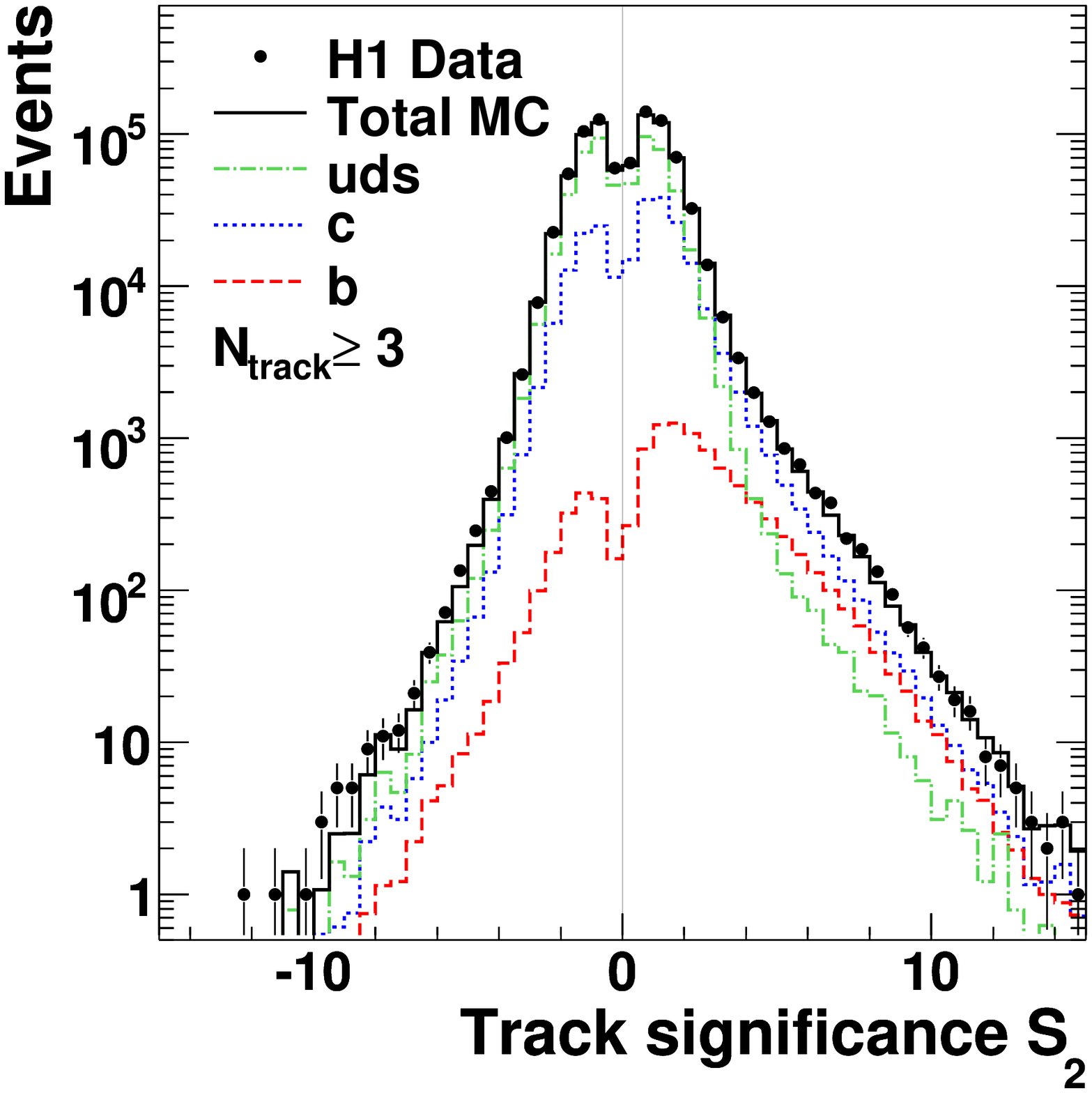}
\includegraphics[width=0.494\textwidth]{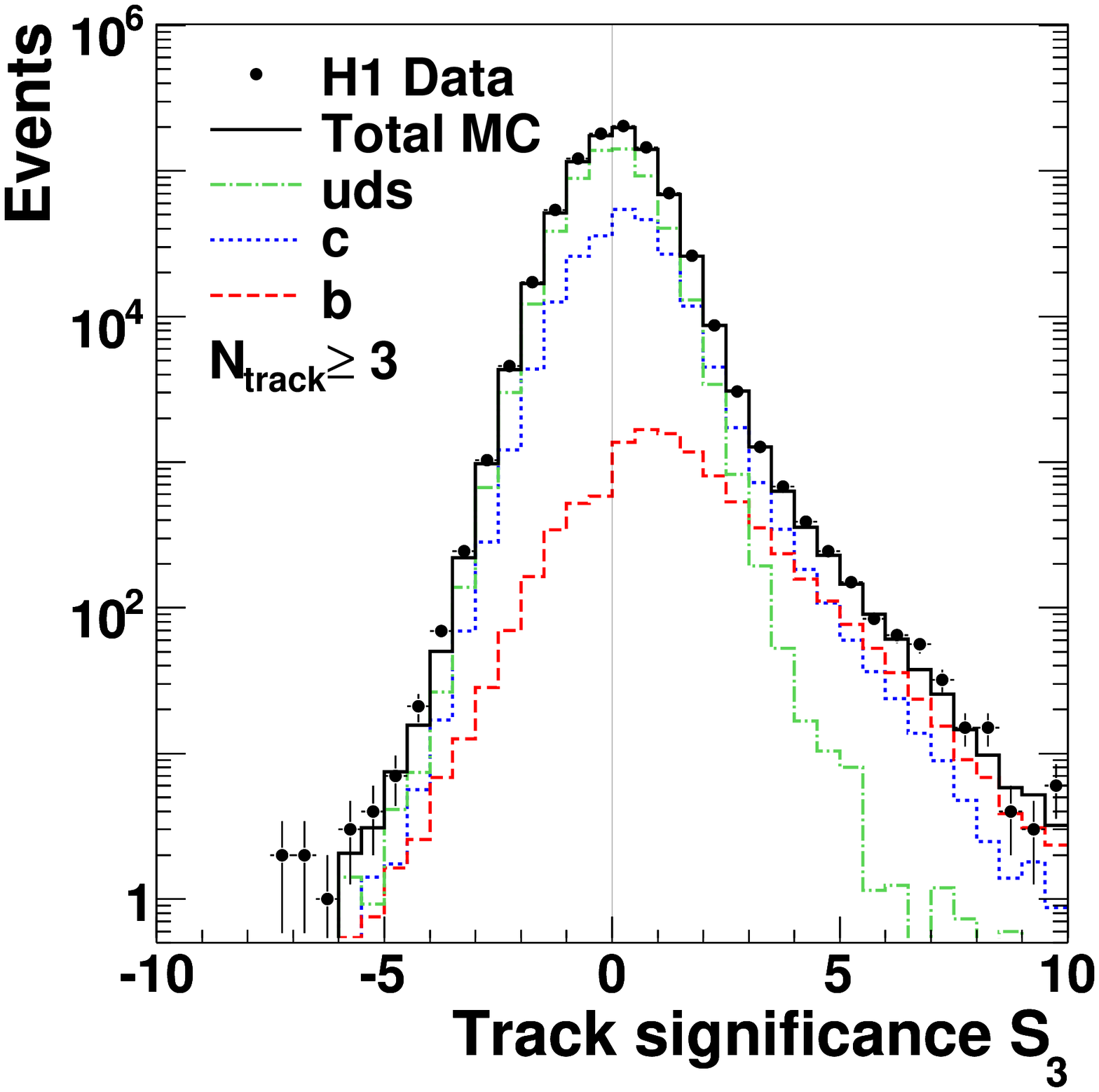}
\includegraphics[width=0.494\textwidth]{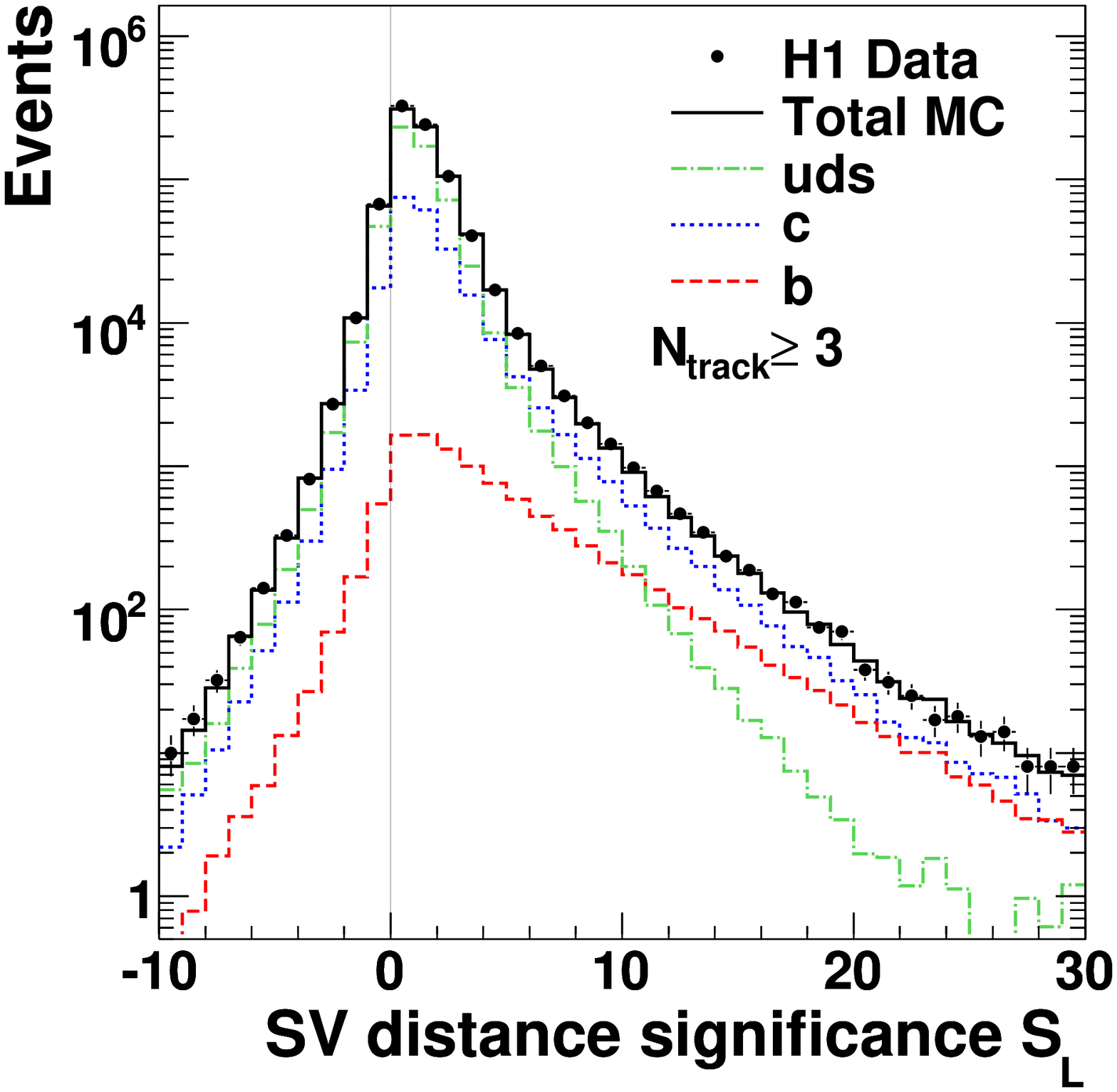}
\end{center}

 \caption{Inputs to the NN  for events with at least 3 CST tracks: $|S_1|$, $|S_2|$, $|S_3|$ and $S_L$.
  Included in the figure is the expectation from the Monte
  Carlo simulation for light, $c$ and $b$ quarks. The contributions from the various
  quark flavours in the Monte Carlo are shown after applying the scale factors $\rho_l$, $\rho_c$ and $\rho_b$ obtained from the fit to the complete data sample.}
  \label{fig:nninputsa}
\end{figure}

\begin{figure}[htb]
\begin{center}
\includegraphics[width=0.494\textwidth]{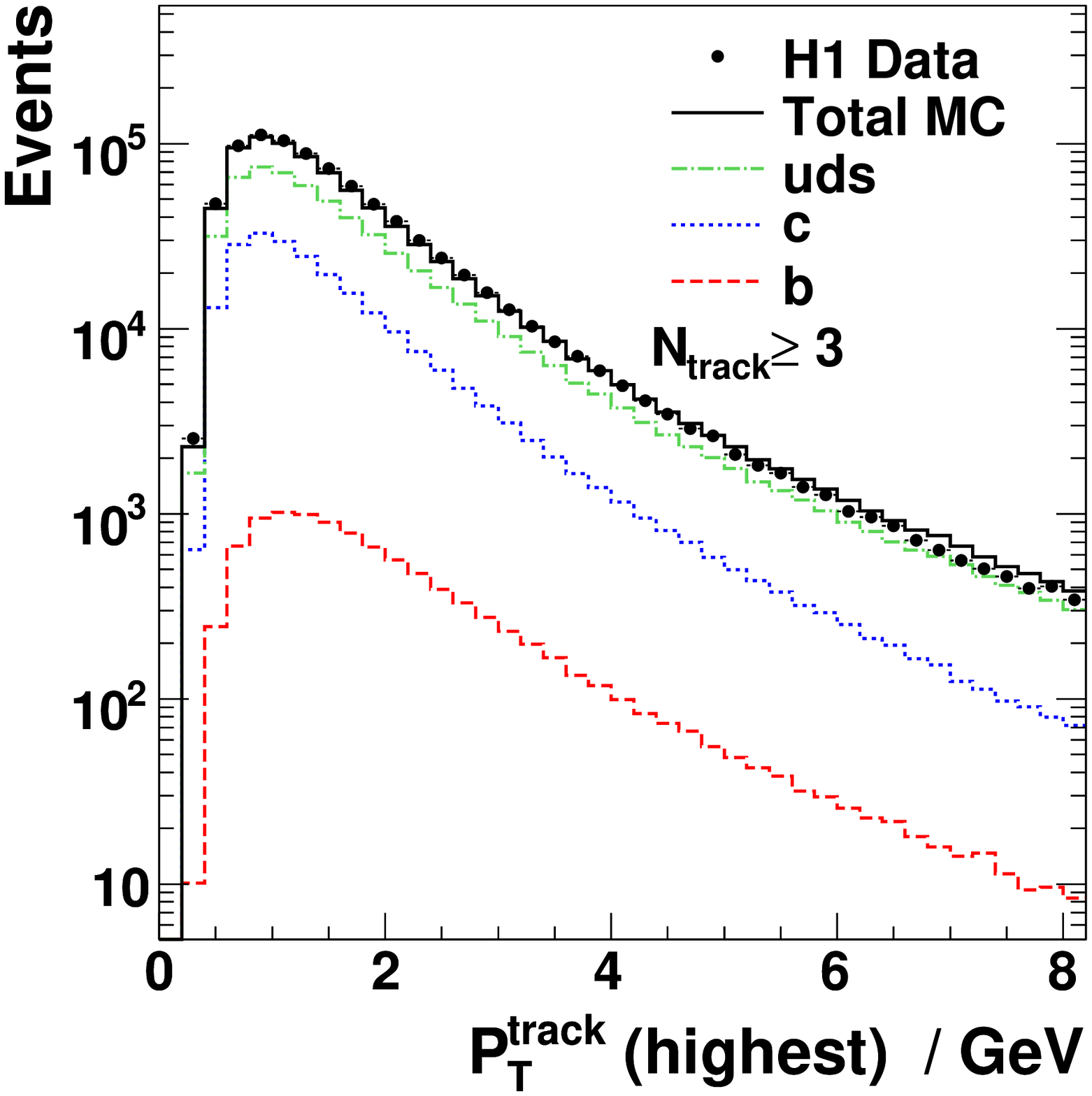}
\includegraphics[width=0.494\textwidth]{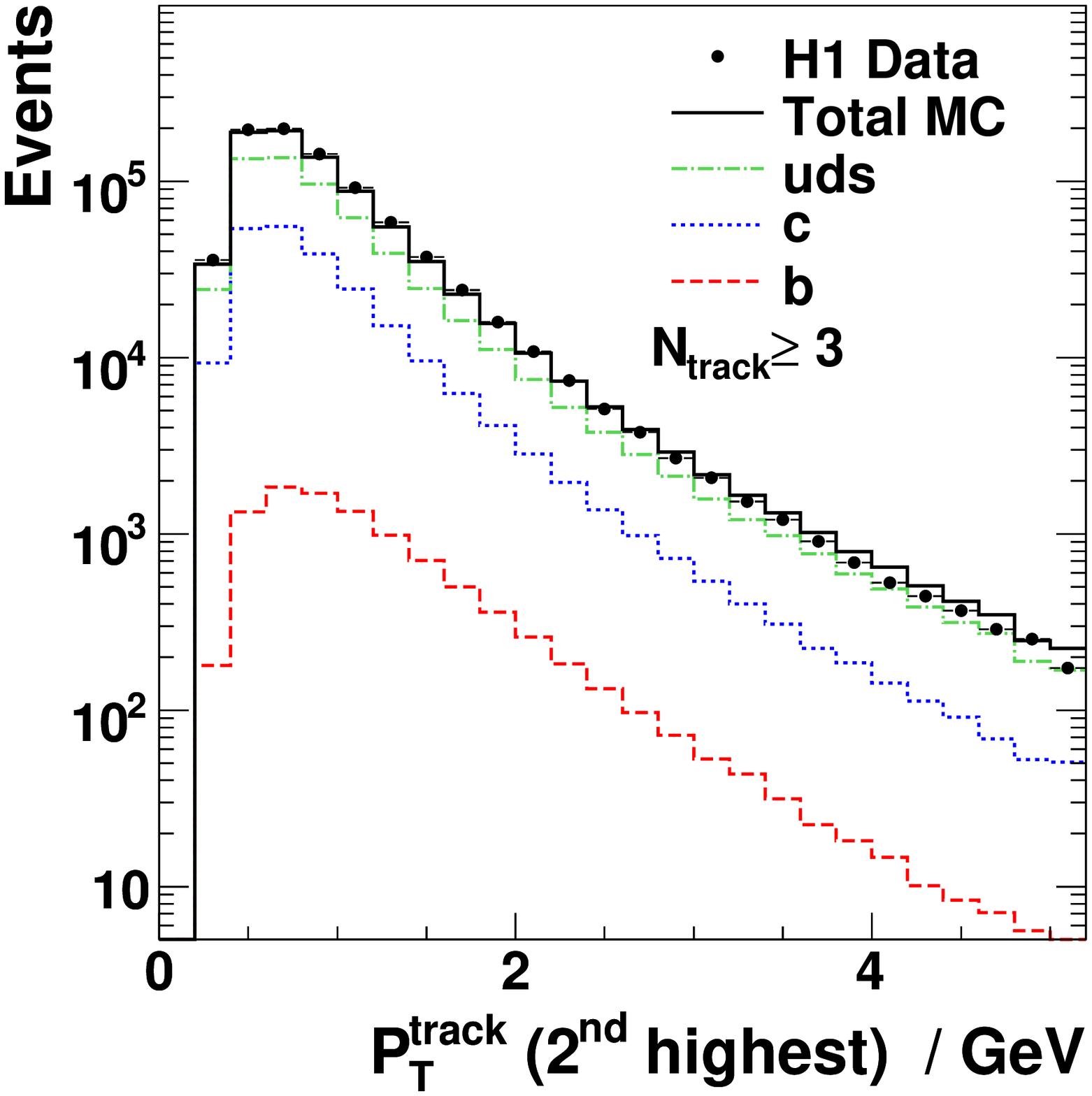}
\includegraphics[width=0.494\textwidth]{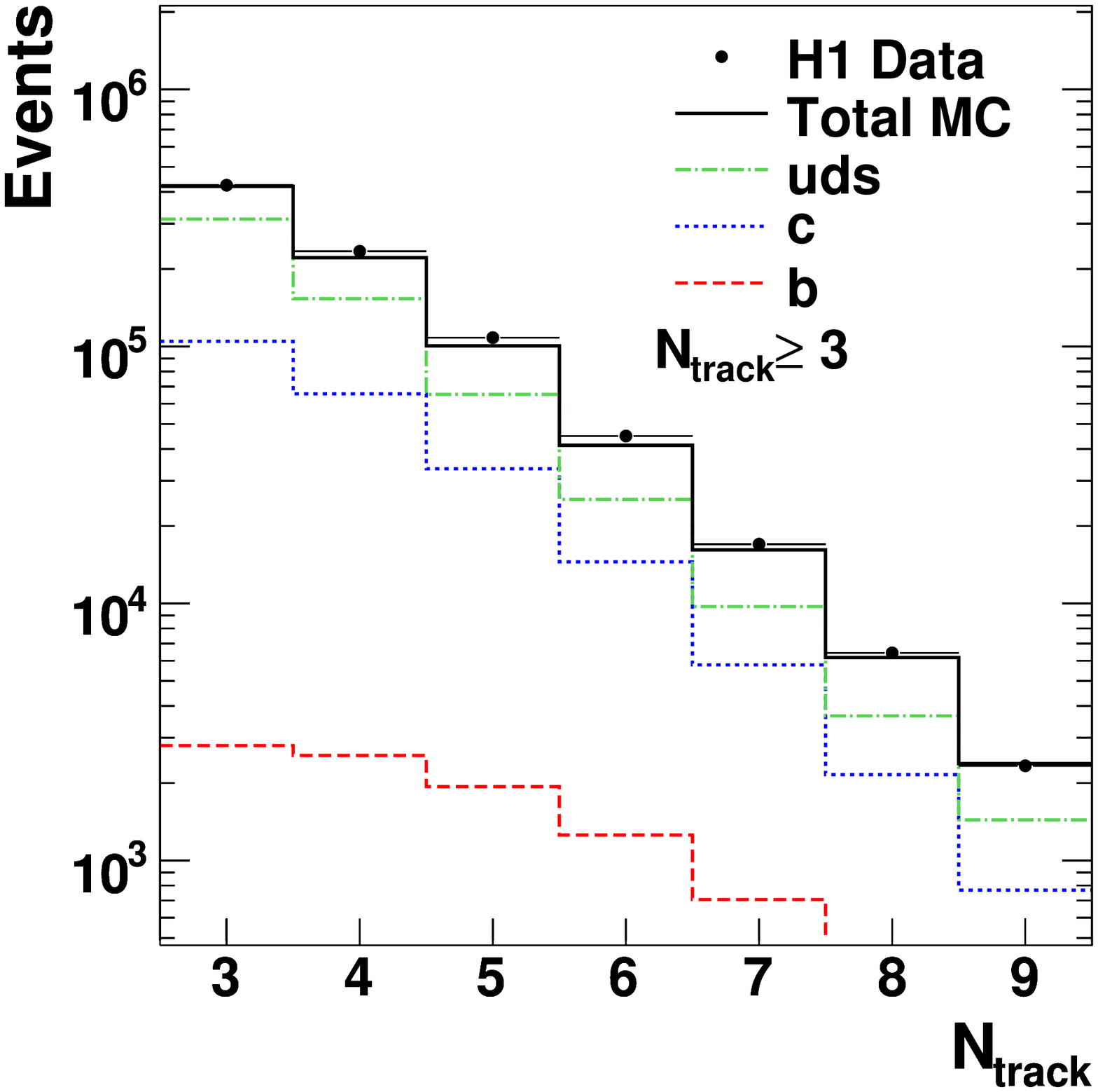}
\includegraphics[width=0.494\textwidth]{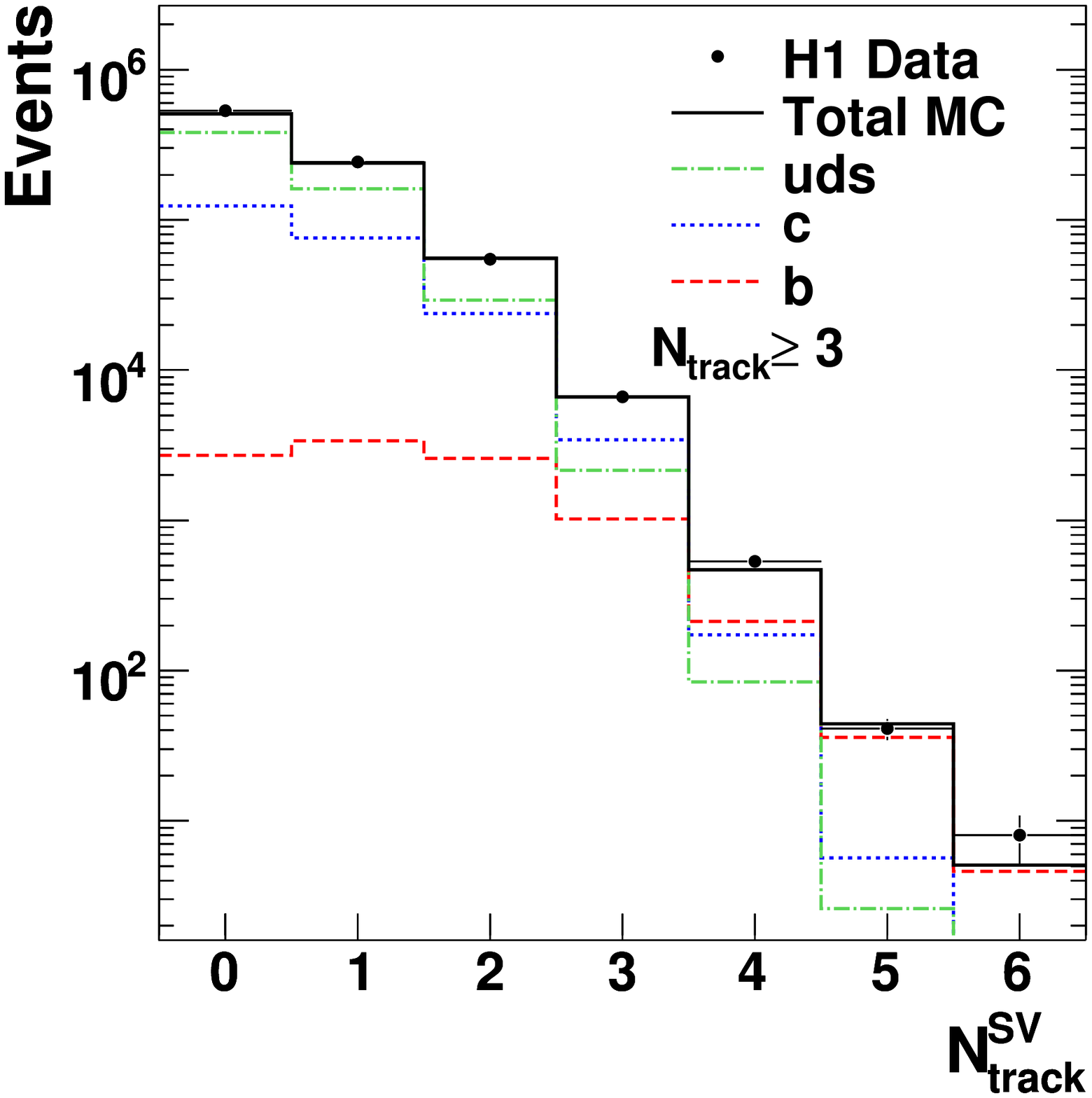}
\end{center}

 \caption{
Inputs to the NN  for events with at least 3 CST tracks:
$P_T^{\rm track}$ of the highest and second highest
transverse momentum  track, $N_{\rm track}$ and 
$N^{\rm SV}_{\rm track}$.
  Included in the figure is the expectation from the Monte
  Carlo simulation for light, $c$ and $b$ quarks. The contributions from the various
  quark flavours in the Monte Carlo are shown after applying the scale factors  $\rho_l$, $\rho_c$ and $\rho_b$ obtained from the fit to the complete data sample.}
  \label{fig:nninputsb}
\end{figure}

\begin{figure}[htb]
\begin{center}
\includegraphics[width=0.494\textwidth]{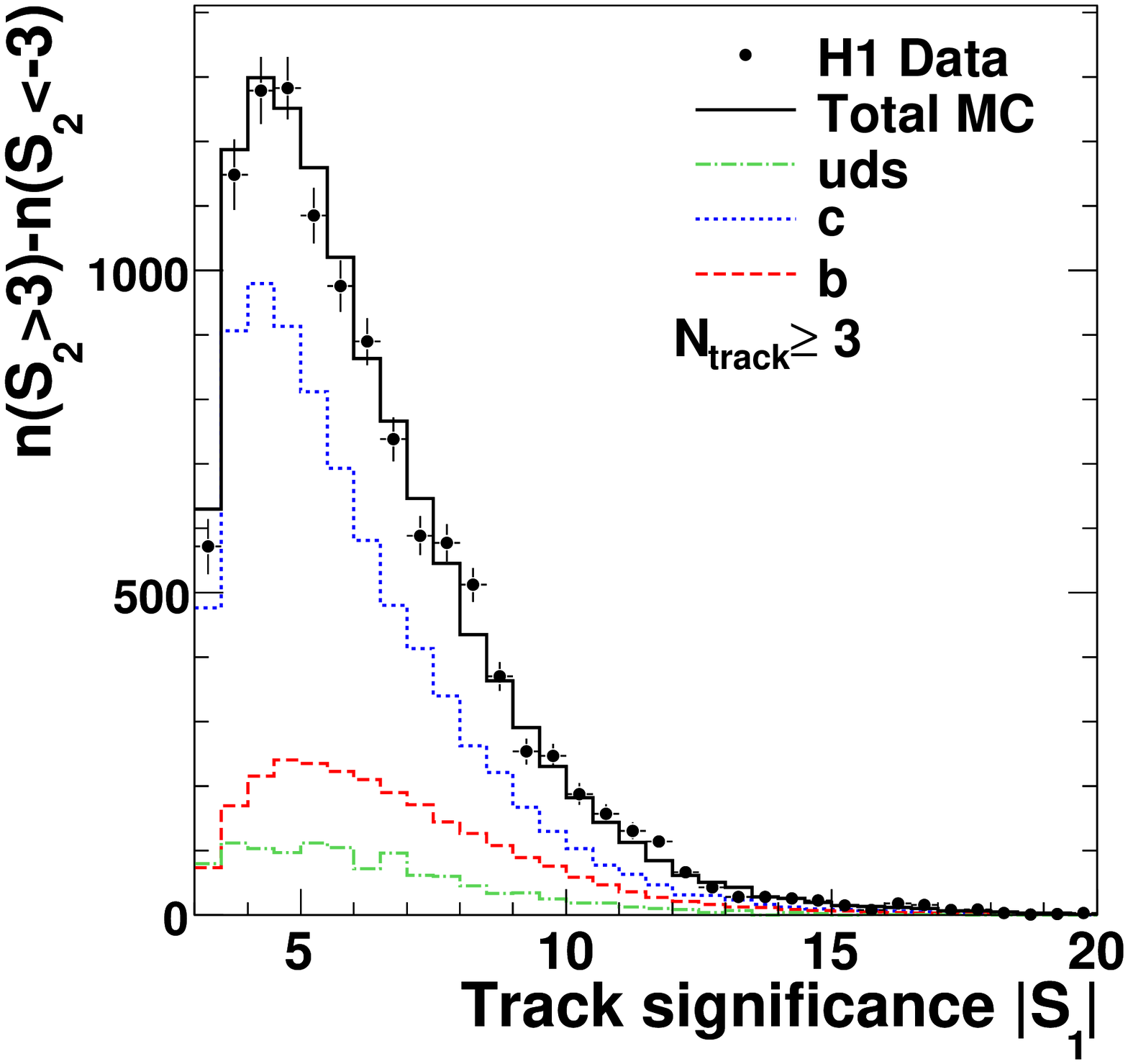}
\includegraphics[width=0.494\textwidth]{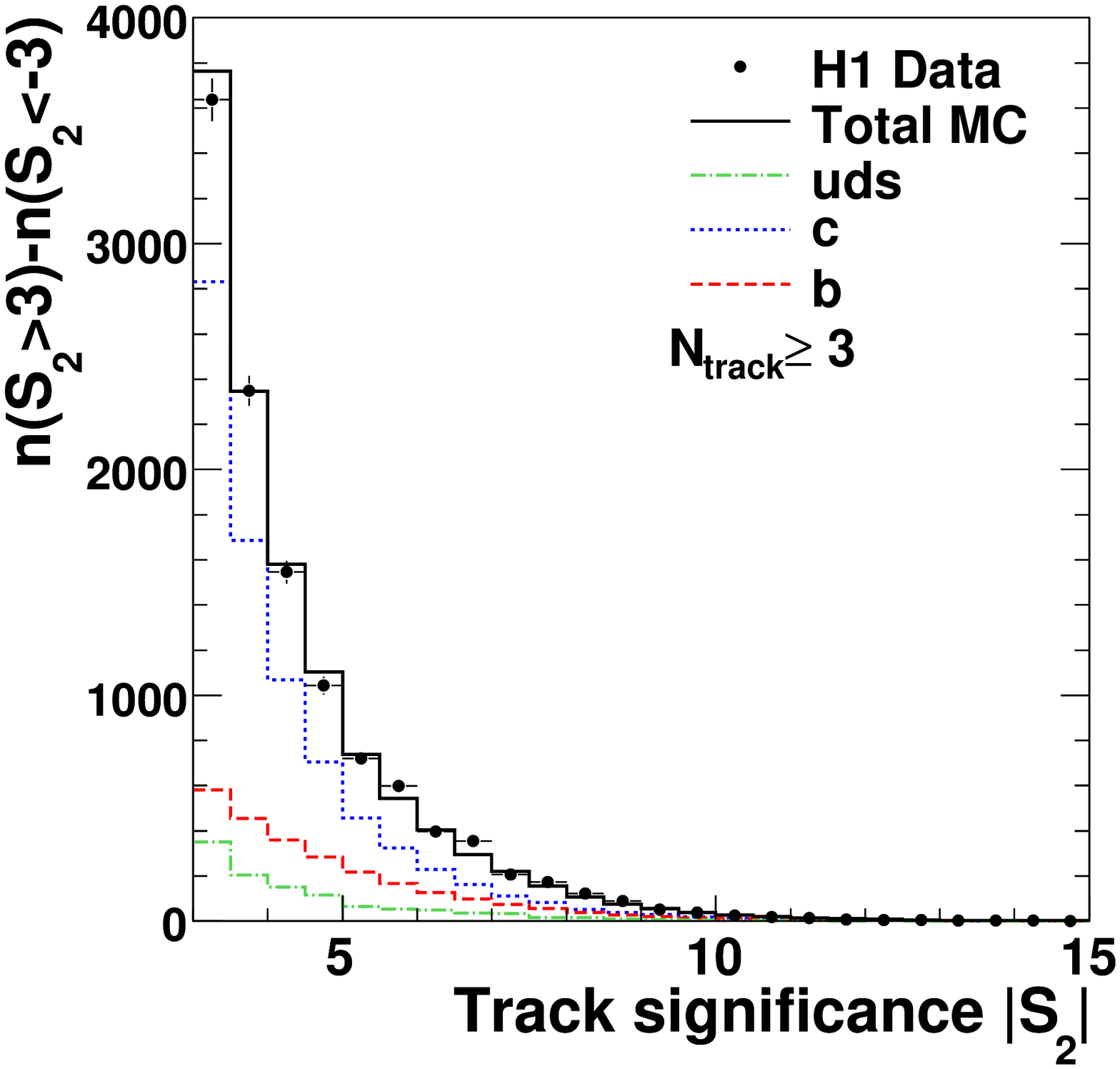}
\includegraphics[width=0.494\textwidth]{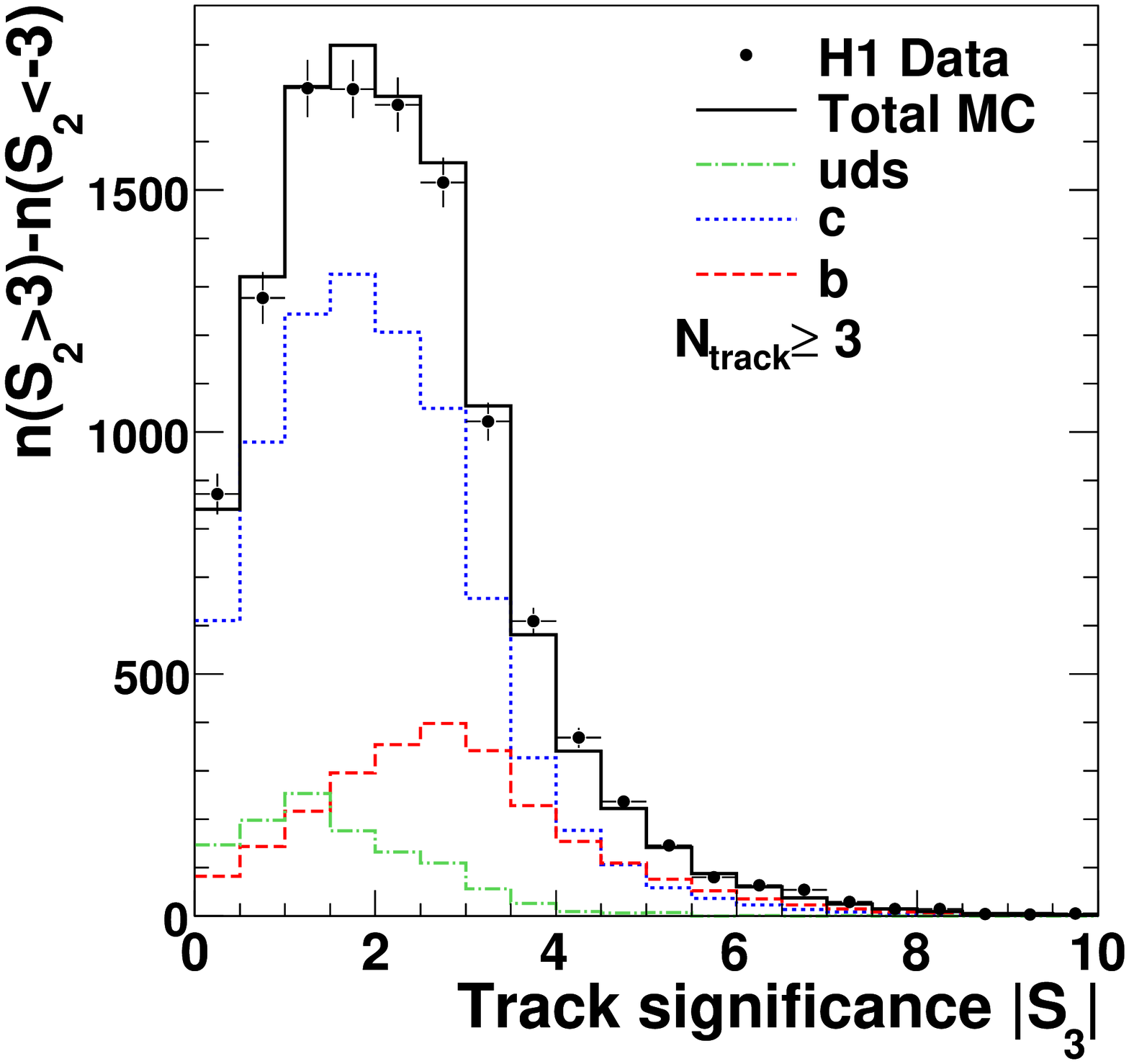}
\includegraphics[width=0.494\textwidth]{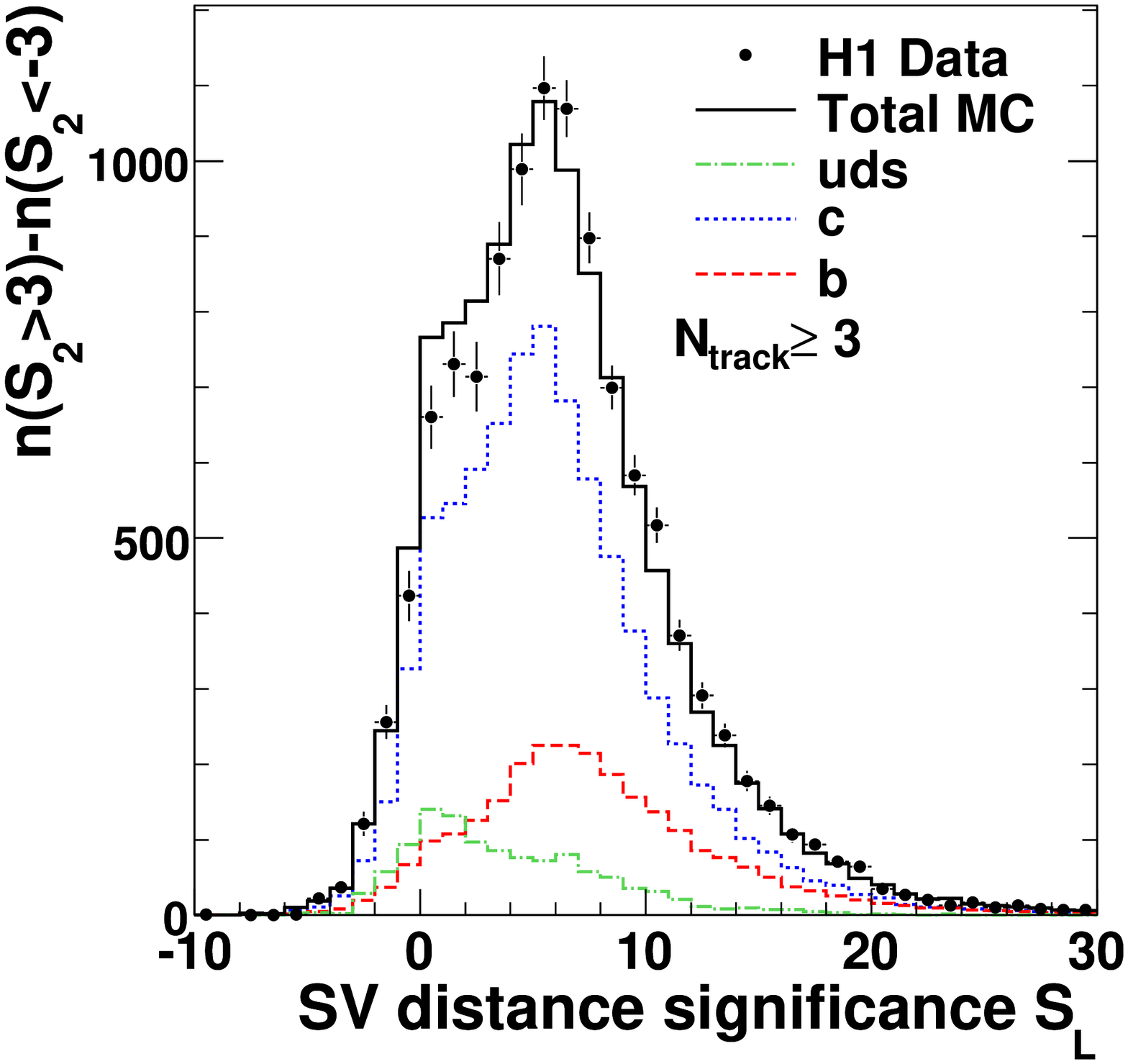}
\end{center}

 \caption{The subtracted
 distributions for $|S_1|$, $|S_2|$, $|S_3|$ and $S_L$. 
Each plot shows the difference
between the distributions for those events
with $S_2>3$ and the corresponding distribution with $S_2<-3$.
Included in the figure is the expectation from the Monte
Carlo simulation for light, $c$ and $b$ quarks. The contributions from the various
  quark flavours in the Monte Carlo are shown after applying the scale factors  $\rho_l$, $\rho_c$ and $\rho_b$  obtained from the fit to the complete data sample.}
  \label{fig:nninputsnega}
\end{figure}

\begin{figure}[htb]
\begin{center}
\includegraphics[width=0.494\textwidth]{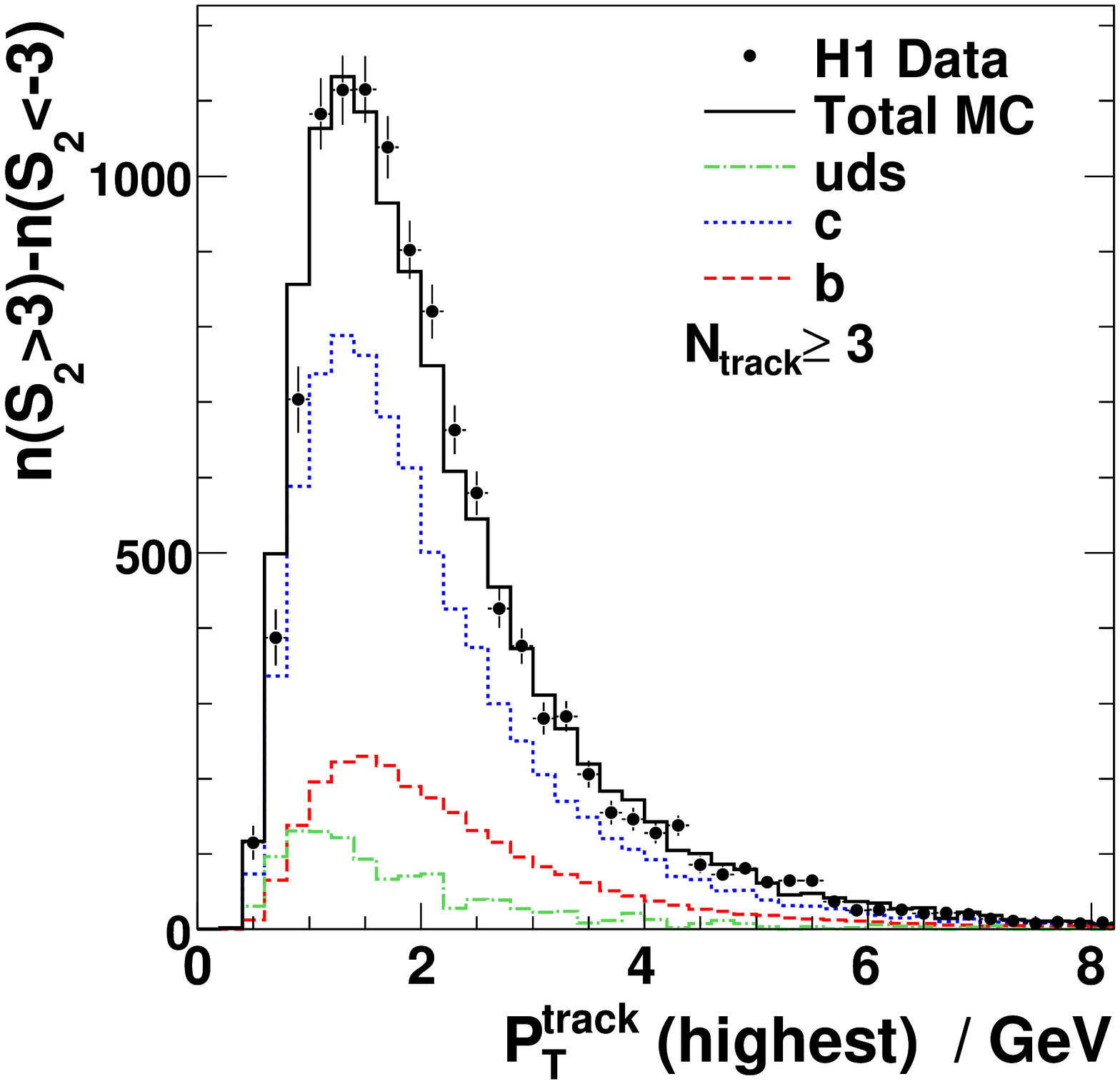}
\includegraphics[width=0.494\textwidth]{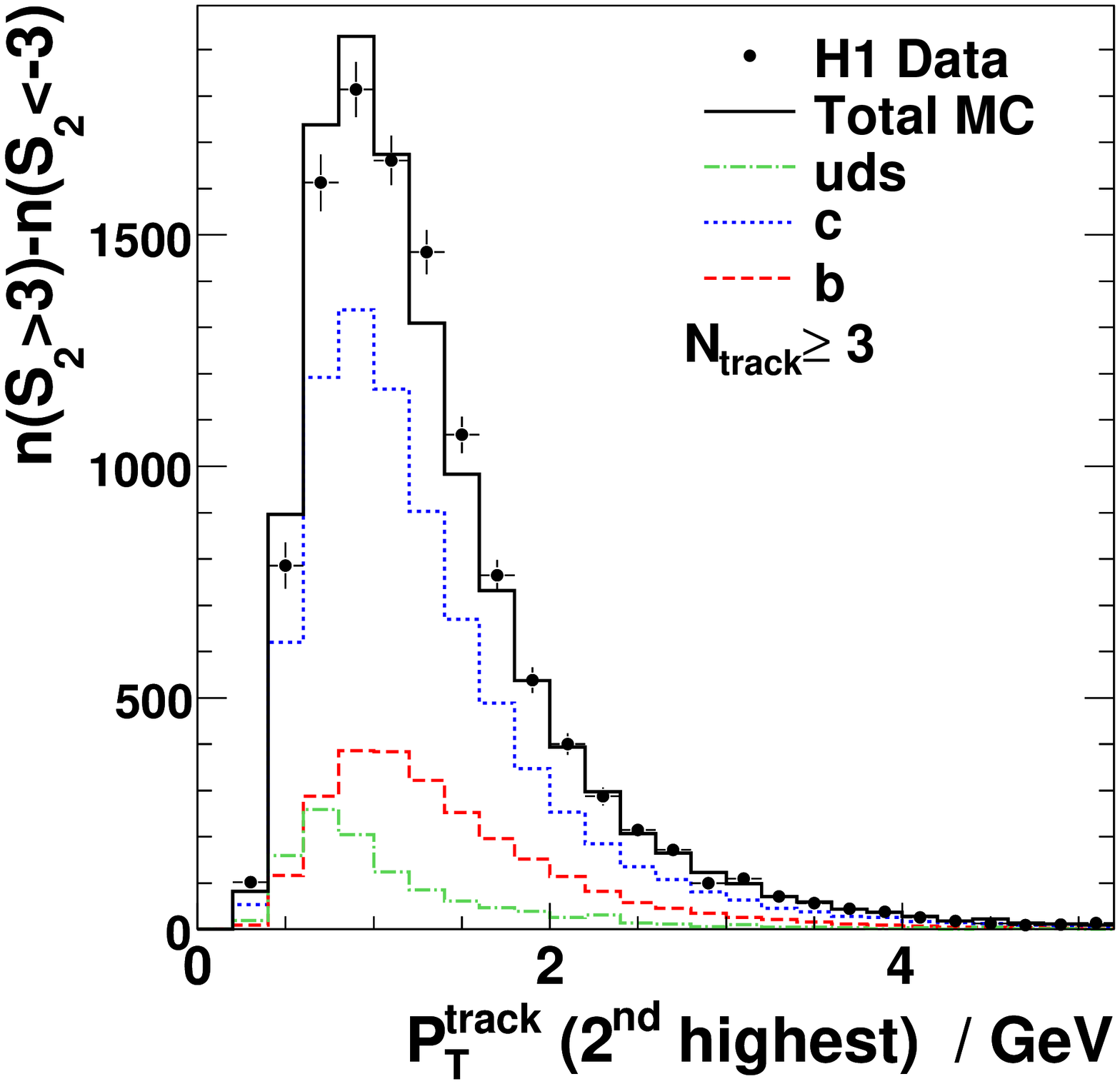}
\includegraphics[width=0.494\textwidth]{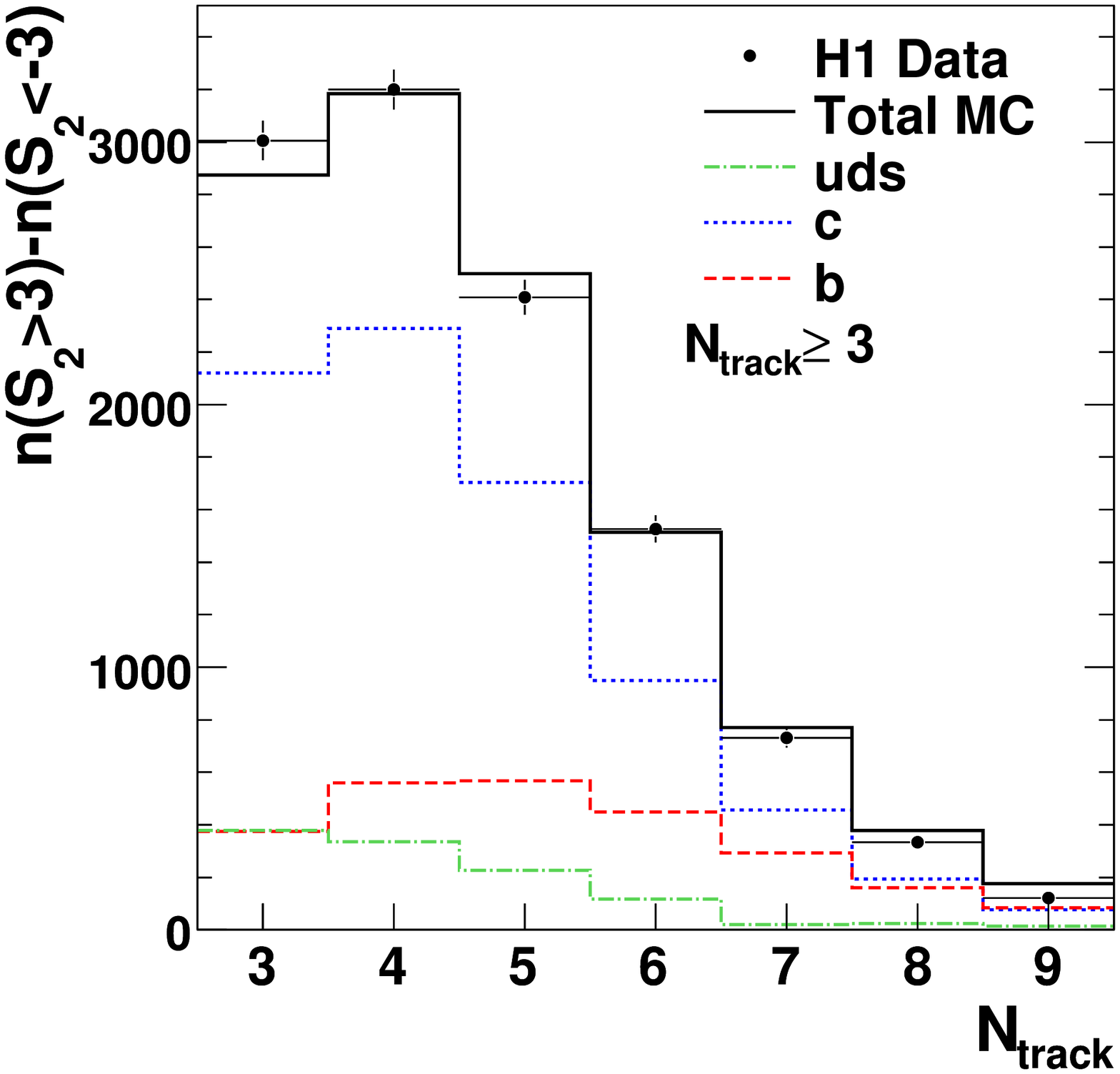}
\includegraphics[width=0.494\textwidth]{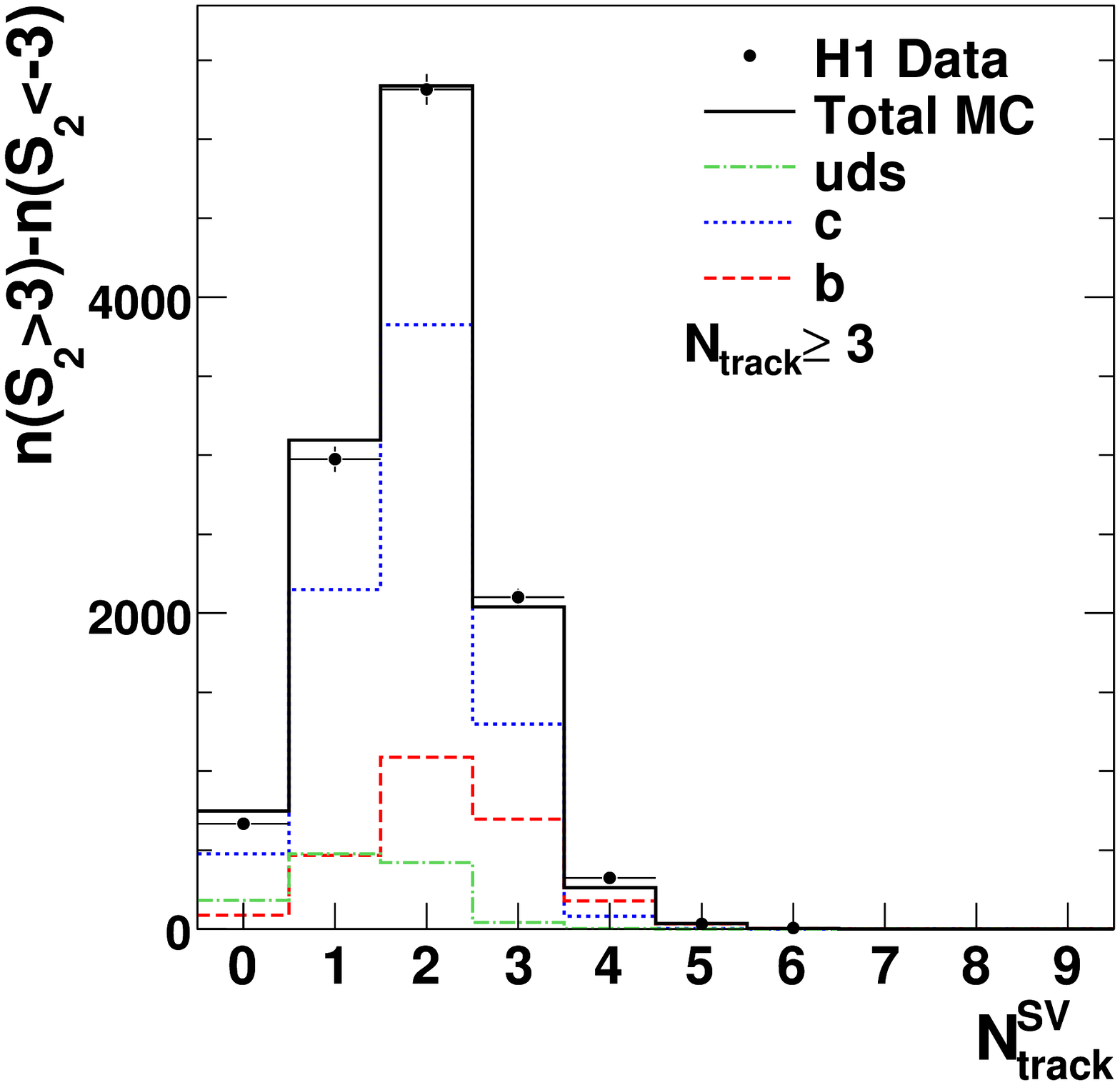}
\end{center}

 \caption{The subtracted
 distributions for $P_T^{\rm track}$ of the highest and second highest
transverse momentum  track, $N_{\rm track}$ and 
$N^{\rm SV}_{\rm track}$. Each plot  shows the difference
between the distributions  for those events
with $S_2>3$ and the corresponding distribution with $S_2<-3$.
  Included in the figure is the expectation from the Monte
  Carlo simulation for light, $c$ and $b$ quarks. The contributions from the various
  quark flavours in the Monte Carlo are shown after applying the scale factors  $\rho_l$, $\rho_c$ and $\rho_b$  obtained from the fit to the complete data sample.}
  \label{fig:nninputsnegb}
\end{figure}

The NN has inputs of $S_1$, $S_2$, $S_3$, $S_L$, $N_{\rm track}$, 
$N^{\rm SV}_{\rm track}$ and $P_T^{\rm track}$ of the CST tracks with the
highest and second highest transverse momentum. The NN has one hidden
layer with $5$ nodes. It was trained using a sample of around $5000$ Monte
Carlo $b$ events as `signal' and a similar number of Monte Carlo $c$
events as `background'. No attempt is made to discriminate against the
light quark events since their impact is minimized by the
subtraction procedure described below.  The same NN is used for all
$y$--$Q^2$ bins.  The distributions of the NN inputs are shown in
figures~\ref{fig:nninputsa} and
\ref{fig:nninputsb}. These distributions are dominated by light quark
events. It can be seen that the Monte Carlo simulation gives a
reasonable description of these distributions.  It is also apparent
that these distributions have good separation power between light, $c$
and $b$ events. The decrease in events around zero for the $S_1$ and
$S_2$ distributions is due to the requirement that $|S_1|> |S_2|>
|S_3|$.  In order to see how well the Monte Carlo simulation describes
the heavy flavour contribution of the NN input distributions, the
distribution for those events with $S_2>3$ is taken and the
distribution for those events with $S_2<-3$ is subtracted from
it. This has the effect of greatly reducing the light quark
distribution which is almost symmetric in $S_2$. This subtraction
method can be used for any distribution.  The subtracted distributions
of the NN inputs are shown in figures~\ref{fig:nninputsnega}
and~\ref{fig:nninputsnegb}.  The Monte Carlo simulation gives a good
description of these distributions. It can be seen that $b$ events
tend to have a higher track multiplicity and more tracks at higher
$P_T$. Other distributions are checked in a similar way. Distributions
of variables related to the struck quark,  $P_T^q$ and $\eta^q$, as well as
the kinematic variables $\log Q^2$ and  $\log x$ are shown in
figure~\ref{fig:controlneg} for $N_{\rm track} \ge 2$. The Monte Carlo
simulation gives a good description of these distributions.

\begin{figure}[htb]
\includegraphics[width=0.9\textwidth]{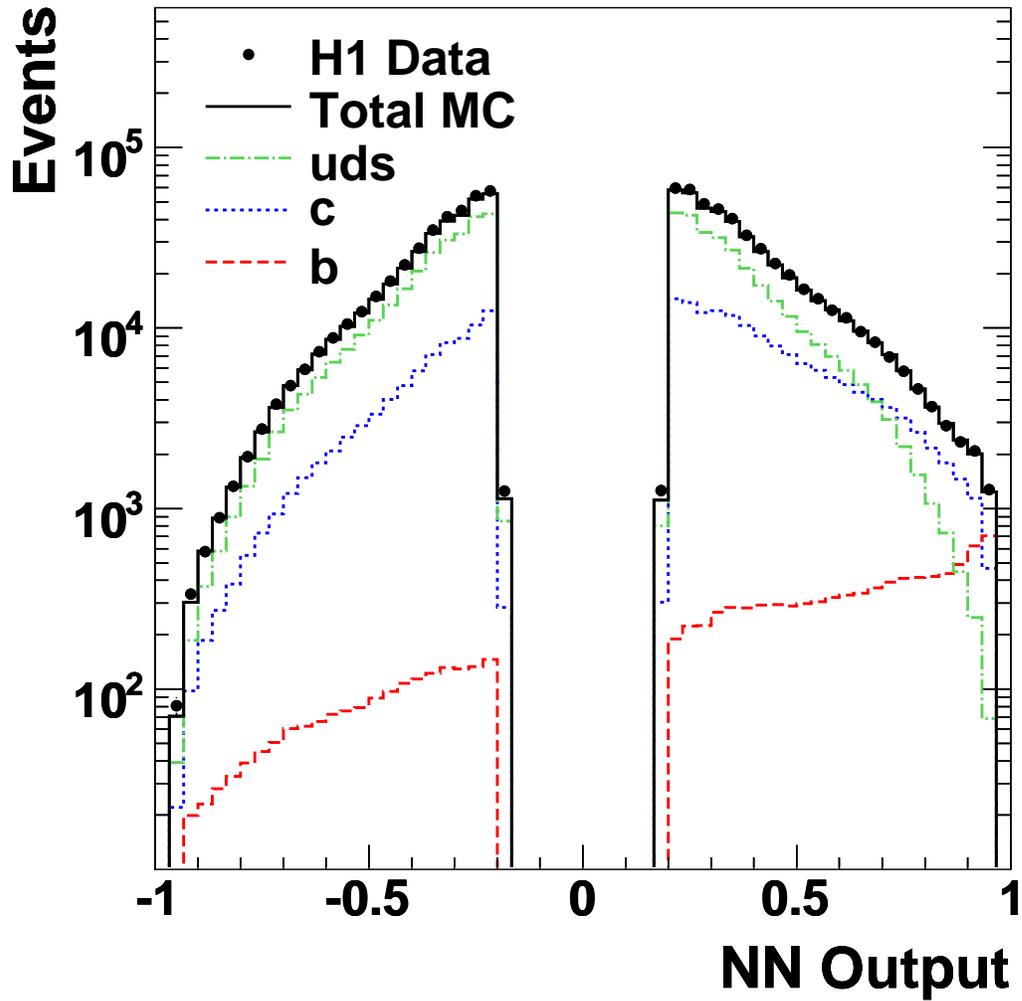}

 \caption{The output of  the NN.
  Included in the figure is the expectation from the Monte
  Carlo simulation for light, $c$ and $b$ quarks. The contributions from the various
  quark flavours in the Monte Carlo are shown after applying the scale factors  $\rho_l$, $\rho_c$ and $\rho_b$  obtained from the fit to the complete data sample.}
  \label{fig:nnoutput}
\end{figure}

The output of the NN is shown in figure~\ref{fig:nnoutput}. It gives
output values in the range from about $0.2$ to $0.95$.  The NN output
is signed according to the sign of $S_1$. It can be seen that: the
light quark distribution is approximately symmetric and peaks towards
low absolute values; the $c$ and $b$ distributions are asymmetric with
more positive than negative entries; the $b$ events are peaked towards
$1$, whereas the $c$ events are peaked towards $0$. The Monte Carlo
simulation gives a good description of the distribution. 

\begin{figure}[htb]

{  \includegraphics[width=0.494\textwidth]{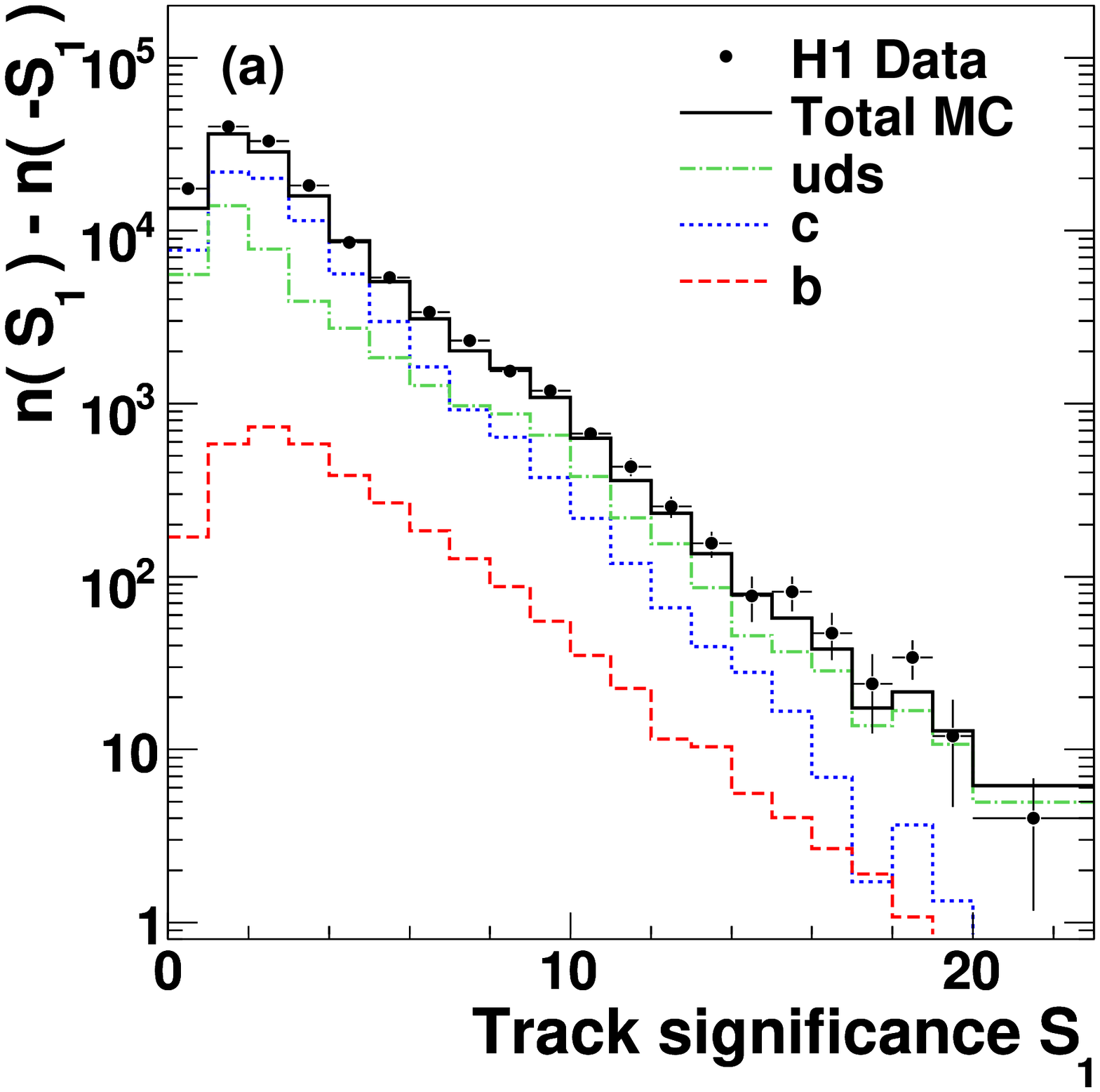}}
{  \includegraphics[width=0.494\textwidth]{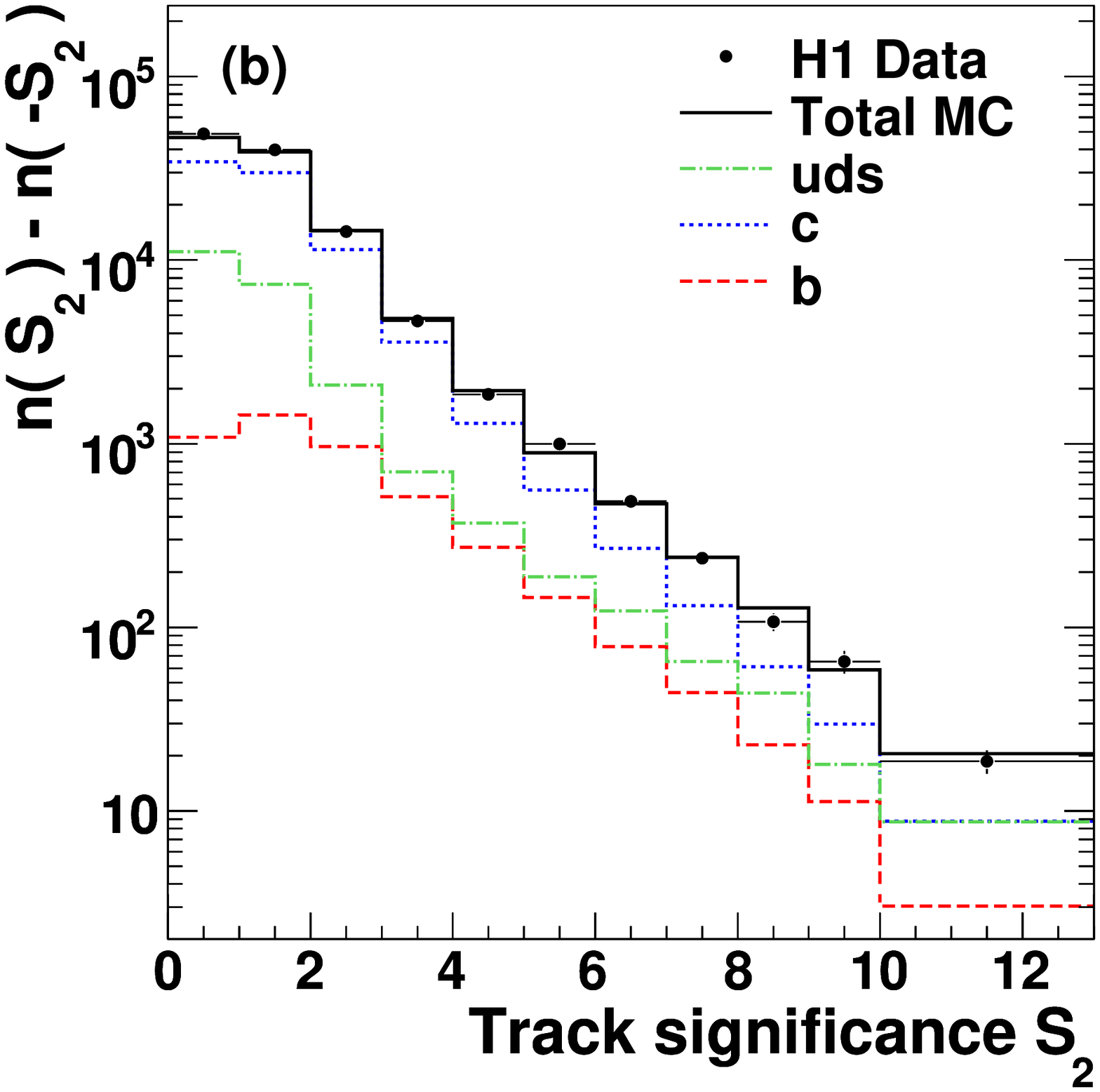}}
{  \includegraphics[width=0.494\textwidth]{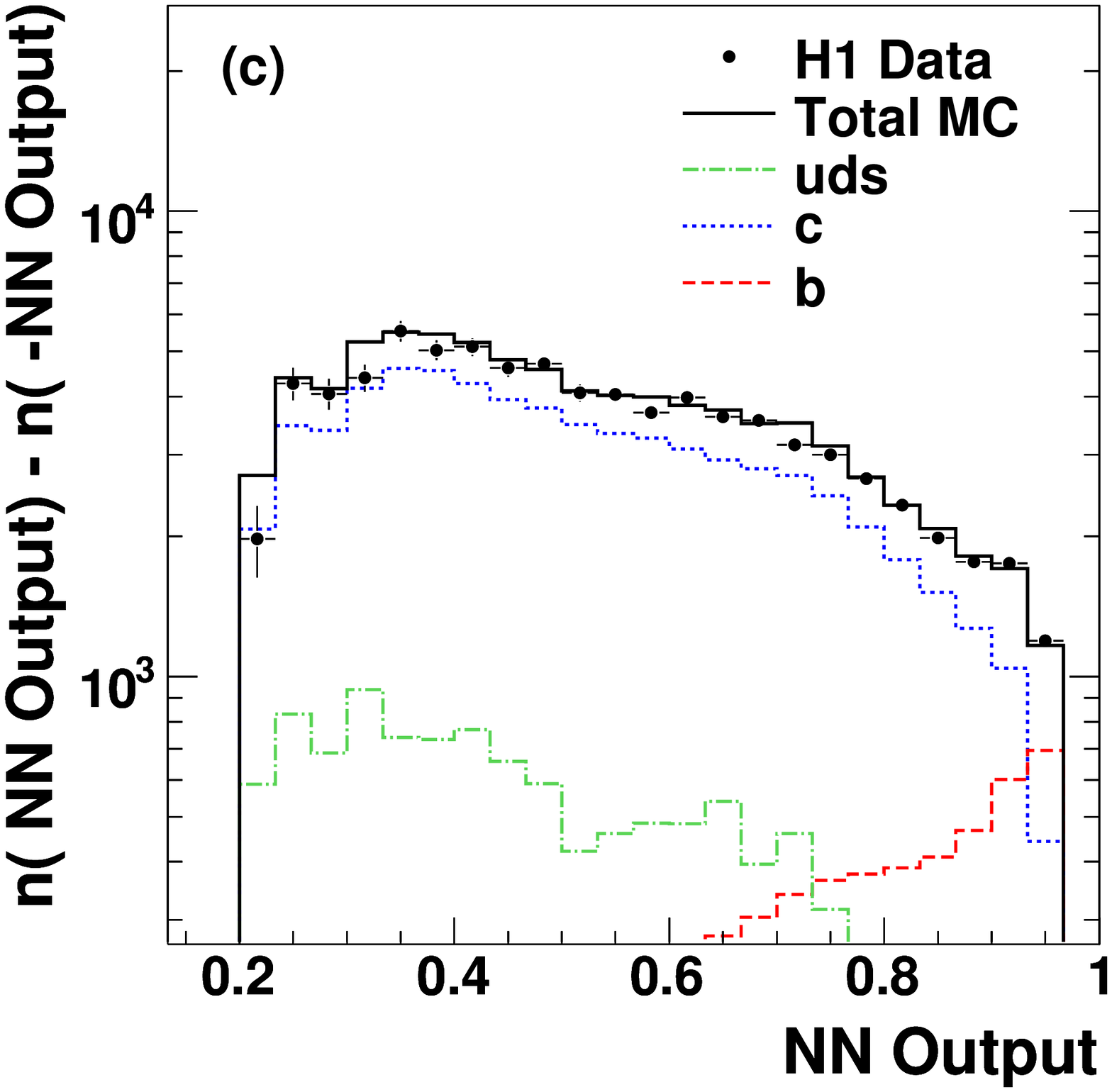}}

 \caption{The subtracted
  distributions of (a) $S_1$ and (b) $S_2$ and (c) NN output.
  Included in the figure is
  the result from the fit to the complete data sample of the Monte Carlo distributions
of the various quark flavours to obtain the scale factors $\rho_l$, $\rho_c$ and $\rho_b$.}

  \label{fig:s1s2nnnegsub}

\end{figure}

Since the light quark $S_1$, $S_2$ and NN distributions are nearly
symmetric around zero the sensitivity to the modelling of the light
quarks can be reduced by subtracting the contents of the negative bins
from the contents of the corresponding positive bins. The subtracted
distributions are shown in figure~\ref{fig:s1s2nnnegsub}. The
resulting distributions are dominated by $c$ quark events, with a $b$
quark fraction increasing towards the upper end of the
distributions. The light quarks contribute a small fraction, although
this fraction is larger than in~\cite{Aktas:2005iw}, mainly due to the
lower $Q^2$ and $P_T^{\rm track}$ selections in the present analysis.

The fractions of $c$, $b$ and light quarks of the data are extracted
in each $y$--$Q^2$ interval using a least squares simultaneous fit to
the subtracted $S_1$, $S_2$ and NN distributions (as in
figure~\ref{fig:s1s2nnnegsub}) and the total number of inclusive
events before any CST track selection. Only those bins in the
significance distributions which have at least $25$ events before
subtraction are considered in the fit, since Gaussian errors are assumed.  
The last fitted bin of the significance distributions, which
usually has the lowest statistics, is made $3$ times as wide as the
other bins (see figure~\ref{fig:s1s2nnnegsub}).  If any of the bins
before subtraction within the NN output range contain less than $25$ events
the bin size is doubled, which ensures all bins contain at least
 $25$ events.

The light, $c$ and $b$   Monte Carlo simulation samples are used as
templates.  The Monte Carlo light, $c$ and $b$ contributions in each
$y$--$Q^2$ interval are scaled by factors $\rho_l$, $\rho_c$ and
$\rho_b$, respectively, to give the best fit to the observed
subtracted $S_1$, $S_2$ and NN distributions and the total number of
inclusive events in each $y-Q^2$ interval. Only the statistical errors 
of the data and Monte
Carlo simulation are considered in the fit.  The fit to the subtracted
significance and NN distributions mainly constrains $\rho_c$ and
$\rho_b$, whereas the overall normalisation constrains $\rho_l$.

Since the error on $\rho_c$ is much smaller than that of $\rho_b$ the
$c$ cross section is measured in more $y$--$Q^2$ intervals
than the $b$ cross section. Therefore two sets of fits are performed,
one with a fine binning of $29$ bins and the other with a coarse
binning of $12$ bins. The two sets of fits are performed in an
identical manner. The results of the two sets of fits are listed in
tables~\ref{tab:rhotabledeffine} and~\ref{tab:rhotabledefcoarse}.
Also included in the tables are the $\chi^2/$n.d.f. evaluated using
statistical errors only. Acceptable values are obtained for all
bins. The fit is also performed to the complete data sample and shown in
 figure~\ref{fig:s1s2nnnegsub}. The stability of the method is checked
 by repeating
 the fit to the complete data in a variety of ways: fitting the $e^+p$
 and $e^-p$ data separately; using $S_3$ instead of the NN; using
 $S_L$ instead of the NN; using the NN alone without $S_1$ and $S_2$;
 using $S_1$ and $S_2$ without the NN; using the NN alone without
 negative subtraction.  All give consistent results within statistical
 and systematic errors.

The results of the fit in each $y$--$Q^2$ interval are converted to a
measurement of the `reduced $c$ cross section'
defined from
the differential cross section as
\begin{equation}
\tilde{\sigma}^{c\bar{c}} (x, Q^2) = \frac{{\rm d}^2\sigma^{c\bar{c}} }{{\rm d} x\,{\rm d} Q^2}  \frac {x Q^4 } {2 \pi \alpha^2 (1+ (1-y)^2)},
\end{equation}

\noindent using:
\begin{equation}
\tilde{\sigma}^{c\bar{c}} (x, Q^2) = 
\tilde{\sigma} (x, Q^2) \frac{\rho_c N^{\rm MC gen}_c}{\rho_l N^{\rm MC gen}_l+\rho_c N^{\rm MC gen}_c+\rho_b N^{\rm MC gen}_b}  
\delta_{\rm BCC},
\label{eqn:cross}
\end{equation}
where $\alpha$ is the fine structure constant evaluated with scale
$Q^2$, and $N^{\rm MC gen}_l$, $N^{\rm MC gen}_c$, and $N^{\rm MC
gen}_b$ are the number of light, $c$, and $b$ quark events generated
from the Monte Carlo simulation in each bin. The inclusive reduced
cross section $\tilde{\sigma} (x, Q^2)$ is taken from H1 measurements:
Tables~17 and 19 from \cite{h1newf2paper1}, Tables~10 and 11 from
\cite{h1newf2paper2} and Table~11  from \cite{H19900NCCC}.  A bin
centre correction $\delta_{\rm BCC}$ is applied using a NLO QCD
expectation for $\tilde{\sigma}^{c\bar{c}}$ and $\tilde{\sigma}$ to
convert the bin averaged measurement into a measurement at a given
$x$--$Q^2$ point.  This NLO QCD expectation is calculated from the
results of a fit similar to that performed in~\cite{Adloff:1999ah} but
using the massive scheme to generate heavy flavours.

The reduced cross section is corrected using the Monte Carlo simulation for
pure QED radiative effects. The photoproduction background is not
subtracted, which, due to the method used to calculate the reduced cross
sections, means that the fraction of $c$ and $b$ events in the
photoproduction background is assumed to be the same as in the DIS
data. In most of the $y$ range the background from photoproduction
events is negligible and does not exceed $4\%$ in any $y$--$Q^2$
interval used in this analysis.  Events that contain $c$ hadrons via
the decay of $b$ hadrons are not included in the definition of the reduced $c$
cross section. The reduced $b$ cross section is evaluated in the
same manner.

\section{Systematic Uncertainties}
\begin{figure}[htb]
\begin{center}
\includegraphics[width=0.494\textwidth]{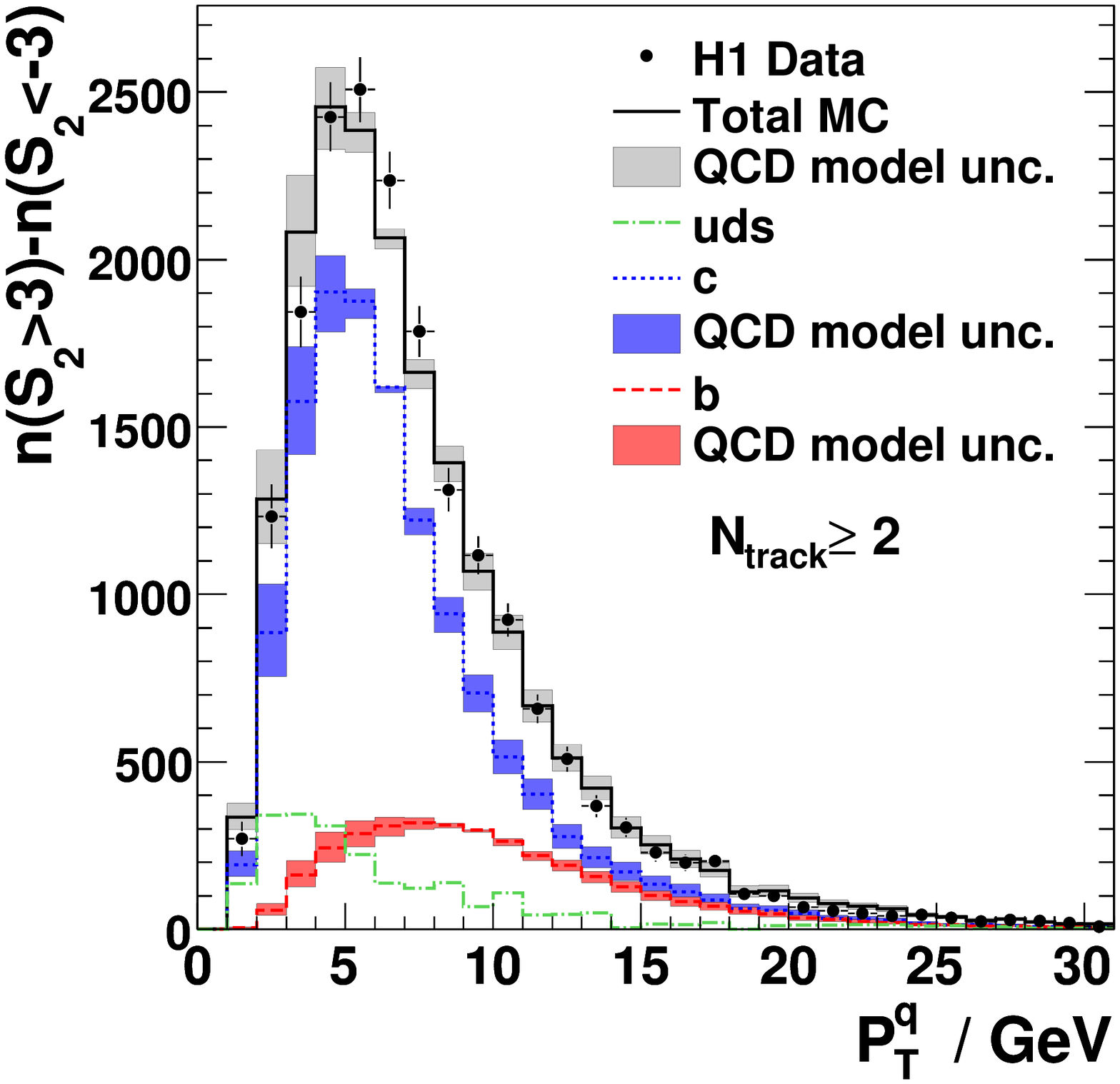}
\includegraphics[width=0.494\textwidth]{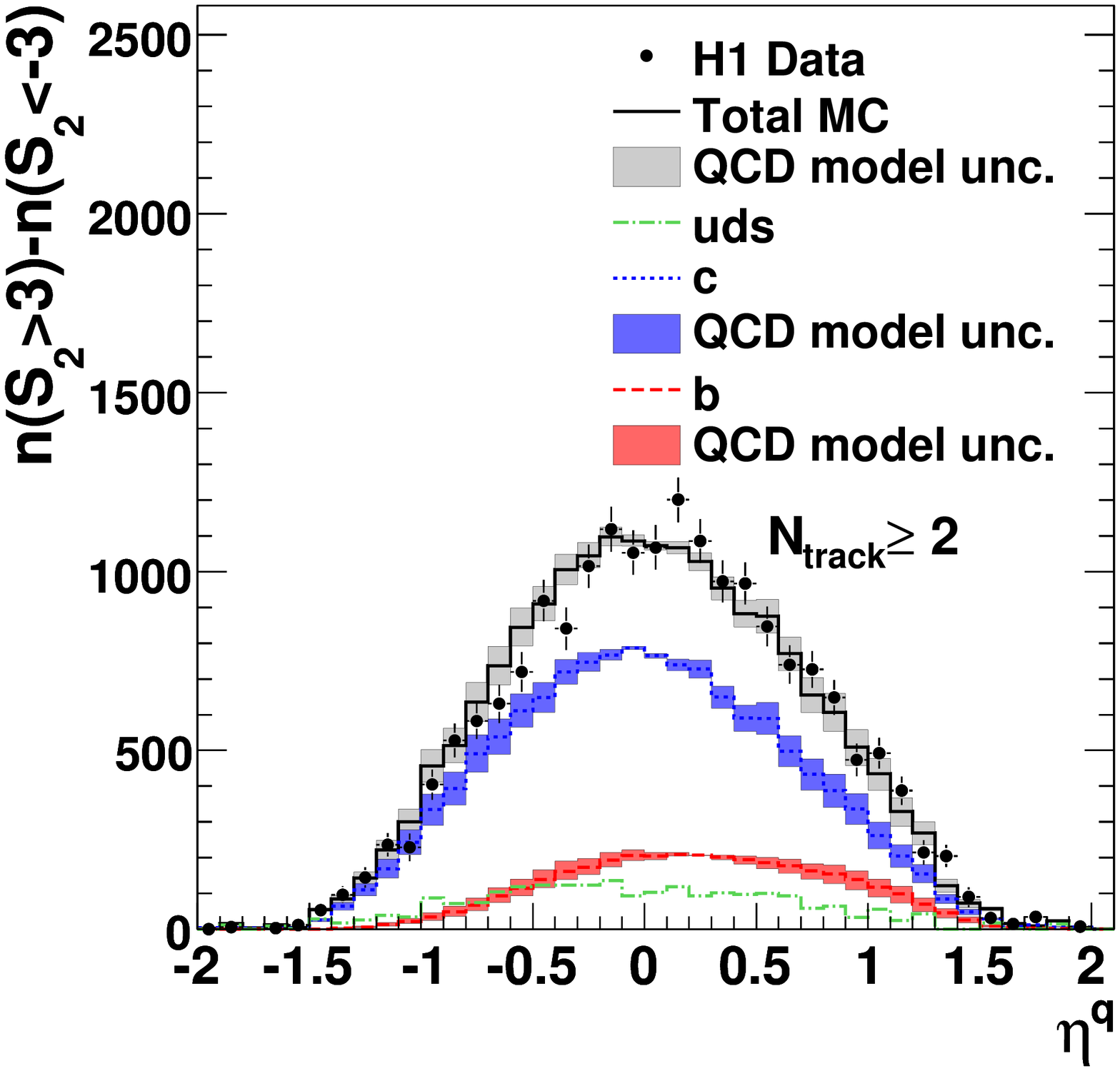}
\includegraphics[width=0.494\textwidth]{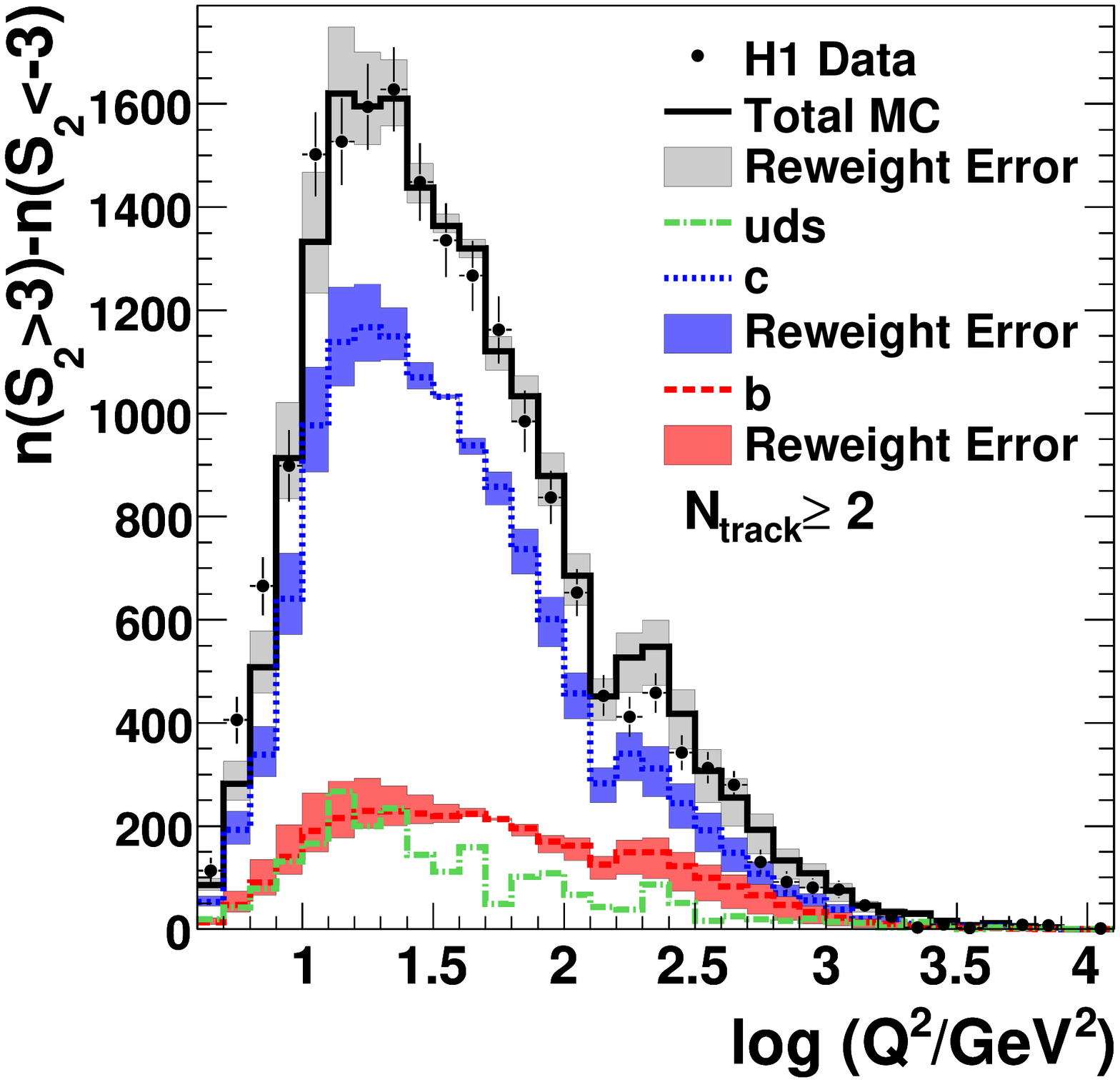}
\includegraphics[width=0.494\textwidth]{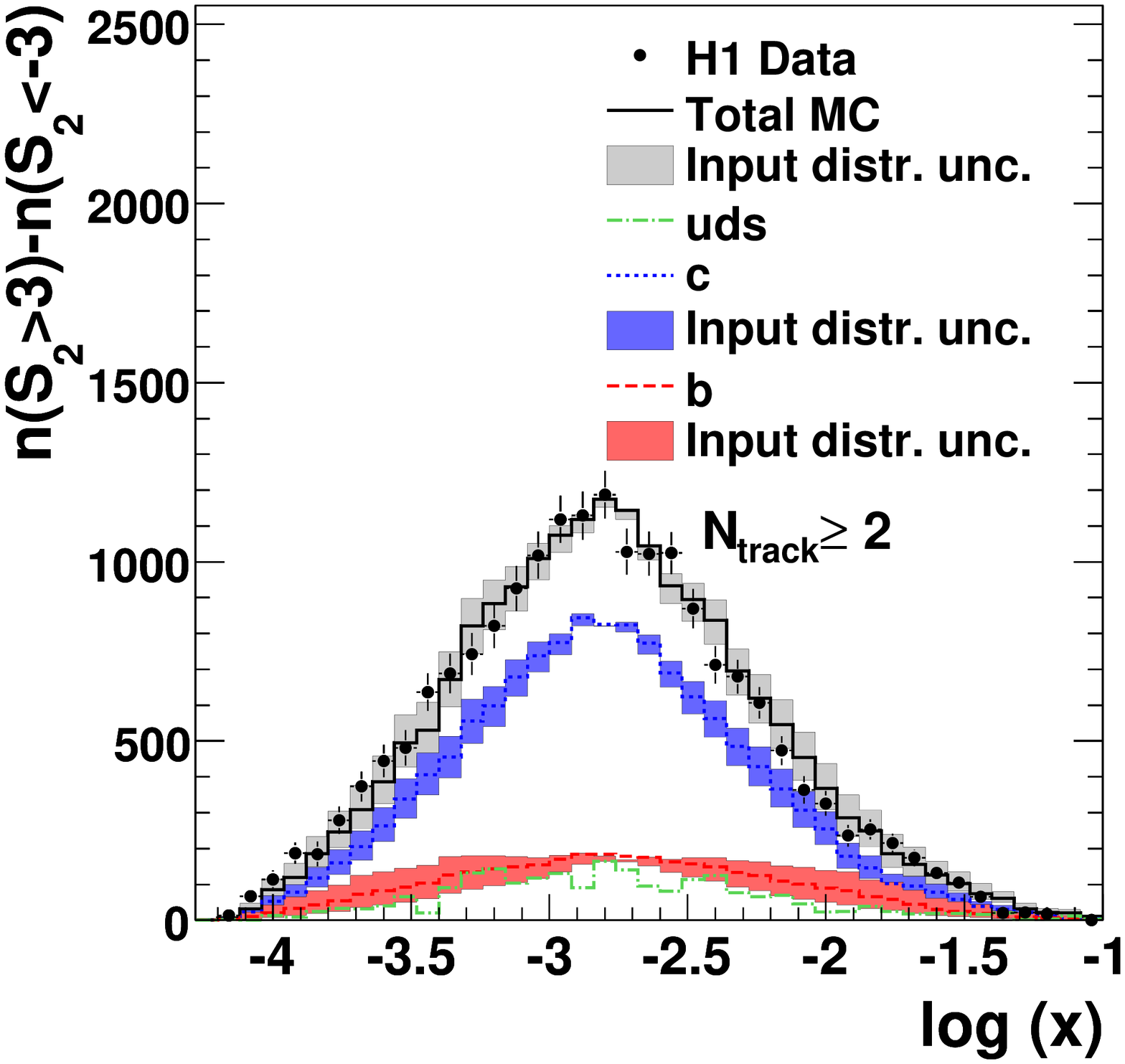}
\end{center}

 \caption{The subtracted distributions for 
 $P_T^q$, $\eta^q$, $\log Q^2$, $\log x$. Each plot shows the
 difference between the 
 distributions for those events with $S_2>3$ and the corresponding
 distribution with $S_2<-3$.  Included in the figure is the
 expectation from the Monte Carlo simulation for light, $c$ and $b$
 quarks. The contributions from the various quark flavours in the
 Monte Carlo are shown after applying the scale factors $\rho_l$,
 $\rho_c$ and $\rho_b$ obtained from the fit to the complete data
 sample. Also shown are the variations in $P_T^q$ and $\eta^q$
 resulting from uncertainties in the QCD model of heavy quark
 production, as well as the variations in $\log Q^2$ and $\log x$,
 resulting from uncertainties on the input heavy quark structure
 functions.} \label{fig:controlneg}
\end{figure}

The following uncertainties are taken into account in order to evaluate
the systematic error.
\label{sec:systematics}
\begin{itemize}
\item An uncertainty in the $\delta$ resolution of the tracks is estimated
 by varying the resolution by an amount that encompasses
any difference between the data and simulation evaluated from figure~\ref{fig:dca}.  This was achieved by applying
an additional Gaussian smearing in the Monte Carlo simulation of
$200$~$\mu{\rm m}$ to $5\%$ of randomly selected tracks and
$12$~$\mu{\rm m}$ to the rest.

\item A  track efficiency uncertainty is assigned of $1\%$ due to the CTD and
of $2\%$ due to the CST.

\item The uncertainties on the various $D$ and $B$ meson lifetimes,
  decay branching fractions and mean charge multiplicities are
  estimated by varying the input values of the Monte Carlo simulation
  by the errors on the world average measurements.  For the branching
  fractions of $b$ quarks to hadrons and the lifetimes of the $D$ and
  $B$ mesons the central values and errors on the world averages are
  taken from\cite{pdg2006}. For the branching fractions of $c$
  quarks to hadrons the values and uncertainties are taken
  from\cite{Gladilin:1999pj}, which are consistent with measurements
  made in DIS at HERA\cite{Aktas:2004ka}. For the mean charged track multiplicities the
  values and uncertainties for $c$ and $b$ quarks are taken from
  MarkIII\cite{Coffman:1991ud} and LEP/SLD\cite{lepjetmulti}
  measurements, respectively. 

\item 
The uncertainty on the fragmentation function of the heavy quarks is
estimated by re\-weighting the events according to the longitudinal
string momentum fraction $z$ carried by the heavy hadron in the Lund
model using weights of $(1 \mp 0.7)\cdot(1-z) + z \cdot (1 \pm 0.7)$
for charm quarks and of $(1 \mp 0.5)\cdot(1-z) + z \cdot (1 \pm 0.5)$
for beauty quarks. The variations for the charm fragmentation 
are motivated by comparison of the Monte Carlo simulation with 
H1 data~\cite{h1frag}.

\item An uncertainty on the QCD model of heavy quark production
 is estimated by reweighting the jet transverse momentum and
 pseudorapidity by $(P^{\rm jet}_T/(10 ~{\rm GeV}))^{\pm 0.2}$ and $(1
 \pm \eta^{\rm jet})^{\pm 0.15}$ for charm jets and $(P^{\rm
 jet}_T/(10 ~{\rm GeV}))^{\pm 0.3}$ and $(1 \pm \eta^{\rm jet})^{\pm
 0.3}$ for bottom jets.  These values are obtained by comparing these
 variations with the measured reduced cross section for $b$ and $c$
 jets\cite{rahmatthesis}.  The effects of each of these uncertainties
 on the subtracted reconstructed distributions of $P_T^q$ and $\eta_q$
 are shown in figure~\ref{fig:controlneg}, where the data is seen to
 be consistent with the Monte Carlo simulation within the
 uncertainties.

\item The uncertainty on the asymmetry of the light quark $\delta$
  distribution is estimated by repeating the fits with the subtracted
  light quark  distributions
  (figure~\ref{fig:s1s2nnnegsub}) changed by $\pm30\%$. The light
  quark asymmetry was checked to be within this uncertainty by comparing the
  asymmetry of Monte Carlo simulation events to that of the data
  for $K^0$ candidates, in the
  region $0.1<|\delta|<0.5~{\rm cm}$, where the light quark asymmetry
  is enhanced.

\item An error on $\phi^q$ is estimated by shifting $\phi^q$ by $2^\circ$($5^\circ$) for events with (without) a
  reconstructed jet. These shifts were estimated by comparing the
  difference between $\phi^q$ and the track azimuthal angle
  in data and Monte Carlo simulation.
\item The uncertainty on the hadronic energy scale is estimated by changing
the hadronic energy by $\pm 2\%$.

\item The uncertainty in the photoproduction background is taken
as $100\%$ of the fraction of photoproduction events in each bin,
for events with $N_{\rm track} \ge 1$. 

\item Uncertainties on the acceptance and bin centre correction due to
  the input heavy quark structure functions used are estimated by
  reweighting the input $\tilde{\sigma}^{c\bar{c}}$ distribution by
  $x^{\pm0.1}$ and $1 \pm 0.2 \ln [Q^2/(10~{\rm GeV}^2)]$ and
  $\tilde{\sigma}^{b\bar{b}}$ by $x^{\pm 0.3}$ and $1 \pm 0.4 \ln
  [Q^2/(10~{\rm GeV}^2)]$. The range of variation of the input
  structure functions was estimated by comparing to the measured
  values obtained in this analysis. The effects of each of these
  uncertainties on the subtracted distributions of $\log Q^2$ and
  $\log x$ are shown in figure~\ref{fig:controlneg}, where the data
  is seen to be consistent with the Monte Carlo simulation within
  the uncertainties.

\end{itemize}

The above systematic uncertainties are evaluated by making the changes
described above to the Monte Carlo simulation and repeating the
procedure to evaluate the reduced $c$ and $b$ cross sections, including the
fits.  The uncertainties are evaluated separately for each $x$--$Q^2$
measurement.

Additional contributions to the systematic error due to the inclusive
cross section are taken from the corresponding $x$--$Q^2$ bin in
\cite{h1newf2paper1, h1newf2paper2, H19900NCCC}, since the
measurements are normalised to the inclusive cross section measurements (see
equation~\ref{eqn:cross}).  The error due to the inclusive DIS selection
includes a $1.1$--$1.5\%$ uncertainty on the luminosity measurement; an
uncertainty on the scattered electron polar angle of $0.2$--$3.0~{\rm
mrad}$ and energy of $0.2$--$2.0\%$ depending on the energy and angle;
typically $<1\%$ combined error due to trigger and scattered electron
identification efficiency; and a $0.5$--$1.0\%$ uncertainty on the reduced cross
section evaluation due to QED radiative corrections.

A detailed list of the systematic effect
on each reduced cross section measurement is given in tables~\ref{tab:sig}
and~\ref{tab:sysmod}. 
The errors of $\delta$ resolution and track efficiency are
considered uncorrelated between the HERA~I and HERA~II datasets.
 All other errors are assumed $100\%$ correlated.

\section{Results}
\label{results}

\subsection{Comparison and Combination of Data}
\begin{figure}[htb]
  \begin{center}
  \includegraphics[width=0.85\textwidth]{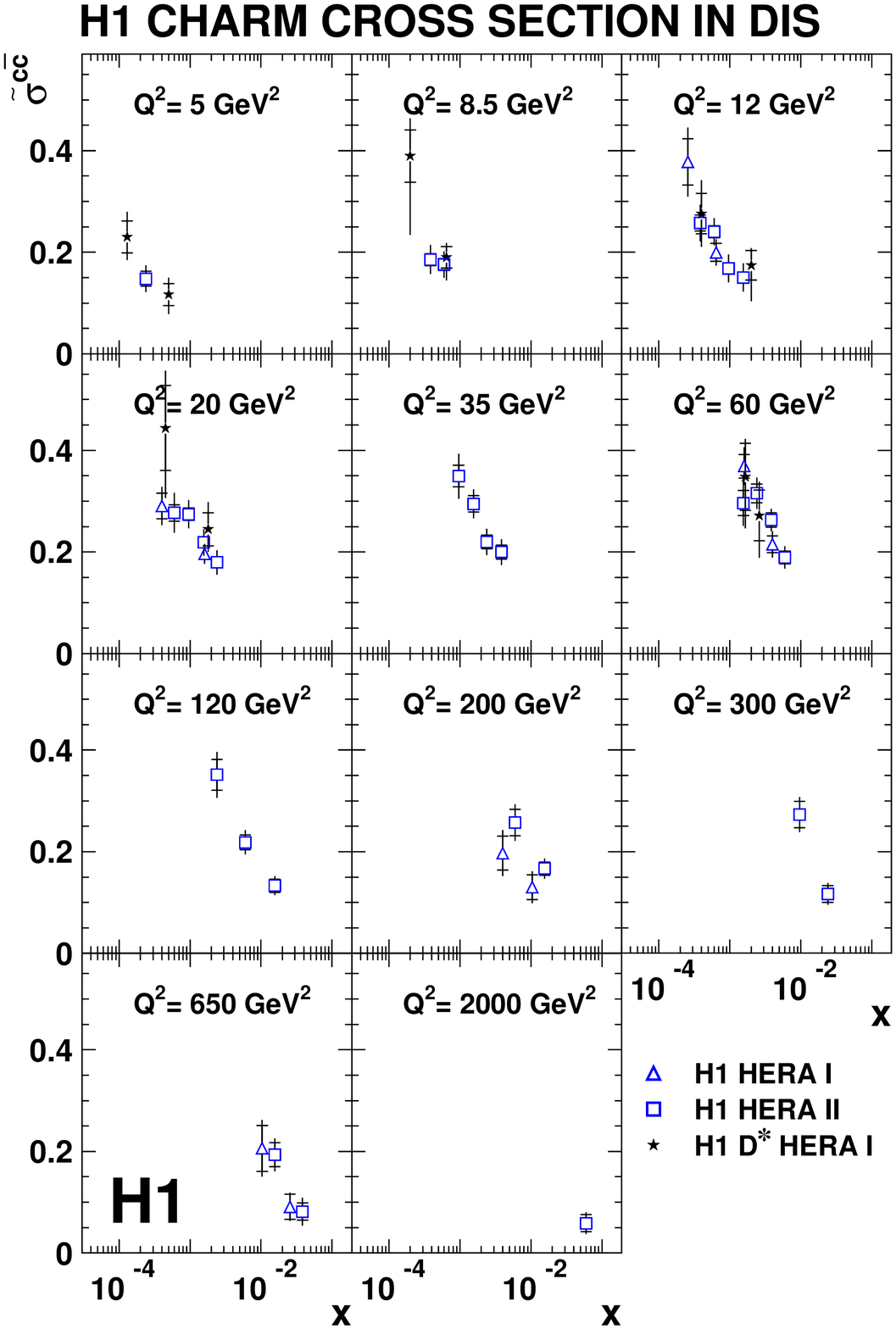} \caption{The
  measured reduced cross section $\tilde{\sigma}^{c\bar{c}}$ shown as
  a function of $x$ for different $Q^2$ values.  The inner error
  bars show the statistical error, the outer error bars represent the
  statistical and systematic errors added in quadrature. 
  The  \mbox{HERA II} measurements
  are compared with those from \mbox{HERA I}.   The measurements obtained
  from $D^*$ mesons by H1~\cite{H1Dstar} using  \mbox{HERA I} data 
  are also shown. The $x$ values of the  \mbox{HERA I}  data are shifted 
  for visual clarity.
}  
\label{fig:f2cc} \end{center}
\end{figure}

\begin{figure}[htb]
  \begin{center}
  \includegraphics[width=0.75\textwidth]{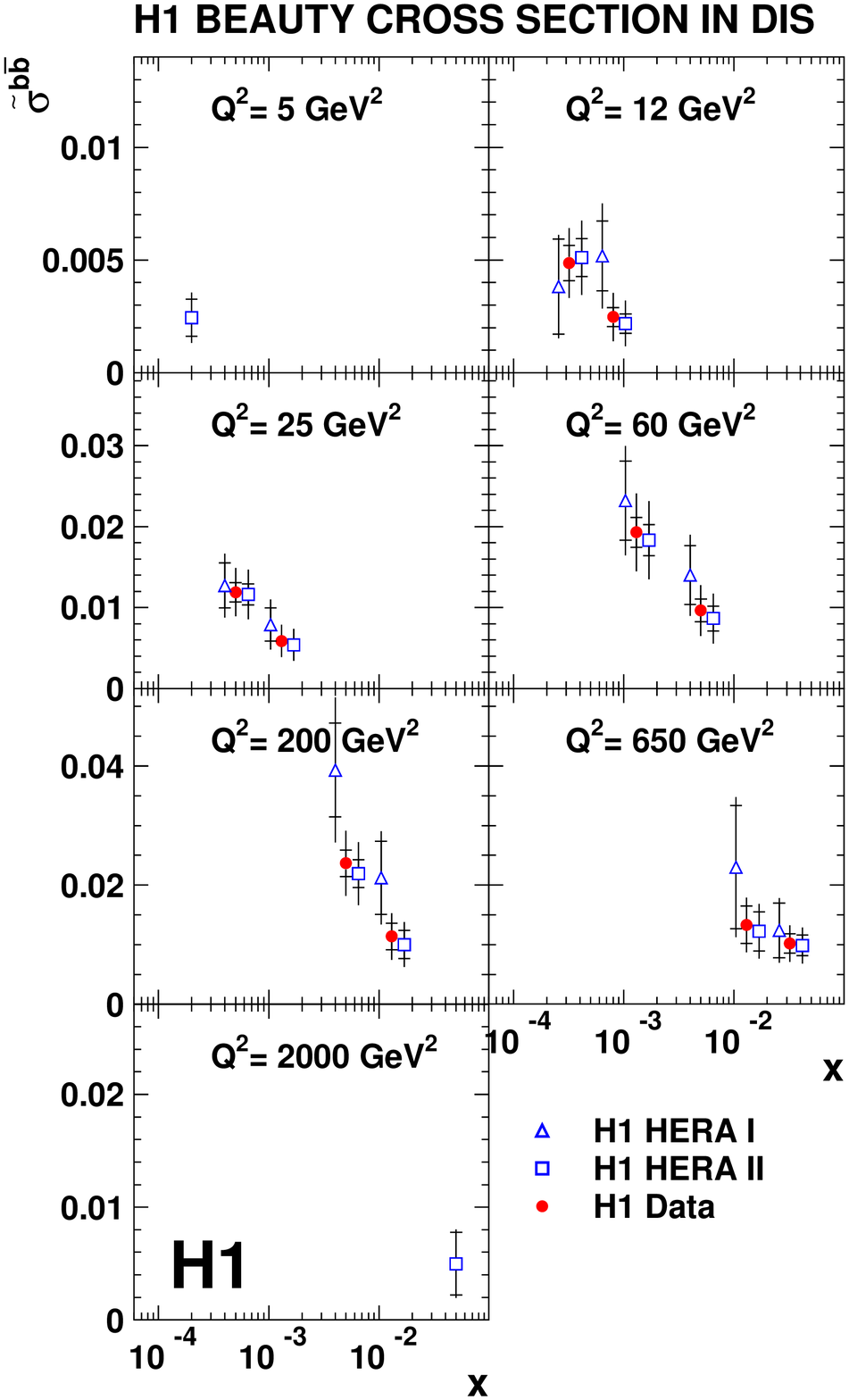}
  \caption{The measured reduced cross section
  $\tilde{\sigma}^{b\bar{b}}$ shown as a function of $x$ for
  different $Q^2$ values.  The inner error bars show the statistical
  error, the outer error bars represent the statistical and systematic
  errors added in quadrature.  The  \mbox{HERA II} measurements are compared with
  those from  \mbox{HERA I}. The combined  H1 data are also shown.
The $x$ values of the  \mbox{HERA I} and  \mbox{HERA II} data are shifted for visual clarity.}
\label{fig:f2bb} \end{center}
\end{figure}

The measurements of $\tilde{\sigma}^{c\bar{c}}$ and
$\tilde{\sigma}^{b\bar{b}}$ are shown as a function of $x$ for fixed
values of $Q^2$ in figures~\ref{fig:f2cc} and \ref{fig:f2bb}
respectively. Also shown in these figures are the HERA~I data
extracted using measurements based on the displacement of
tracks\cite{Aktas:2004az,Aktas:2005iw}. The HERA~I measurements for
$Q^2 \le 60 \ {\rm GeV^2}$ are normalised to the H1 inclusive
cross sections measurements from~\cite{h1newf2paper1} 
and~\cite{h1newf2paper2}.  
The $\tilde{\sigma}^{c\bar{c}}$ and $\tilde{\sigma}^{b\bar{b}}$ data from
HERA~I and HERA~II show good agreement for all measured $x$ and $Q^2$
values. In figure~\ref{fig:f2cc}  the
$\tilde{\sigma}^{c\bar{c}}$ data are also compared with those extracted from
$D^*$ meson measurements by H1~\cite{H1Dstar},
which were obtained using a NLO program~\cite{Harris:1997zq} based on DGLAP 
evolution to
extrapolate the measurements outside the visible $D^*$ range.  The $D^*$ data 
agree well with  the measurements  from the present analysis.

The HERA~I and HERA~II datasets are combined for each $x$-$Q^2$ point
where there are two measurements.  The combination is performed using
a weighted mean using those errors considered uncorrelated between the
two data sets (the statistical errors and the systematic errors of
$\delta$ resolution and track efficiency, see
section~\ref{sec:systematics}):
\begin{eqnarray}
\sigma_{\rm comb}= w_I\sigma_{\rm I}+ w_{II}\sigma_{\rm II},
\end{eqnarray}
\begin{eqnarray}
{\rm with} \,\,\,\,
w_I=\frac{\delta^2_{\rm II}}{\delta^2_{\rm I}+\delta^2_{\rm II}} \,\,\,\, {\rm and} \,\,\,\,
w_{II}=\frac{\delta^2_{\rm I}}{\delta^2_{\rm I}+\delta^2_{\rm II}},
\end{eqnarray}
where $\sigma_{\rm comb}$ is the combined measurement and $\sigma_{\rm
I}$ and $\sigma_{\rm II}$ are the HERA~I and HERA~II measurements
respectively, with their respective errors of $\delta_{\rm I}=\sum_i
\delta^i_{\rm I}$ and $\delta_{\rm II}=\sum_i \delta^i_{\rm II}$,
where $i$ denotes the statistical error and each source of
uncorrelated error between the two data sets. These contributions to
the combined error are evaluated as:
\begin{eqnarray}
(\delta^i_{\rm comb})^2= (w_I \delta^i_{\rm I})^2+ (w_{II} \delta^i_{\rm II})^2.
\end{eqnarray}
For those systematics $j$ considered correlated between the two datasets
(all apart from the errors of $\delta$ resolution and track
efficiency) the systematic error as evaluated from HERA~II is taken,
$\delta^j_{\rm II}$. The total error on the combined measurement is
therefore evaluated from:
\begin{eqnarray}
(\delta^{\rm tot}_{\rm comb})^2=\sum_i (\delta^i_{\rm comb})^2+\sum_j (\delta^j_{\rm II})^2.
\end{eqnarray}

All data that is subsequently shown is from the combined dataset.
Tables~\ref{tab:sig} and~\ref{tab:sysmod} list the combined results
for $c$ and $b$.  The results listed in these tables supersede the
HERA~I measurements published in~\cite{Aktas:2004az}
and~\cite{Aktas:2005iw}. If fits are performed with the data listed in
these tables it should be noted that, although each $x$--$Q^2$
measurement is statistically independent, there is a correlation
between the $c$ and $b$ measurements. Since the $c$ and $b$ binning is
very different this correlation may be neglected for most bins. However, for
the first and last bin the binning is identical so the correlation
coefficient $C_{bc}$ listed in table~\ref{tab:rhotabledeffine} should be used.

\subsection{Comparison with QCD}

The leading contribution to heavy flavour production 
in the region $Q^2 \lapprox M^2$ is
given by the massive boson-gluon fusion matrix
element~\cite{massive,massivenlo} convoluted with the gluon density of the
proton. In the region where $Q^2$ is much larger than $M^2$ the
massive approach may be a poor approximation due to the large
logarithms $\log Q^2/M^2$ which are not resummed~\cite{VFNS1}.  Here, 
the heavy quarks can be treated as massless partons with the leading order
contribution coming from the quark parton model and the heavy quark
parton densities.
In QCD fits to global hard-scattering data the parton density functions are
usually extracted using the general mass variable flavour number 
scheme (GM VFNS)~\cite{collins,VFNS1,VFNS2,VFNS3} for heavy quarks which
interpolates from the massive approach at low scales to the massless
approach at high scales.

The data are compared 
with recent QCD predictions based on the GM VFNS from
MSTW\cite{mstw} (at NLO and NNLO), CTEQ\cite{cteqvfns} (at NLO)
and from H1\cite{h1newf2paper2} (at NLO).
The MSTW predictions use the MSTW08 PDFs which have $m_c=1.4~{\rm GeV}$ and 
$m_b=4.75~{\rm GeV}$, and the 
renormalization and factorization scales are set to $\mu_r = \mu_f = Q$.  
The CTEQ predictions use the CTEQ6.6 PDFs where $m_c=1.3~{\rm GeV}$, 
$m_b=4.5~{\rm GeV}$  and 
$\mu_r = \mu_f = \sqrt{Q^2+M^2}$.  
The predictions from H1 use the H1PDF 2009 PDFs and
the same heavy flavour treatment, including the quark masses and perturbative
scales, as for the
MSTW08 NLO predictions~\cite{mstw}.

The data are also compared with predictions based on CCFM\cite{ccfm2} parton 
evolution and 
massive heavy flavour production.  
The CCFM predictions use the A0 
PDF set~\cite{a0} with $m_c=1.4~{\rm GeV}$, $m_b=4.75~{\rm GeV}$
and $\mu_r = \mu_f = \sqrt{\hat{s}+Q_{T}^2}$, where $\hat{s}$ is the
square of the partonic centre of mass energy and $Q_{T}$ is the transverse
momentum of the heavy quark pair.

\begin{figure}[htb]
 \begin{center}
  \includegraphics[width=0.89\textwidth]{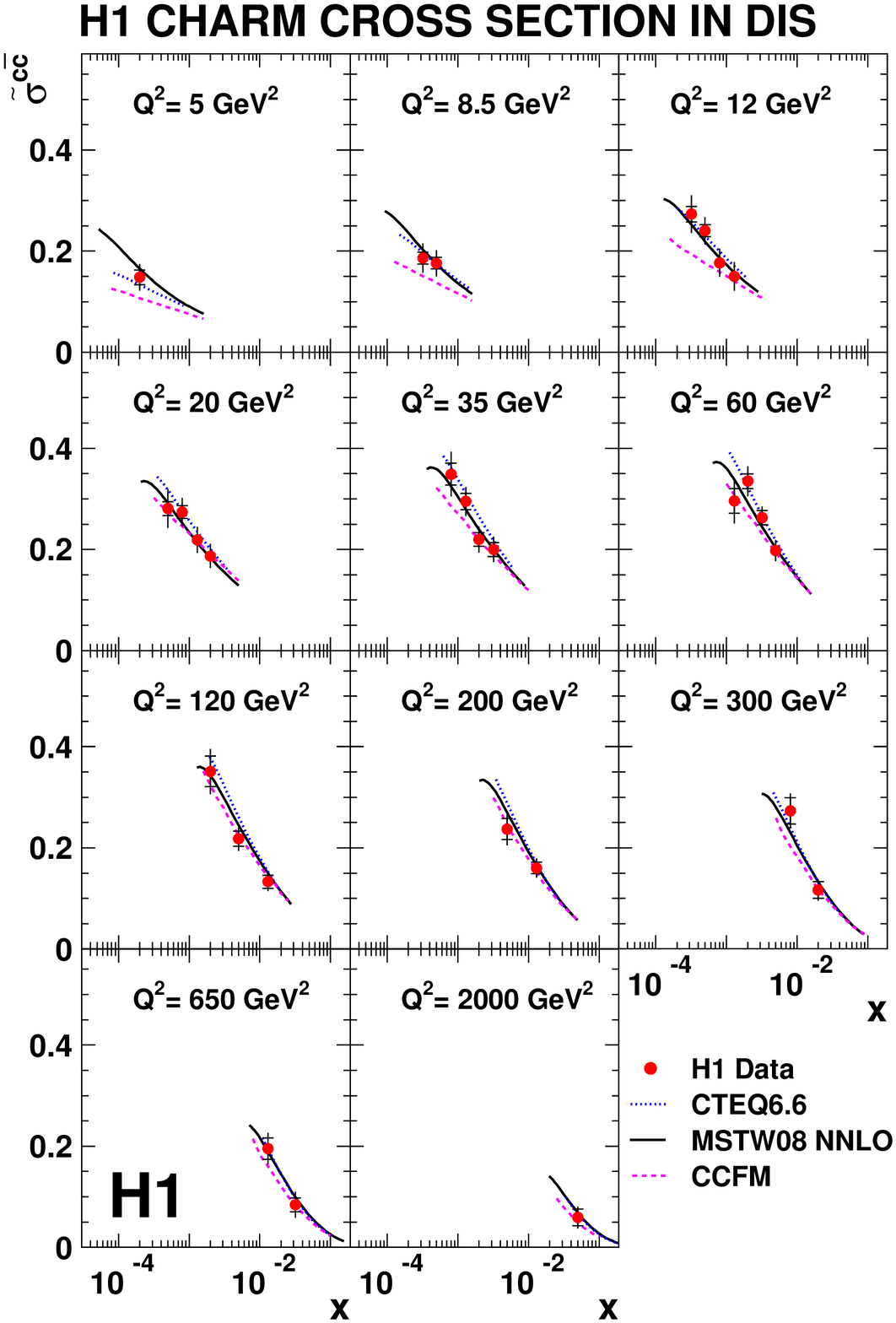} \caption{
  The combined reduced cross section
  $\tilde{\sigma}^{c\bar{c}}$ shown as a function of $x$ for 
  different $Q^2$ values.  The inner error bars show the statistical
  error, the outer error bars represent the statistical and systematic
  errors added in quadrature.  
  The predictions of QCD calculations are also shown.}  
\label{fig:f2ccav} \end{center}
\end{figure}

\begin{figure}[htb]
 \begin{center}
  \includegraphics[width=0.89\textwidth]{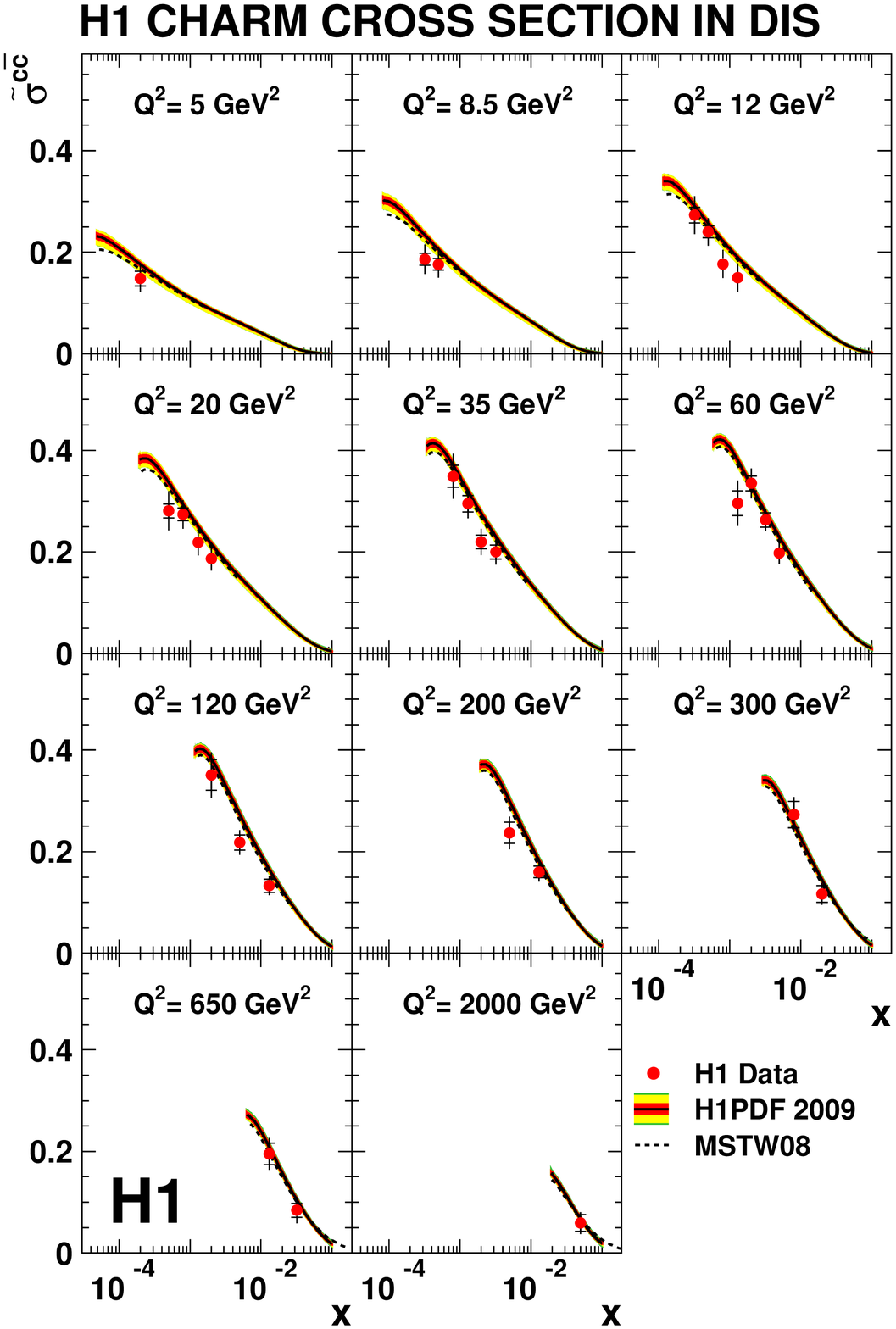} \caption{
  The combined reduced cross section
  $\tilde{\sigma}^{c\bar{c}}$ (as in figure~\ref{fig:f2ccav}).  
  The predictions of the H1PDF 2009 and 
  MSTW08 NLO QCD fits are also shown. For the H1 QCD fit the inner error bands
  show the experimental uncertainty, the middle error bands include the
  theoretical model uncertainties of the fit assumptions, and the 
  outer error band represents the total uncertainty including the 
  parameterisation uncertainty.}  
  \label{fig:f2ccav2} 
  \end{center}
\end{figure}

The  $\tilde{\sigma}^{c\bar{c}}$ data as a function of $x$ for fixed
values of $Q^2$ are compared in figure~\ref{fig:f2ccav} with
the QCD predictions from CCFM, CTEQ and MSTW at NNLO, and 
in figure~\ref{fig:f2ccav2} with the predictions from H1 and MSTW
at NLO.  
In figure~\ref{fig:f2ccav} the GM VFNS predictions from CTEQ and MSTW
at NNLO are observed to be 
similar with the size of the largest differences between the two being at the 
level of the total experimental errors on the data. The CTEQ and MSTW 
predictions provide a reasonable description of the rise 
of the data with decreasing $x$ across the whole of the measured kinematic 
range thus supporting the validity of PDFs extracted using the GM VFNS.
The predictions based on CCFM evolution tend to undershoot the data at the 
lowest values of $Q^2$ and $x$ but also provide a reasonable description for 
the rest of the measured phase space.
In figure~\ref{fig:f2ccav2} the GM VFNS predictions from H1 and MSTW at NLO,
which implement the same heavy flavour treatment~\cite{mstw}, 
are similar and 
also provide a reasonable description of the data. The H1 predictions
are shown with uncertainty bands representing the experimental and theoretical
uncertainties~\cite{h1newf2paper2}. 
The inner error band describes the experimental fit uncertainty, the 
middle error band represents the experimental and model uncertainties added
in quadrature and the outer error band represents the fit parameterisation
uncertainty added in quadrature with all the other uncertainties.
The largest contribution to the uncertainty comes from the model
which is dominated by the variation of the charm quark mass, which is varied
from $1.38$ to $1.47\ {\rm GeV}$. The total
uncertainties on the H1PDF 2009 reduced charm cross section predictions are 
generally smaller than those on the data.

\begin{figure}[htb]
  \begin{center}
  \includegraphics[width=0.79\textwidth]{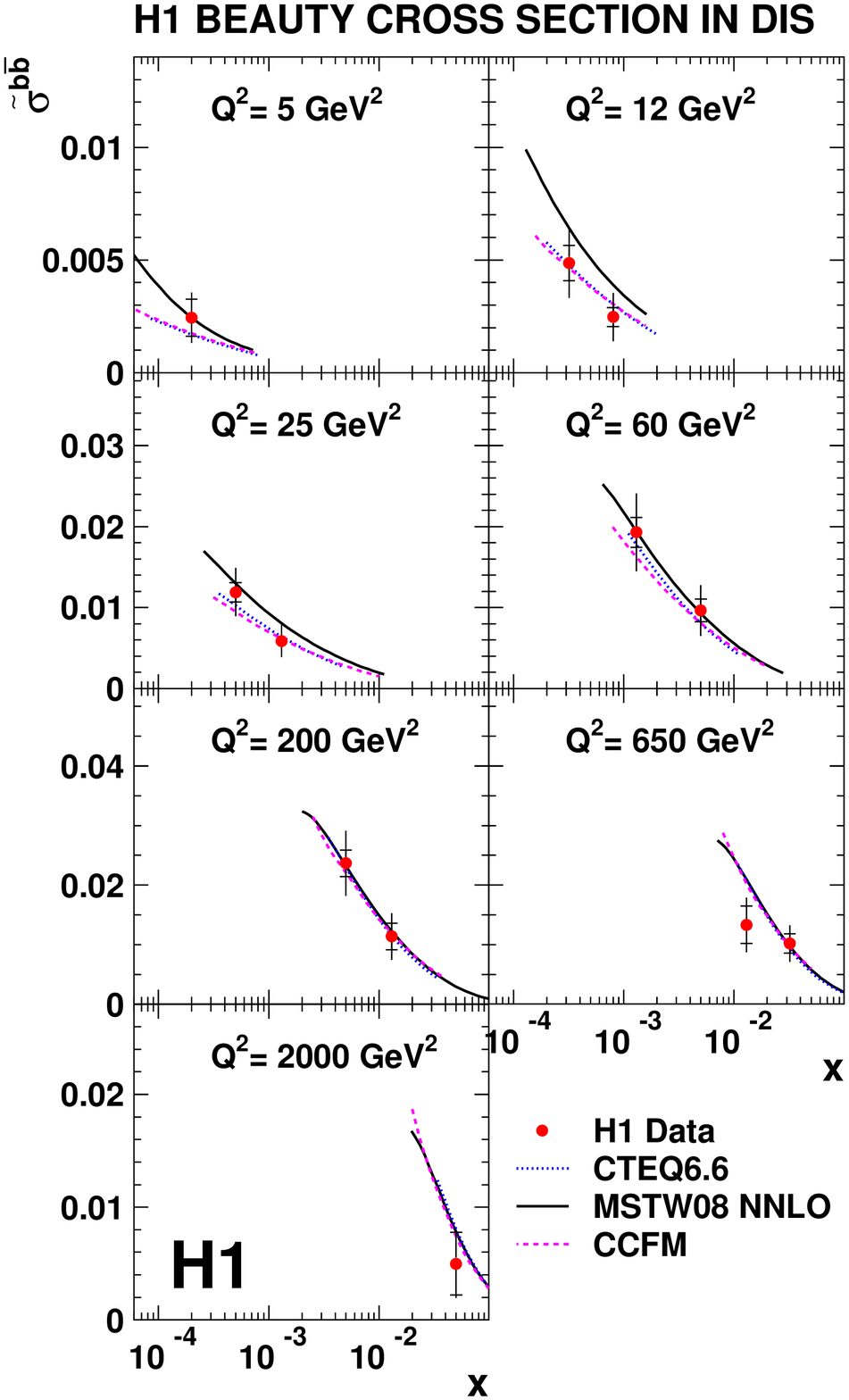} \caption{The
  combined reduced cross section
  $\tilde{\sigma}^{b\bar{b}}$ shown as a function of $x$ for 
  different $Q^2$ values.  The inner error bars show the statistical
  error, the outer error bars represent the statistical and systematic
  errors added in quadrature.  
  The predictions of QCD calculations are also shown.}
  \label{fig:f2bbav} \end{center}
\end{figure}

\begin{figure}[htb]
  \begin{center}
  \includegraphics[width=0.79\textwidth]{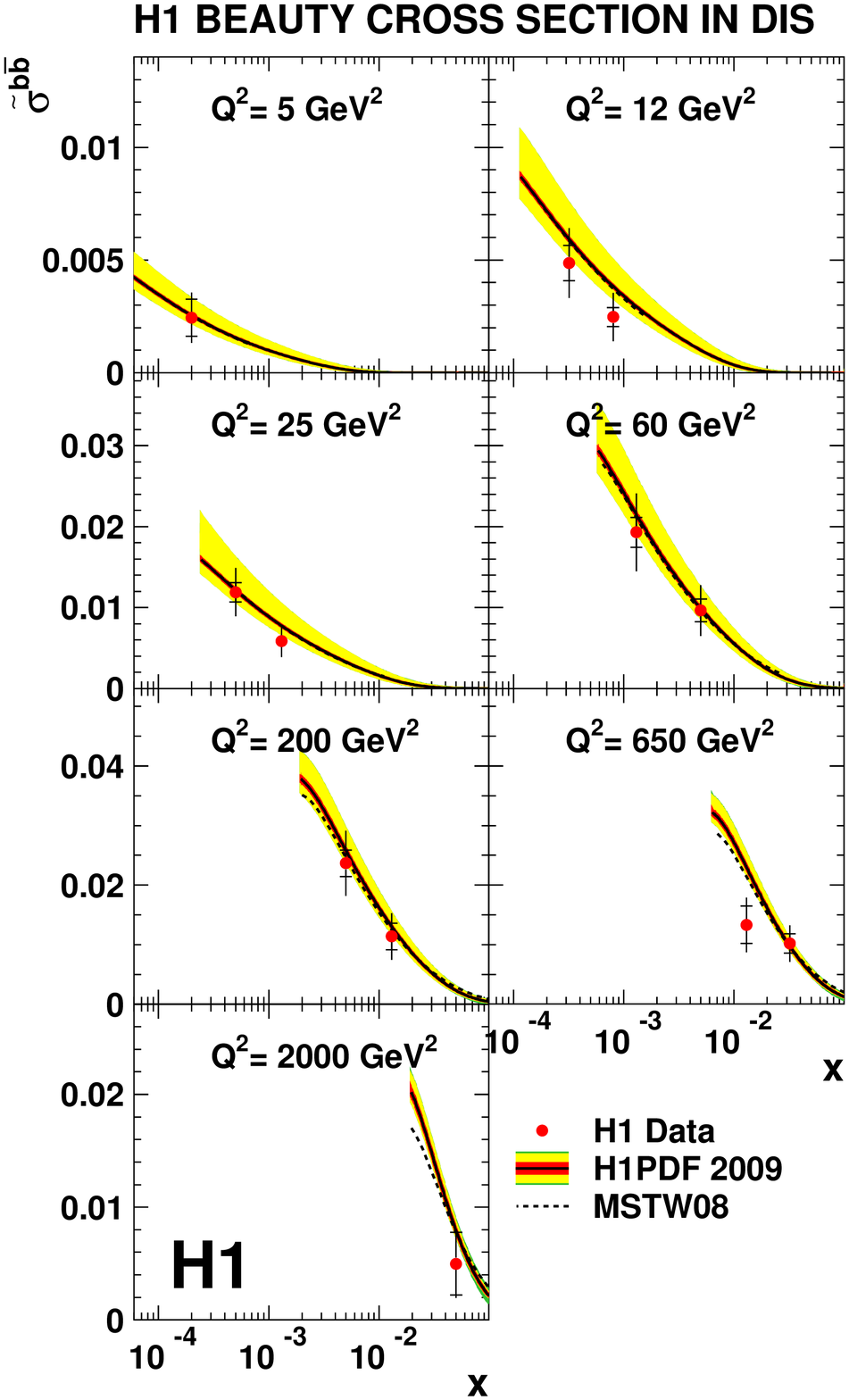} 
\caption{
  The
  combined reduced cross section
  $\tilde{\sigma}^{b\bar{b}}$ (as in figure~\ref{fig:f2bbav}).
  The predictions of the H1PDF 2009 and 
  MSTW08 NLO QCD fits are also shown. For the H1 QCD fit the inner error bands
  show the experimental uncertainty, the middle error bands include the
  theoretical model uncertainties of the fit assumptions, and the 
  outer error band represents the total uncertainty including the 
  parameterisation uncertainty.
}
  \label{fig:f2bbav2} 
  \end{center}
\end{figure}

The $\tilde{\sigma}^{b\bar{b}}$ data as a function of $x$ for fixed
values of $Q^2$ are compared with the QCD predictions in 
figures~\ref{fig:f2bbav} and~\ref{fig:f2bbav2}.
In figure~\ref{fig:f2bbav} the CTEQ and CCFM predictions are seen to be 
very similar across the whole range of the measurements. The MSTW NNLO 
predictions are around $35\%$ higher than CTEQ and CCFM at low values of 
$Q^2$, with the difference decreasing with increasing values of $Q^2$.
The differences between the theory predictions for the $b$ cross  
section at low $Q^2$ are much reduced compared
with the theoretical status at the time of the HERA~I publication where
there was a factor $2$ 
difference at $Q^2 = 12 \ {\rm GeV^2}$~\cite{Aktas:2005iw}. 
In figure~\ref{fig:f2bbav2} the MSTW and H1 NLO QCD predictions 
for $\tilde{\sigma}^{b\bar{b}}$ are observed, 
as for the case of $\tilde{\sigma}^{c\bar{c}}$, to be very similar.  
The uncertainty on the H1
predictions is again dominated by the model uncertainty due to the
variation of the $b$ quark mass,  which is varied from $4.3$ to $5.0\ {\rm GeV}$. At lower values of $Q^2$ the uncertainties on 
the H1 PDF predictions are larger than those on the data.
The reduced $b$ cross section data, including the points in the newly measured regions,
 are well described by all the present QCD predictions. 
  
The structure function $F_2^{c\bar{c}}$ is evaluated from the reduced cross section
\begin{equation}
\tilde{\sigma}^{c\bar{c}} =   F_2^{c\bar{c}}   - \frac{y^2 }{1+ (1-y)^2}  F_L^{c\bar{c}},
\label{eq:sigcc}
\end{equation}
where the longitudinal structure function $F_L^{c\bar{c}}$ is
estimated from the same NLO QCD expectation as used for the bin centre
correction. The correction due to $F_L^{c\bar{c}}$ is negligible for most bins but contributes up to
$6.7$\% of the reduced cross section at the highest value of $y$.
The structure function $F_2^{b\bar{b}}$ is evaluated in
the same manner.

\begin{figure}[htb]
  \begin{center}
  \includegraphics[width=0.85\textwidth]{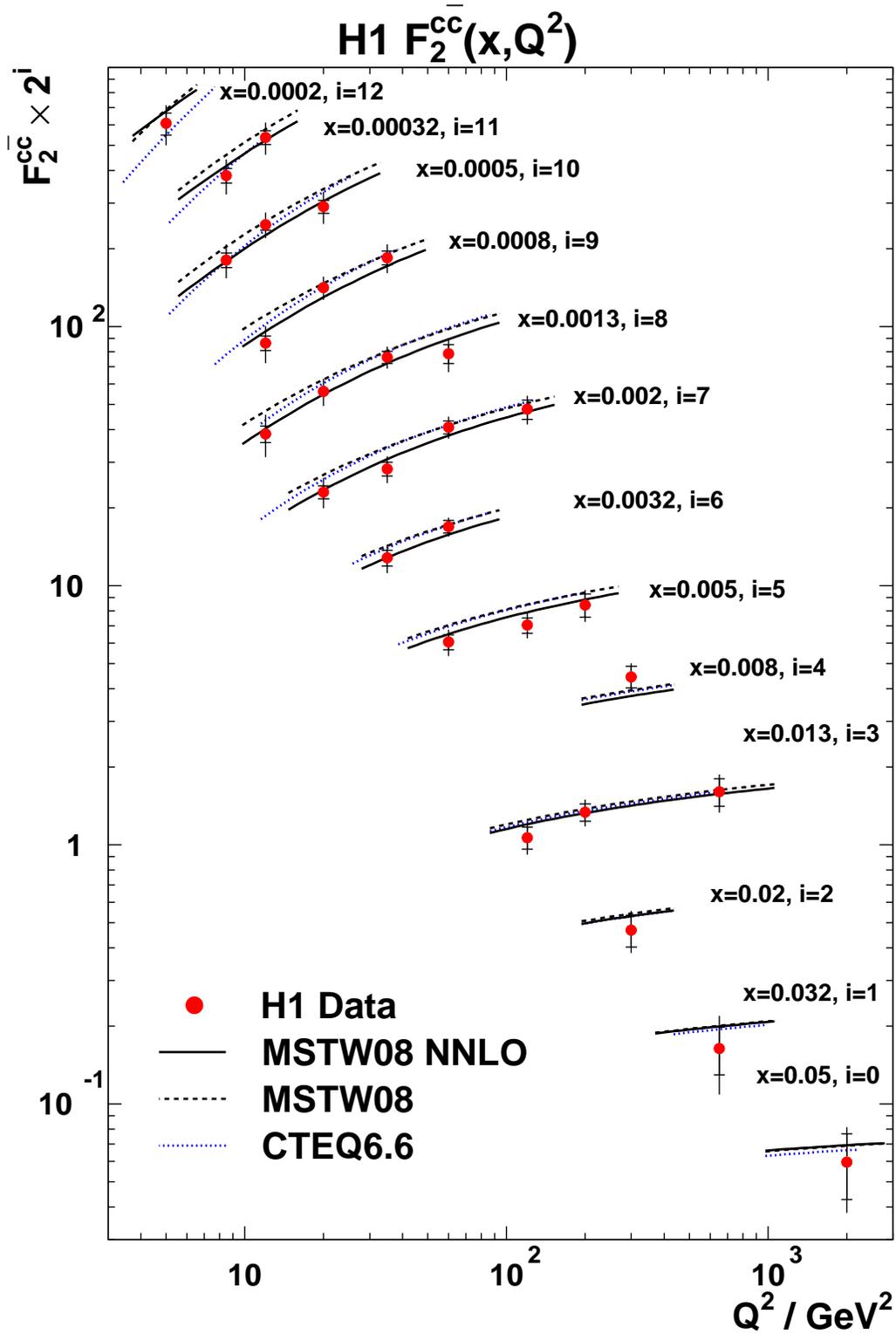}
  \caption{The combined $F_2^{c\bar{c}}$ shown
  as a function of $Q^2$ for various $x$ values.  The inner error bars
  show the statistical error, the outer error bars represent the
  statistical and systematic errors added in quadrature.  
  The predictions of QCD calculations are also shown.}  \label{fig:f2ccq2} \end{center}
\end{figure}

\begin{figure}[htb]
  \begin{center}
  \includegraphics[width=0.9\textwidth]{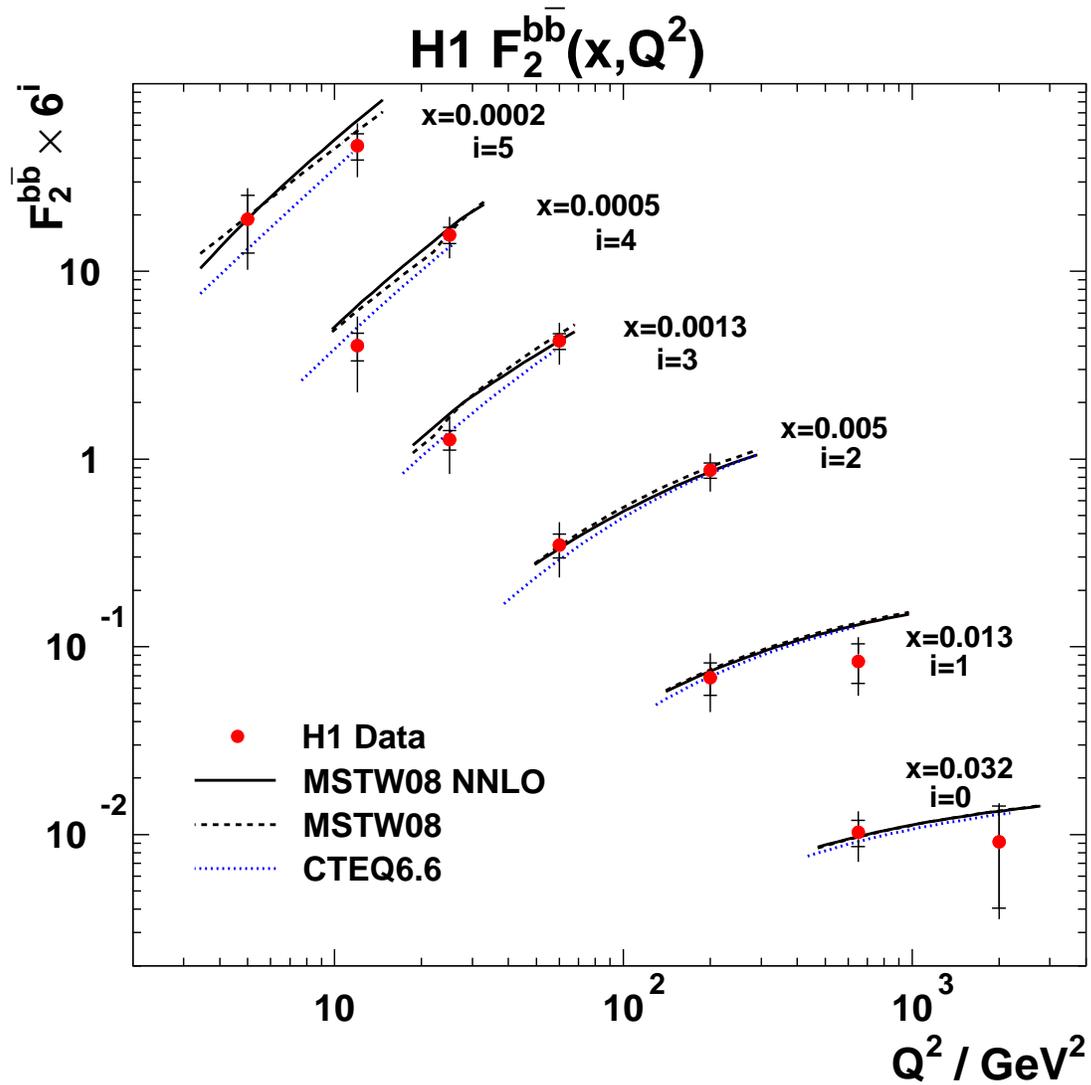}
  \caption{The combined  $F_2^{b\bar{b}}$ shown
  as a function of $Q^2$ for various $x$ values.  The inner error bars
  show the statistical error, the outer error bars represent the
  statistical and systematic errors added in quadrature. The
  predictions of QCD calculations are also shown. Note that some points have been interpolated in $x$ for visual clarity.}  
\label{fig:f2bbq2} 
\end{center}
\end{figure}

The measurements of
$F_2^{c\bar{c}}$ and 
$F_2^{b\bar{b}}$ are presented in table~\ref{tab:sig} and shown as a
function of $Q^2$ in 
figure~\ref{fig:f2ccq2} and
figure~\ref{fig:f2bbq2} respectively. 
The data are compared with the GM VFNS QCD predictions
from CTEQ\cite{cteqvfns} at NLO and from
MSTW at NLO and NNLO\cite{mstw}.
The description of the charm data by the MSTW QCD 
calculations is reasonable, with the NNLO being somewhat better than NLO.
The CTEQ NLO prediction also gives a reasonable description of the data. 

\begin{figure}[htb]
  \begin{center} \includegraphics[width=0.88\textwidth]{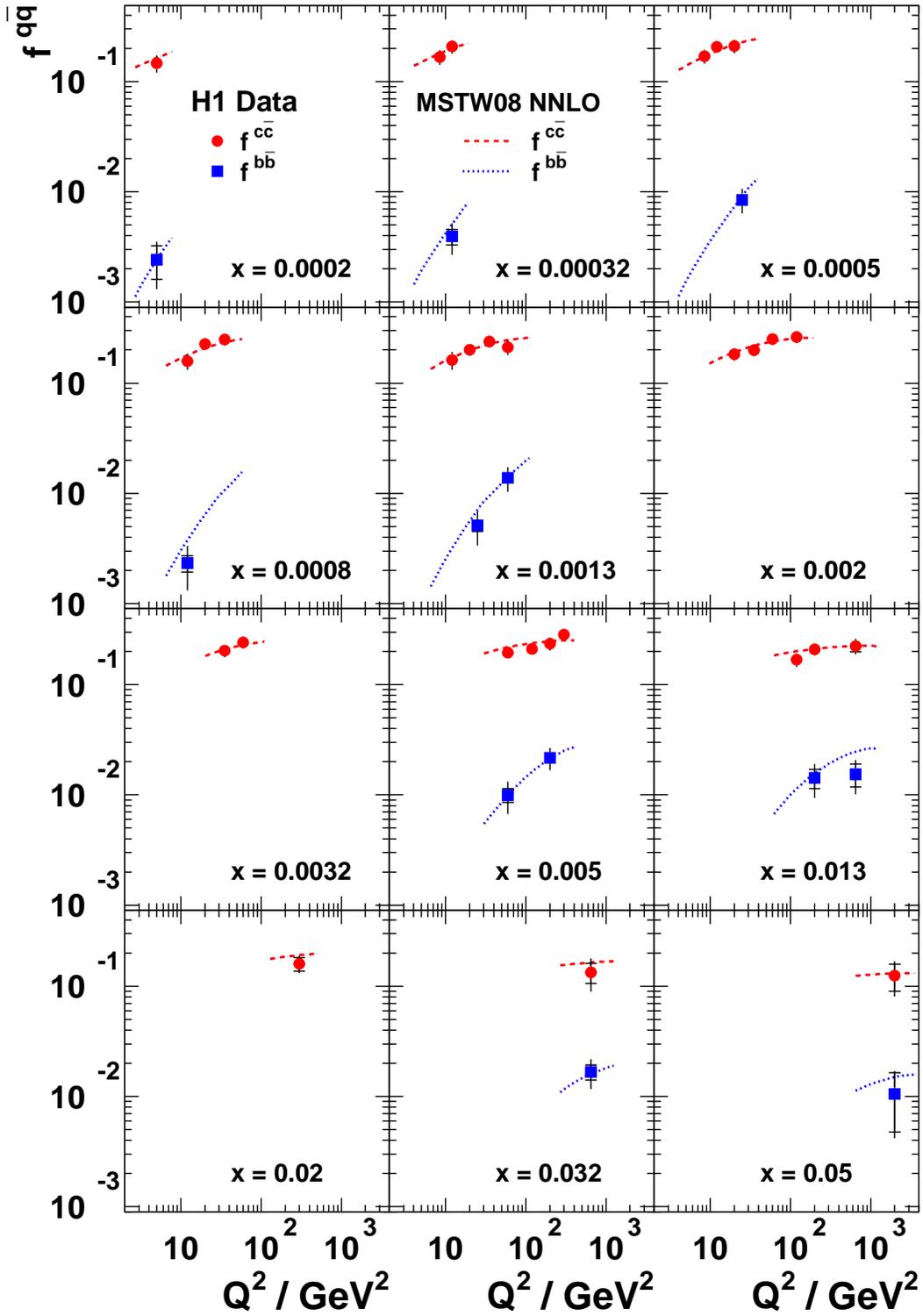} \caption{The
  charm and beauty contributions to the total cross section, $f^{c\bar{c}}$ and
  $f^{b\bar{b}}$, shown as a function of $Q^2$ for different $x$
  values.  The inner error bars show the statistical error, the outer
  error bars represent the statistical and systematic errors added in
  quadrature.  
  A prediction of NNLO QCD is also shown. The charm data point at $x=0.005$ and
$Q^2=300 \ {\rm GeV^2}$ has been interpolated from $x=0.008$.}
  \label{fig:frac} \end{center}
\end{figure}

The measurements are presented in
figure~\ref{fig:frac} in the form of the fractional contribution to
the total $ep$ cross section
\begin{equation}
f^{c\bar{c}} =  \frac{{\rm d}^2 \sigma^{c\bar{c}}} {{\rm d} x\, {\rm d} Q^2}
/
\frac{
{\rm d}^2 \sigma}{ {\rm d} x\, {\rm d} Q^2
}.
\end{equation}
The $b$ fraction $f^{b\bar{b}}$ is defined in the same manner.  
In the present kinematic range the value of 
$f^{c\bar{c}}$ is around $17\%$ on average and 
increases slightly with increasing $Q^2$ and decreasing $x$.
The value of $f^{b\bar{b}}$ increases rapidly with
$Q^2$ from about $0.2\%$ at $Q^2 = 5 \ {\rm GeV^2}$ to around $1\%$ for 
$Q^2 \gapprox 60 \ {\rm GeV^2}$. 
The NNLO QCD predictions of MSTW shown in figure~\ref{fig:frac} are found to
describe the data well.

\section{Conclusion}

The reduced charm and beauty cross sections in deep inelastic
scattering are measured for a wide range of $Q^2$ and Bjorken $x$
using the HERA~II data.  The analysis was performed using several
variables including the significance (the impact parameter divided by
its error) and the position of the secondary vertex as reconstructed
from the vertex detector. For selected track multiplicities of $1$ or $2$ the
highest and second highest significance distributions are used to
evaluate the charm and beauty content of the data. For selected track
multiplicities $\ge 3$ several variables are combined using an
artificial neural network.

The reduced cross sections agree with previous measurements using a similar technique, 
but have reduced errors and cover an extended $Q^2$ range. 
HERA~I and  HERA~II data are combined 
resulting in more precise reduced cross section and structure function
measurements.  The charm and
beauty fractional contributions to the total $ep$ cross section are
also  measured.  In this kinematic range the charm cross section
contributes on average $17\%$ and the beauty fraction increases from
about $0.2\%$ at $Q^2 = 5 \ {\rm GeV^2}$ to $1.0\%$ for $Q^2 \gapprox
60 \ {\rm GeV^2}$.  The measurements are described by predictions
using perturbative QCD in the general mass variable flavour number
scheme at NLO and NNLO.

\section*{Acknowledgements}

We are grateful to the HERA machine group whose outstanding efforts
have made this experiment possible.  We thank the engineers and
technicians for their work in constructing and maintaining the H1
detector, our funding agencies for financial support, the DESY
technical staff for continual assistance and the DESY directorate for
support and for the hospitality which they extend to the non-DESY
members of the collaboration.


\newpage

\begin{table}[htb]
\tiny
\begin{center}
 \renewcommand{\arraystretch}{1.15}
\begin{tabular}{|r|r|c|c|c|c|c|c|c|c|c|} \hline
 \multicolumn{1}{|c|}{bin} & \multicolumn{1}{c|}{$Q^2$} & $x$ & $y$ & $\rho_l$ &  $\rho_c$ & $\rho_b$ & $\chi^2/{\rm n.d.f.}$ & $C_{lc}$  & $C_{lb}$ &  $C_{bc}$ \bigstrut[t] \\  \hline
 $   1 $ & $   5.0 $ & $ 0.00020 $ & $ 0.246 $ & $    1.27  \pm     0.02  $ & $     1.39 \pm     0.13 $ & $     1.72 \pm    0.58 $ & $  19.9 /  24 $ & $   -0.99 $ & $    0.56 $ & $    -0.61 $ \\ 
 $   2 $ & $   8.5 $ & $ 0.00050 $ & $ 0.167 $ & $    1.15  \pm     0.01  $ & $     1.27 \pm     0.08 $ & $     0.58 \pm    0.56 $ & $  26.4 /  24 $ & $   -0.99 $ & $    0.57 $ & $    -0.64 $ \\ 
 $   3 $ & $   8.5 $ & $ 0.00032 $ & $ 0.262 $ & $    1.15  \pm     0.01  $ & $     1.07 \pm     0.07 $ & $     0.86 \pm    0.22 $ & $  35.9 /  39 $ & $   -0.99 $ & $    0.52 $ & $    -0.57 $ \\ 
 $   4 $ & $  12.0 $ & $ 0.00130 $ & $ 0.091 $ & $    1.11  \pm     0.01  $ & $     1.13 \pm     0.08 $ & $     0.94 \pm    0.92 $ & $  25.6 /  24 $ & $   -0.98 $ & $    0.62 $ & $    -0.73 $ \\ 
 $   5 $ & $  12.0 $ & $ 0.00080 $ & $ 0.148 $ & $    1.14  \pm     0.01  $ & $     0.99 \pm     0.06 $ & $     1.29 \pm    0.41 $ & $  67.3 /  40 $ & $   -0.98 $ & $    0.59 $ & $    -0.67 $ \\ 
 $   6 $ & $  12.0 $ & $ 0.00050 $ & $ 0.236 $ & $    1.09  \pm     0.01  $ & $     1.17 \pm     0.06 $ & $     0.48 \pm    0.24 $ & $  34.9 /  38 $ & $   -0.99 $ & $    0.52 $ & $    -0.59 $ \\ 
 $   7 $ & $  12.0 $ & $ 0.00032 $ & $ 0.369 $ & $    1.15  \pm     0.02  $ & $     1.11 \pm     0.07 $ & $     0.88 \pm    0.21 $ & $  64.3 /  39 $ & $   -0.99 $ & $    0.55 $ & $    -0.60 $ \\ 
 $   8 $ & $  20.0 $ & $ 0.00200 $ & $ 0.098 $ & $    1.12  \pm     0.01  $ & $     1.10 \pm     0.06 $ & $     1.77 \pm    0.72 $ & $  23.3 /  26 $ & $   -0.97 $ & $    0.62 $ & $    -0.77 $ \\ 
 $   9 $ & $  20.0 $ & $ 0.00130 $ & $ 0.151 $ & $    1.14  \pm     0.01  $ & $     1.10 \pm     0.05 $ & $     0.97 \pm    0.30 $ & $  52.3 /  39 $ & $   -0.98 $ & $    0.57 $ & $    -0.67 $ \\ 
 $  10 $ & $  20.0 $ & $ 0.00080 $ & $ 0.246 $ & $    1.15  \pm     0.01  $ & $     1.14 \pm     0.05 $ & $     0.60 \pm    0.21 $ & $  36.5 /  38 $ & $   -0.98 $ & $    0.52 $ & $    -0.61 $ \\ 
 $  11 $ & $  20.0 $ & $ 0.00050 $ & $ 0.394 $ & $    1.25  \pm     0.02  $ & $     0.98 \pm     0.06 $ & $     1.00 \pm    0.15 $ & $  33.2 /  40 $ & $   -0.99 $ & $    0.56 $ & $    -0.61 $ \\ 
 $  12 $ & $  35.0 $ & $ 0.00320 $ & $ 0.108 $ & $    1.12  \pm     0.02  $ & $     1.09 \pm     0.07 $ & $     0.93 \pm    0.72 $ & $  27.2 /  26 $ & $   -0.96 $ & $    0.62 $ & $    -0.78 $ \\ 
 $  13 $ & $  35.0 $ & $ 0.00200 $ & $ 0.172 $ & $    1.18  \pm     0.02  $ & $     0.97 \pm     0.06 $ & $     1.52 \pm    0.35 $ & $  43.5 /  39 $ & $   -0.97 $ & $    0.58 $ & $    -0.70 $ \\ 
 $  14 $ & $  35.0 $ & $ 0.00130 $ & $ 0.265 $ & $    1.19  \pm     0.02  $ & $     1.08 \pm     0.06 $ & $     0.67 \pm    0.22 $ & $  58.2 /  39 $ & $   -0.98 $ & $    0.54 $ & $    -0.63 $ \\ 
 $  15 $ & $  35.0 $ & $ 0.00080 $ & $ 0.431 $ & $    1.25  \pm     0.02  $ & $     1.04 \pm     0.06 $ & $     1.02 \pm    0.17 $ & $  46.5 /  39 $ & $   -0.99 $ & $    0.57 $ & $    -0.64 $ \\ 
 $  16 $ & $  60.0 $ & $ 0.00500 $ & $ 0.118 $ & $    1.15  \pm     0.02  $ & $     0.99 \pm     0.06 $ & $     2.17 \pm    0.49 $ & $  50.6 /  40 $ & $   -0.95 $ & $    0.59 $ & $    -0.76 $ \\ 
 $  17 $ & $  60.0 $ & $ 0.00320 $ & $ 0.185 $ & $    1.16  \pm     0.02  $ & $     1.10 \pm     0.06 $ & $     0.66 \pm    0.25 $ & $  61.6 /  37 $ & $   -0.96 $ & $    0.52 $ & $    -0.66 $ \\ 
 $  18 $ & $  60.0 $ & $ 0.00200 $ & $ 0.295 $ & $    1.15  \pm     0.02  $ & $     1.00 \pm     0.06 $ & $     0.90 \pm    0.20 $ & $  38.6 /  39 $ & $   -0.97 $ & $    0.52 $ & $    -0.64 $ \\ 
 $  19 $ & $  60.0 $ & $ 0.00130 $ & $ 0.454 $ & $    1.32  \pm     0.02  $ & $     0.80 \pm     0.07 $ & $     1.06 \pm    0.18 $ & $  76.8 /  38 $ & $   -0.98 $ & $    0.57 $ & $    -0.66 $ \\ 
 $  20 $ & $ 120.0 $ & $ 0.01300 $ & $ 0.091 $ & $    1.13  \pm     0.02  $ & $     0.91 \pm     0.09 $ & $     1.37 \pm    0.46 $ & $  32.4 /  23 $ & $   -0.95 $ & $    0.61 $ & $    -0.77 $ \\ 
 $  21 $ & $ 120.0 $ & $ 0.00500 $ & $ 0.236 $ & $    1.20  \pm     0.02  $ & $     0.88 \pm     0.06 $ & $     1.27 \pm    0.20 $ & $  37.0 /  38 $ & $   -0.96 $ & $    0.55 $ & $    -0.69 $ \\ 
 $  22 $ & $ 120.0 $ & $ 0.00200 $ & $ 0.591 $ & $    1.25  \pm     0.03  $ & $     0.92 \pm     0.08 $ & $     0.94 \pm    0.17 $ & $  42.5 /  37 $ & $   -0.97 $ & $    0.54 $ & $    -0.65 $ \\ 
 $  23 $ & $ 200.0 $ & $ 0.01300 $ & $ 0.151 $ & $    1.14  \pm     0.02  $ & $     1.03 \pm     0.08 $ & $     0.65 \pm    0.21 $ & $  40.7 /  36 $ & $   -0.95 $ & $    0.53 $ & $    -0.67 $ \\ 
 $  24 $ & $ 200.0 $ & $ 0.00500 $ & $ 0.394 $ & $    1.19  \pm     0.03  $ & $     0.83 \pm     0.08 $ & $     0.75 \pm    0.16 $ & $  41.1 /  36 $ & $   -0.97 $ & $    0.53 $ & $    -0.65 $ \\ 
 $  25 $ & $ 300.0 $ & $ 0.02000 $ & $ 0.148 $ & $    1.08  \pm     0.03  $ & $     0.83 \pm     0.12 $ & $     1.16 \pm    0.27 $ & $  32.7 /  33 $ & $   -0.95 $ & $    0.52 $ & $    -0.67 $ \\ 
 $  26 $ & $ 300.0 $ & $ 0.00800 $ & $ 0.369 $ & $    1.12  \pm     0.03  $ & $     0.99 \pm     0.09 $ & $     0.54 \pm    0.17 $ & $  35.8 /  34 $ & $   -0.96 $ & $    0.51 $ & $    -0.64 $ \\ 
 $  27 $ & $ 650.0 $ & $ 0.03200 $ & $ 0.200 $ & $    1.08  \pm     0.03  $ & $     0.72 \pm     0.15 $ & $     0.75 \pm    0.29 $ & $  28.2 /  32 $ & $   -0.96 $ & $    0.52 $ & $    -0.65 $ \\ 
 $  28 $ & $ 650.0 $ & $ 0.01300 $ & $ 0.492 $ & $    1.09  \pm     0.03  $ & $     0.85 \pm     0.10 $ & $     0.82 \pm    0.16 $ & $  37.0 /  35 $ & $   -0.96 $ & $    0.54 $ & $    -0.66 $ \\ 
 $  29 $ & $ 2000.0 $ & $ 0.05000 $ & $ 0.394 $ & $    1.07  \pm     0.04  $ & $     0.73 \pm     0.20 $ & $     0.67 \pm    0.37 $ & $  25.5 /  25 $ & $   -0.96 $ & $    0.55 $ & $    -0.67 $ \\ 
\hline 
 \end{tabular} 

\end{center}
\normalsize
    \caption{The fit parameters $\rho_l$, $\rho_c$ and $\rho_b$, along
    with their errors, the $\chi^2$ per degree of freedom and the
    correlation coefficients of the fit parameters for each bin in
    $Q^2$ and $x$.  The parameters are shown for the fine binning
    scheme used to evaluate the reduced $c$ cross section.}
\label{tab:rhotabledeffine}
\end{table}

\begin{table}[htb]
\tiny
\begin{center}
 \renewcommand{\arraystretch}{1.15}
 \begin{tabular}{|r|r|c|c|c|c|c|c|c|c|c|} \hline  
 \multicolumn{1}{|c|}{bin} & \multicolumn{1}{c|}{$Q^2$} & $x$ & $y$ & $\rho_l$ &  $\rho_c$ & $\rho_b$ & $\chi^2/{\rm n.d.f.}$ & $C_{lc}$  & $C_{lb}$ &  $C_{bc}$ \bigstrut[t] \\  \hline  
 $   1 $ & $   5.0 $ & $ 0.00020 $ & $ 0.246 $ & $    1.27  \pm     0.02  $ & $     1.39 \pm     0.13 $ & $     1.72 \pm    0.58 $ & $  19.9 /  24 $ & $   -0.99 $ & $    0.56 $ & $    -0.61 $ \\ 
 $   2 $ & $  12.0 $ & $ 0.00032 $ & $ 0.369 $ & $    1.17  \pm     0.01  $ & $     1.08 \pm     0.06 $ & $     1.06 \pm    0.17 $ & $  49.4 /  40 $ & $   -0.99 $ & $    0.55 $ & $    -0.60 $ \\ 
 $   3 $ & $  12.0 $ & $ 0.00080 $ & $ 0.148 $ & $    1.12  \pm     0.01  $ & $     1.12 \pm     0.03 $ & $     0.75 \pm    0.15 $ & $  61.0 /  46 $ & $   -0.99 $ & $    0.54 $ & $    -0.61 $ \\ 
 $   4 $ & $  25.0 $ & $ 0.00050 $ & $ 0.492 $ & $    1.29  \pm     0.02  $ & $     0.95 \pm     0.05 $ & $     1.15 \pm    0.13 $ & $  56.3 /  42 $ & $   -0.99 $ & $    0.57 $ & $    -0.63 $ \\ 
 $   5 $ & $  25.0 $ & $ 0.00130 $ & $ 0.189 $ & $    1.15  \pm     0.01  $ & $     1.08 \pm     0.02 $ & $     0.93 \pm    0.13 $ & $  67.3 /  45 $ & $   -0.98 $ & $    0.55 $ & $    -0.65 $ \\ 
 $   6 $ & $  60.0 $ & $ 0.00130 $ & $ 0.454 $ & $    1.20  \pm     0.01  $ & $     0.91 \pm     0.03 $ & $     1.00 \pm    0.10 $ & $  61.1 /  43 $ & $   -0.97 $ & $    0.54 $ & $    -0.65 $ \\ 
 $   7 $ & $  60.0 $ & $ 0.00500 $ & $ 0.118 $ & $    1.13  \pm     0.01  $ & $     1.01 \pm     0.04 $ & $     1.16 \pm    0.20 $ & $  51.3 /  44 $ & $   -0.96 $ & $    0.55 $ & $    -0.71 $ \\ 
 $   8 $ & $ 200.0 $ & $ 0.00500 $ & $ 0.394 $ & $    1.14  \pm     0.02  $ & $     0.83 \pm     0.04 $ & $     0.88 \pm    0.09 $ & $  35.6 /  40 $ & $   -0.97 $ & $    0.55 $ & $    -0.67 $ \\ 
 $   9 $ & $ 200.0 $ & $ 0.01300 $ & $ 0.151 $ & $    1.19  \pm     0.02  $ & $     1.06 \pm     0.07 $ & $     1.01 \pm    0.24 $ & $  35.3 /  39 $ & $   -0.95 $ & $    0.53 $ & $    -0.68 $ \\ 
 $  10 $ & $ 650.0 $ & $ 0.01300 $ & $ 0.492 $ & $    1.07  \pm     0.03  $ & $     0.91 \pm     0.09 $ & $     0.50 \pm    0.13 $ & $  42.9 /  36 $ & $   -0.97 $ & $    0.55 $ & $    -0.66 $ \\ 
 $  11 $ & $ 650.0 $ & $ 0.03200 $ & $ 0.200 $ & $    1.08  \pm     0.02  $ & $     0.74 \pm     0.09 $ & $     1.12 \pm    0.19 $ & $  48.8 /  37 $ & $   -0.96 $ & $    0.52 $ & $    -0.65 $ \\ 
 $  12 $ & $ 2000.0 $ & $ 0.05000 $ & $ 0.394 $ & $    1.07  \pm     0.04  $ & $     0.73 \pm     0.20 $ & $     0.67 \pm    0.37 $ & $  25.5 /  25 $ & $   -0.96 $ & $    0.55 $ & $    -0.67 $ \\ 
\hline 
 \end{tabular} 

\end{center}
\normalsize
    \caption{The fit parameters $\rho_l$, $\rho_c$ and $\rho_b$, along
    with their errors, the $\chi^2$ per degree of freedom and the
    correlation coefficients of the fit parameters for each bin in $Q^2$ and $x$.
    The parameters are shown for the coarse binning scheme used to evaluate
    the reduced $b$ cross section.}
\label{tab:rhotabledefcoarse}
\end{table}

\begin{sidewaystable}[h!]
\tiny
\begin{center}
 \renewcommand{\arraystretch}{1.15}
 \begin{tabular}{|c|d|c|c|c|c|d|d|d|d|d|d|d|d|d|d|d|d|d|d|d|d|d|} \hline 
 bin & Q^2 & $x$ & $y$ & $\tilde{\sigma}^{q\bar{q}}$ & $F^{q\bar{q}}_2$ & \multicolumn{1}{r|}{$\delta_{\rm stat}$} & \multicolumn{1}{r|}{$\delta_{\rm sys}$} & \multicolumn{1}{r|}{$\delta_{\rm tot}$} & \multicolumn{1}{r|}{$\delta_{\rm unc}$} & \multicolumn{1}{r|}{$\delta_{\rm res}$} & \multicolumn{1}{r|}{$\delta_{\rm eff CJC}$} & \multicolumn{1}{r|}{$\delta_{\rm eff CST}$}  & \multicolumn{1}{r|}{$\delta_{\rm frag C}$} & \multicolumn{1}{r|}{$\delta_{\rm frag B}$} & \multicolumn{1}{r|}{$\delta_{uds}$} & \multicolumn{1}{r|}{$\delta_{\phi^q}$} & \multicolumn{1}{r|}{$\delta_{\rm had E}$}  & \multicolumn{1}{r|}{$\delta_{\rm gp}$} &   \multicolumn{1}{r|}{$\delta_{\rm F_2}$}  \bigstrut[t] \\ 
  &  \multicolumn{1}{r|}{(GeV$^2$)} &  & & & & \multicolumn{1}{r|}{($\%$)} & \multicolumn{1}{r|}{($\%$)} & \multicolumn{1}{r|}{($\%$)} & \multicolumn{1}{r|}{($\%$)} & \multicolumn{1}{r|}{($\%$)} & \multicolumn{1}{r|}{($\%$)} & \multicolumn{1}{r|}{($\%$)} & \multicolumn{1}{r|}{($\%$)} & \multicolumn{1}{r|}{($\%$)} & \multicolumn{1}{r|}{($\%$)} & \multicolumn{1}{r|}{($\%$)}  & \multicolumn{1}{r|}{($\%$)}  & \multicolumn{1}{r|}{($\%$)} & \multicolumn{1}{r|}{($\%$)} \\ \hline
 $c$   1 & $   5.0  $ & $ 0.00020 $ & $ 0.246 $ & $   0.148 $ & $   0.149 $ &   9.8 &  14.6 &  17.6 &   3.3 &   1.9 &  -1.9 &  -2.8 &  -3.5 &   0.3 &  -7.7 &   4.6 &  -1.2 &   3.1 &   1.0 \\ 
 $c$   2 & $   8.5  $ & $ 0.00050 $ & $ 0.167 $ & $   0.176 $ & $   0.176 $ &   6.5 &  13.3 &  14.8 &   2.0 &   1.9 &  -1.3 &  -2.0 &  -2.8 &   0.0 & -10.0 &   1.8 &  -1.2 &   0.4 &   1.0 \\ 
 $c$   3 & $   8.5  $ & $ 0.00032 $ & $ 0.262 $ & $   0.186 $ & $   0.187 $ &   6.4 &  14.1 &  15.5 &   3.8 &   1.1 &  -1.0 &  -1.5 &  -5.8 &   0.0 &  -7.2 &   2.1 &  -0.2 &   3.6 &   1.0 \\ 
 $c$   4 & $  12.0  $ & $ 0.00130 $ & $ 0.091 $ & $   0.150 $ & $   0.150 $ &   7.3 &  17.2 &  18.7 &   1.2 &   1.5 &  -0.6 &  -0.9 &  -2.5 &   0.1 & -15.7 &   1.7 &   1.7 &   0.0 &   1.0 \\ 
 $c$   5 & $  12.0  $ & $ 0.00080 $ & $ 0.148 $ & $   0.177 $ & $   0.177 $ &   5.2 &  15.0 &  15.9 &   1.3 &   1.0 &  -0.8 &  -1.3 &  -2.5 &   0.0 & -13.5 &   1.2 &  -0.7 &   0.2 &   1.1 \\ 
 $c$   6 & $  12.0  $ & $ 0.00050 $ & $ 0.236 $ & $   0.240 $ & $   0.242 $ &   4.9 &  10.1 &  11.2 &   1.3 &   2.0 &  -1.0 &  -1.4 &  -3.0 &   0.0 &  -5.6 &   1.7 &  -1.7 &   0.6 &   1.0 \\ 
 $c$   7 & $  12.0  $ & $ 0.00032 $ & $ 0.369 $ & $   0.273 $ & $   0.277 $ &   5.6 &  12.6 &  13.8 &   1.4 &   2.1 &  -0.8 &  -1.3 &  -5.6 &   0.1 &  -5.8 &   1.6 &  -0.3 &   3.1 &   1.1 \\ 
 $c$   8 & $  20.0  $ & $ 0.00200 $ & $ 0.098 $ & $   0.187 $ & $   0.188 $ &   4.0 &  12.0 &  12.7 &   1.6 &   1.4 &  -0.4 &  -0.8 &  -2.2 &   0.2 &  -9.9 &   1.2 &   2.2 &   0.0 &   1.1 \\ 
 $c$   9 & $  20.0  $ & $ 0.00130 $ & $ 0.151 $ & $   0.219 $ & $   0.219 $ &   4.6 &  11.0 &  11.9 &   1.2 &   1.2 &  -0.9 &  -1.4 &  -2.2 &   0.0 &  -8.7 &   1.6 &   0.7 &   0.2 &   1.1 \\ 
 $c$  10 & $  20.0  $ & $ 0.00080 $ & $ 0.246 $ & $   0.274 $ & $   0.276 $ &   4.5 &   9.2 &  10.2 &   1.2 &   1.5 &  -1.0 &  -1.4 &  -2.6 &   0.0 &  -5.8 &   1.4 &  -0.8 &   0.5 &   1.0 \\ 
 $c$  11 & $  20.0  $ & $ 0.00050 $ & $ 0.394 $ & $   0.281 $ & $   0.287 $ &   4.9 &  12.9 &  13.8 &   1.5 &   1.6 &  -0.5 &  -0.7 &  -5.9 &   0.2 &  -6.7 &   1.7 &  -1.3 &   3.2 &   1.1 \\ 
 $c$  12 & $  35.0  $ & $ 0.00320 $ & $ 0.108 $ & $   0.200 $ & $   0.200 $ &   6.9 &  10.7 &  12.7 &   1.9 &   2.4 &  -0.6 &  -0.8 &  -1.8 &   0.3 &  -8.8 &   1.2 &   3.0 &   0.1 &   1.1 \\ 
 $c$  13 & $  35.0  $ & $ 0.00200 $ & $ 0.172 $ & $   0.220 $ & $   0.220 $ &   6.1 &  10.1 &  11.8 &   1.5 &   1.9 &  -0.7 &  -1.0 &  -1.9 &   0.0 &  -7.9 &   2.5 &  -0.3 &   0.2 &   1.0 \\ 
 $c$  14 & $  35.0  $ & $ 0.00130 $ & $ 0.265 $ & $   0.295 $ & $   0.297 $ &   5.4 &   8.1 &   9.7 &   1.2 &   2.2 &  -0.7 &  -1.0 &  -2.0 &   0.0 &  -5.4 &   1.4 &  -0.4 &   0.3 &   1.0 \\ 
 $c$  15 & $  35.0  $ & $ 0.00080 $ & $ 0.431 $ & $   0.349 $ & $   0.360 $ &   6.1 &  11.1 &  12.7 &   1.3 &   0.6 &  -0.5 &  -0.8 &  -4.6 &   0.1 &  -6.4 &   2.0 &  -1.0 &   1.8 &   1.1 \\ 
 $c$  16 & $  60.0  $ & $ 0.00500 $ & $ 0.118 $ & $   0.198 $ & $   0.199 $ &   5.1 &   9.5 &  10.8 &   1.8 &   1.9 &  -0.5 &  -0.8 &  -1.7 &   0.7 &  -6.8 &   2.3 &   1.2 &   0.1 &   1.1 \\ 
 $c$  17 & $  60.0  $ & $ 0.00320 $ & $ 0.185 $ & $   0.263 $ & $   0.264 $ &   5.5 &   6.3 &   8.4 &   1.1 &   1.5 &  -0.7 &  -1.1 &  -1.5 &   0.0 &  -3.3 &   2.6 &   1.3 &   0.0 &   1.0 \\ 
 $c$  18 & $  60.0  $ & $ 0.00200 $ & $ 0.295 $ & $   0.335 $ & $   0.339 $ &   4.3 &   7.7 &   8.8 &   1.1 &   1.8 &  -0.5 &  -0.8 &  -1.5 &   0.1 &  -5.8 &   1.7 &  -0.3 &   0.1 &   1.0 \\ 
 $c$  19 & $  60.0  $ & $ 0.00130 $ & $ 0.454 $ & $   0.296 $ & $   0.307 $ &   8.3 &  12.6 &  15.1 &   1.2 &   1.6 &  -0.1 &  -0.2 &  -3.7 &   0.2 &  -9.6 &   1.5 &  -1.3 &   1.0 &   1.0 \\ 
 $c$  20 & $ 120.0  $ & $ 0.01300 $ & $ 0.091 $ & $   0.133 $ & $   0.133 $ &   9.7 &  10.2 &  14.1 &   2.7 &   2.6 &  -0.7 &  -1.0 &  -2.2 &   0.9 &  -7.7 &   3.2 &   1.2 &   0.0 &   1.2 \\ 
 $c$  21 & $ 120.0  $ & $ 0.00500 $ & $ 0.236 $ & $   0.218 $ & $   0.220 $ &   6.8 &   8.8 &  11.1 &   1.6 &   2.5 &  -0.3 &  -0.5 &  -1.2 &   0.1 &  -6.6 &   3.1 &   0.7 &   0.0 &   1.1 \\ 
 $c$  22 & $ 120.0  $ & $ 0.00200 $ & $ 0.591 $ & $   0.351 $ & $   0.375 $ &   8.6 &   9.5 &  12.8 &   3.2 &   3.3 &  -0.3 &  -0.4 &  -2.7 &   0.2 &  -5.1 &   1.9 &   0.0 &   0.9 &   2.9 \\ 
 $c$  23 & $ 200.0  $ & $ 0.01300 $ & $ 0.151 $ & $   0.160 $ & $   0.160 $ &   7.1 &   9.5 &  11.9 &   2.8 &   2.1 &  -1.0 &  -1.5 &  -1.8 &   0.2 &  -6.5 &   3.1 &   1.7 &   0.0 &   2.7 \\ 
 $c$  24 & $ 200.0  $ & $ 0.00500 $ & $ 0.394 $ & $   0.237 $ & $   0.243 $ &   8.8 &  10.2 &  13.5 &   3.4 &   3.3 &  -0.1 &  -0.2 &  -2.0 &   0.0 &  -6.2 &   3.7 &   0.3 &   0.5 &   2.9 \\ 
 $c$  25 & $ 300.0  $ & $ 0.02000 $ & $ 0.148 $ & $   0.117 $ & $   0.117 $ &  14.1 &  12.0 &  18.5 &   3.3 &   2.6 &  -0.4 &  -0.6 &  -1.3 &   0.4 &  -9.3 &   3.8 &   1.9 &   0.0 &   2.9 \\ 
 $c$  26 & $ 300.0  $ & $ 0.00800 $ & $ 0.369 $ & $   0.273 $ & $   0.278 $ &   9.5 &   8.4 &  12.7 &   3.3 &   3.6 &  -0.2 &  -0.3 &  -1.1 &   0.0 &  -3.4 &   4.8 &  -0.1 &   0.5 &   2.9 \\ 
 $c$  27 & $ 650.0  $ & $ 0.03200 $ & $ 0.200 $ & $   0.084 $ & $   0.085 $ &  16.7 &  26.0 &  30.9 &   3.8 &   5.2 &  -2.2 &  -3.3 &  -3.4 &   0.2 & -20.8 &  10.7 &   4.5 &   0.0 &   3.4 \\ 
 $c$  28 & $ 650.0  $ & $ 0.01300 $ & $ 0.492 $ & $   0.195 $ & $   0.203 $ &  10.8 &  12.0 &  16.2 &   3.7 &   3.0 &  -0.1 &  -0.2 &  -1.0 &   0.2 &  -6.9 &   6.9 &  -0.2 &   0.4 &   3.3 \\ 
 $c$  29 & $ 2000.0  $ & $ 0.05000 $ & $ 0.394 $ & $   0.059 $ & $   0.060 $ &  28.3 &  22.9 &  36.4 &   6.7 &   4.0 &  -1.7 &  -2.6 &  -1.2 &   0.1 & -14.1 &  15.6 &   0.0 &   0.2 &   4.3 \\ 
\hline 
 $b$   1  & $   5.0 $ & $ 0.00020 $ & $ 0.246 $ & $ 0.00244 $ & $ 0.00244 $ &   33.8  &   31.3  &   46.1  &    8.1  &   -2.4  &   -2.8  &   -5.6  &   -3.7  &   -6.6  &    6.0  &    -3.7  &    3.1  &    3.1  &    1.0  \\ 
 $b$   2  & $  12.0 $ & $ 0.00032 $ & $ 0.369 $ & $ 0.00487 $ & $ 0.00490 $ &   16.0  &   27.4  &   31.8  &    8.4  &   -3.4  &   -4.1  &   -8.2  &   -1.1  &   -4.5  &    5.1  &     2.2  &   -0.8  &    4.1  &    1.1  \\ 
 $b$   3  & $  12.0 $ & $ 0.00080 $ & $ 0.148 $ & $ 0.00247 $ & $ 0.00248 $ &   16.7  &   40.2  &   43.5  &    9.6  &   -8.1  &   -7.4  &  -14.8  &   -1.6  &   -5.3  &   17.8  &     2.2  &    6.6  &    0.6  &    1.1  \\ 
 $b$   4  & $  25.0 $ & $ 0.00050 $ & $ 0.492 $ & $ 0.01189 $ & $ 0.01206 $ &   10.1  &   23.0  &   25.1  &    5.0  &   -4.0  &   -3.8  &   -7.5  &    0.2  &   -4.5  &    3.5  &     0.8  &    0.3  &    3.1  &    1.3  \\ 
 $b$   5  & $  25.0 $ & $ 0.00130 $ & $ 0.189 $ & $ 0.00586 $ & $ 0.00587 $ &   12.1  &   31.9  &   34.1  &    7.8  &   -8.7  &   -5.6  &  -11.2  &   -0.8  &   -4.9  &   10.5  &     0.9  &    3.5  &    0.2  &    1.0  \\ 
 $b$   6  & $  60.0 $ & $ 0.00130 $ & $ 0.454 $ & $ 0.01928 $ & $ 0.01969 $ &    9.5  &   23.1  &   25.0  &   10.9  &   -4.6  &   -4.0  &   -8.1  &   -0.1  &   -2.9  &    4.1  &     1.4  &    0.7  &    0.8  &    1.0  \\ 
 $b$   7  & $  60.0 $ & $ 0.00500 $ & $ 0.118 $ & $ 0.00964 $ & $ 0.00965 $ &   14.6  &   29.1  &   32.6  &    4.3  &   -7.0  &   -5.2  &  -10.4  &   -0.5  &   -7.9  &    2.3  &    -2.8  &    8.6  &    0.0  &    1.1  \\ 
 $b$   8  & $ 200.0 $ & $ 0.00500 $ & $ 0.394 $ & $ 0.02365 $ & $ 0.02422 $ &    9.4  &   21.2  &   23.2  &    7.9  &   -4.6  &   -4.2  &   -8.5  &    0.2  &   -2.7  &    1.1  &     1.0  &    0.5  &    1.9  &    2.9  \\ 
 $b$   9  & $ 200.0 $ & $ 0.01300 $ & $ 0.151 $ & $ 0.01139 $ & $ 0.01142 $ &   19.5  &   28.4  &   34.4  &    4.6  &   -4.4  &   -4.5  &   -8.9  &    0.2  &  -10.3  &    5.7  &     0.0  &   10.4  &    0.0  &    2.7  \\ 
 $b$  10  & $ 650.0 $ & $ 0.01300 $ & $ 0.492 $ & $ 0.01331 $ & $ 0.01394 $ &   23.6  &   25.4  &   34.7  &   13.4  &   -4.6  &   -5.0  &  -10.1  &   -1.1  &   -2.2  &    5.8  &     4.5  &    1.5  &    0.8  &    3.3  \\ 
 $b$  11  & $ 650.0 $ & $ 0.03200 $ & $ 0.200 $ & $ 0.01018 $ & $ 0.01024 $ &   16.0  &   25.5  &   30.1  &   10.8  &   -3.3  &   -3.3  &   -6.5  &   -0.2  &   -9.0  &    3.6  &     2.1  &    3.1  &    0.0  &    3.4  \\ 
 $b$  12  & $ 2000.0 $ & $ 0.05000 $ & $ 0.394 $ & $ 0.00499 $ & $ 0.00511 $ &   55.7  &   25.1  &   61.1  &   15.0  &   -2.0  &   -4.4  &   -8.7  &    1.7  &   -2.7  &    9.4  &    -3.0  &    1.7  &    0.2  &    4.3  \\ 
\hline 
\end{tabular} 

\end{center}

\normalsize
    \caption{ The measured values and relative errors for the reduced
    DIS cross section ($\tilde{\sigma}^{q\bar{q}}$) and structure
    function $F_2^{q\bar{q}}$ for charm ($c$) and beauty ($b$) quarks
    for the combined datasets. The values for $F_2^{q\bar{q}}$ are
    obtained from the measured reduced cross sections using the NLO QCD fit to
    correct for the contributions from $F_L^{q\bar{q}}$.  The table
    shows the statistical error ($\delta_{\rm stat}$), the systematic
    error ($\delta_{\rm sys}$), the total error ($\delta_{\rm tot}$)
    and the uncorrelated systematic error ($\delta_{\rm unc}$) on the
    reduced cross section. The next ten columns show the effect of a $+ 1
    \sigma$ shift for the correlated systematic error contributions to
    the reduced cross section from: track impact parameter resolution, CJC
    track efficiency, CST track efficiency, $c$ fragmentation, $b$
    fragmentation, light quark contribution, struck quark angle
    $\phi^q$, hadronic energy scale, photoproduction
    background and the DIS event selection.  The $-1 \sigma$ errors
    are taken as the negative of the $+1 \sigma$ errors.  The
    correlated systematic errors are continued in
    table~\ref{tab:sysmod}.
  }
\label{tab:sig}
\end{sidewaystable}

\begin{sidewaystable}
\tiny
\begin{center}
 \renewcommand{\arraystretch}{1.15}
 \begin{tabular}{|c|d|d|d|d|d|d|d|d|d|d|d|d|d|d|} \hline 
 bin  &      \multicolumn{1}{r|}{$\delta_{{\rm mod} Q^2 c}$} & \multicolumn{1}{r|}{$\delta_{{\rm mod} Q^2 b}$}   & \multicolumn{1}{r|}{$\delta_{{\rm mod} x c}$}  & \multicolumn{1}{r|}{$\delta_{{\rm mod} x b}$}   & \multicolumn{1}{r|}{$\delta_{{\rm mod} P_T c}$}  & \multicolumn{1}{r|}{$\delta_{{\rm mod} P_T b}$}   & \multicolumn{1}{r|}{$\delta_{{\rm mod} \eta c}$}  & \multicolumn{1}{r|}{$\delta_{{\rm mod} \eta b}$} & \multicolumn{1}{r|}{$\delta_{{\rm BF} D^+}$} & \multicolumn{1}{r|}{$\delta_{{\rm BF} D^0}$} &  \multicolumn{1}{r|}{$\delta_{{\rm Mult} D^+}$} & \multicolumn{1}{r|}{$\delta_{{\rm Mult} D^0}$} & \multicolumn{1}{r|}{$\delta_{{\rm Mult} D_s}$} & \multicolumn{1}{r|}{$\delta_{{\rm Mult} B}$}  \\ 
 &  \multicolumn{1}{r|}{$(\%)$} & \multicolumn{1}{r|}{$(\%)$} & \multicolumn{1}{r|}{$(\%)$} & \multicolumn{1}{r|}{$(\%)$} & \multicolumn{1}{r|}{$(\%)$} & \multicolumn{1}{r|}{$(\%)$} & \multicolumn{1}{r|}{$(\%)$} & \multicolumn{1}{r|}{$(\%)$} & \multicolumn{1}{r|}{$(\%)$} & \multicolumn{1}{r|}{$(\%)$} & \multicolumn{1}{r|}{$(\%)$} & \multicolumn{1}{r|}{$(\%)$} & \multicolumn{1}{r|}{$(\%)$} & \multicolumn{1}{r|}{$(\%)$} \\ \hline
 $ c    1 $ & $  -0.1 $ & $   0.0 $ & $   2.3 $ & $  -0.4 $ & $  -7.8 $ & $   0.9 $ & $   2.8 $ & $   0.4 $  & $  -0.6 $ & $   0.2 $ & $  -2.0 $ & $  -1.4 $ & $  -1.7 $  & $  -0.7 $ \\ 
 $ c    2 $ & $  -0.7 $ & $   0.0 $ & $   1.4 $ & $  -0.1 $ & $  -5.1 $ & $   0.3 $ & $   3.6 $ & $  -0.1 $  & $  -0.6 $ & $   0.1 $ & $  -1.6 $ & $  -1.6 $ & $  -1.4 $  & $  -0.5 $ \\ 
 $ c    3 $ & $  -0.5 $ & $   0.0 $ & $  -0.1 $ & $   0.0 $ & $  -8.3 $ & $   0.8 $ & $   0.6 $ & $   0.4 $  & $  -0.6 $ & $   0.3 $ & $  -1.8 $ & $  -1.4 $ & $  -1.4 $  & $  -0.9 $ \\ 
 $ c    4 $ & $  -0.3 $ & $   0.0 $ & $   1.3 $ & $   0.0 $ & $  -3.5 $ & $  -0.1 $ & $   3.5 $ & $   0.0 $  & $  -0.6 $ & $   0.0 $ & $  -1.7 $ & $  -1.3 $ & $  -1.7 $  & $   0.4 $ \\ 
 $ c    5 $ & $  -0.2 $ & $   0.0 $ & $   0.8 $ & $   0.0 $ & $  -4.5 $ & $   0.4 $ & $   0.0 $ & $   0.2 $  & $  -0.6 $ & $   0.2 $ & $  -1.7 $ & $  -1.3 $ & $  -1.6 $  & $  -1.0 $ \\ 
 $ c    6 $ & $  -0.2 $ & $   0.0 $ & $   0.4 $ & $   0.0 $ & $  -5.9 $ & $   0.3 $ & $   1.5 $ & $   0.1 $  & $  -0.5 $ & $   0.2 $ & $  -1.7 $ & $  -1.4 $ & $  -1.6 $  & $  -0.3 $ \\ 
 $ c    7 $ & $  -0.2 $ & $   0.0 $ & $  -0.3 $ & $   0.0 $ & $  -7.8 $ & $   1.0 $ & $   0.3 $ & $   0.4 $  & $  -0.5 $ & $   0.3 $ & $  -1.7 $ & $  -1.4 $ & $  -1.7 $  & $  -0.7 $ \\ 
 $ c    8 $ & $  -0.1 $ & $   0.0 $ & $   1.5 $ & $  -0.1 $ & $  -2.6 $ & $   0.3 $ & $   3.6 $ & $   0.0 $  & $  -0.3 $ & $  -0.2 $ & $  -1.2 $ & $  -1.6 $ & $  -1.7 $  & $  -0.6 $ \\ 
 $ c    9 $ & $  -0.1 $ & $   0.0 $ & $   0.9 $ & $   0.0 $ & $  -3.7 $ & $   0.3 $ & $   3.1 $ & $   0.1 $  & $  -0.5 $ & $   0.1 $ & $  -1.6 $ & $  -1.4 $ & $  -1.7 $  & $  -0.5 $ \\ 
 $ c   10 $ & $  -0.1 $ & $   0.0 $ & $   0.3 $ & $   0.0 $ & $  -4.8 $ & $   0.3 $ & $   1.6 $ & $   0.1 $  & $  -0.5 $ & $   0.2 $ & $  -1.7 $ & $  -1.4 $ & $  -1.8 $  & $  -0.5 $ \\ 
 $ c   11 $ & $  -0.2 $ & $   0.0 $ & $  -0.8 $ & $  -0.1 $ & $  -7.2 $ & $   1.3 $ & $   0.6 $ & $   0.6 $  & $  -0.4 $ & $   0.2 $ & $  -1.5 $ & $  -1.5 $ & $  -1.5 $  & $  -1.5 $ \\ 
 $ c   12 $ & $   0.1 $ & $   0.0 $ & $   1.6 $ & $   0.0 $ & $  -1.6 $ & $   0.4 $ & $   0.0 $ & $  -0.1 $  & $  -0.1 $ & $  -0.2 $ & $  -1.0 $ & $  -1.4 $ & $  -1.8 $  & $  -1.1 $ \\ 
 $ c   13 $ & $   0.0 $ & $   0.0 $ & $   0.9 $ & $   0.0 $ & $  -2.3 $ & $   0.6 $ & $   2.9 $ & $   0.2 $  & $  -0.5 $ & $   0.2 $ & $  -1.2 $ & $  -1.4 $ & $  -1.4 $  & $  -1.0 $ \\ 
 $ c   14 $ & $  -0.1 $ & $   0.0 $ & $   0.3 $ & $   0.0 $ & $  -3.4 $ & $   0.5 $ & $   1.8 $ & $   0.2 $  & $  -0.5 $ & $   0.2 $ & $  -1.4 $ & $  -1.5 $ & $  -1.4 $  & $  -0.6 $ \\ 
 $ c   15 $ & $  -0.2 $ & $   0.0 $ & $  -1.0 $ & $   0.0 $ & $  -6.3 $ & $   1.0 $ & $   0.5 $ & $   0.5 $  & $  -0.5 $ & $   0.3 $ & $  -1.8 $ & $  -1.3 $ & $  -1.7 $  & $  -1.1 $ \\ 
 $ c   16 $ & $   0.0 $ & $  -0.1 $ & $   1.4 $ & $  -0.3 $ & $  -1.1 $ & $   0.7 $ & $   3.5 $ & $  -0.4 $  & $  -0.3 $ & $   0.0 $ & $  -1.1 $ & $  -1.1 $ & $  -1.8 $  & $  -1.7 $ \\ 
 $ c   17 $ & $  -0.1 $ & $   0.0 $ & $   0.6 $ & $   0.0 $ & $  -1.6 $ & $   0.3 $ & $   2.4 $ & $   0.0 $  & $  -0.4 $ & $   0.1 $ & $  -1.2 $ & $  -1.4 $ & $  -1.2 $  & $  -0.6 $ \\ 
 $ c   18 $ & $  -0.2 $ & $   0.0 $ & $  -0.4 $ & $  -0.1 $ & $  -2.4 $ & $   0.6 $ & $   1.0 $ & $   0.1 $  & $  -0.4 $ & $   0.1 $ & $  -1.4 $ & $  -1.2 $ & $  -1.5 $  & $  -1.1 $ \\ 
 $ c   19 $ & $  -0.6 $ & $   0.0 $ & $  -1.8 $ & $   0.0 $ & $  -5.3 $ & $   1.3 $ & $   0.0 $ & $   0.5 $  & $  -0.3 $ & $   0.2 $ & $  -1.5 $ & $  -1.3 $ & $  -1.6 $  & $  -1.8 $ \\ 
 $ c   20 $ & $   0.3 $ & $   0.1 $ & $   1.5 $ & $  -0.2 $ & $  -0.1 $ & $   0.5 $ & $   0.0 $ & $  -0.6 $  & $  -0.2 $ & $  -0.2 $ & $  -1.1 $ & $  -0.9 $ & $  -1.3 $  & $  -2.2 $ \\ 
 $ c   21 $ & $   0.3 $ & $   0.0 $ & $  -0.2 $ & $  -0.3 $ & $  -0.9 $ & $   0.7 $ & $   0.9 $ & $  -0.2 $  & $  -0.3 $ & $   0.1 $ & $  -1.2 $ & $  -1.0 $ & $  -1.2 $  & $  -2.4 $ \\ 
 $ c   22 $ & $  -0.9 $ & $   0.1 $ & $  -1.9 $ & $   0.2 $ & $  -4.3 $ & $   0.8 $ & $   0.2 $ & $   0.4 $  & $  -0.4 $ & $   0.3 $ & $  -1.4 $ & $  -1.2 $ & $  -1.1 $  & $  -1.6 $ \\ 
 $ c   23 $ & $  -0.4 $ & $   0.0 $ & $   1.6 $ & $  -0.2 $ & $  -0.7 $ & $   0.2 $ & $   0.0 $ & $  -0.1 $  & $  -0.4 $ & $   0.0 $ & $  -1.3 $ & $  -1.1 $ & $  -1.3 $  & $  -1.1 $ \\ 
 $ c   24 $ & $  -0.1 $ & $   0.1 $ & $  -0.8 $ & $   0.1 $ & $  -2.3 $ & $   0.7 $ & $   0.3 $ & $   0.3 $  & $  -0.3 $ & $   0.1 $ & $  -1.2 $ & $  -1.2 $ & $  -1.5 $  & $  -2.2 $ \\ 
 $ c   25 $ & $  -0.1 $ & $   0.1 $ & $   0.9 $ & $  -0.2 $ & $  -0.4 $ & $   0.6 $ & $   0.0 $ & $  -0.4 $  & $  -0.5 $ & $   0.1 $ & $  -1.5 $ & $  -1.0 $ & $  -2.0 $  & $  -3.2 $ \\ 
 $ c   26 $ & $   0.0 $ & $   0.0 $ & $  -0.5 $ & $   0.1 $ & $  -1.4 $ & $   0.3 $ & $   0.8 $ & $   0.1 $  & $  -0.3 $ & $   0.3 $ & $  -1.1 $ & $  -0.8 $ & $  -1.6 $  & $  -1.6 $ \\ 
 $ c   27 $ & $   0.1 $ & $   0.1 $ & $   1.7 $ & $  -0.2 $ & $   0.1 $ & $   0.2 $ & $   3.5 $ & $  -0.1 $  & $  -0.6 $ & $   0.0 $ & $  -1.3 $ & $  -1.4 $ & $  -2.3 $  & $  -2.5 $ \\ 
 $ c   28 $ & $   0.1 $ & $   0.3 $ & $  -0.7 $ & $   0.3 $ & $  -2.0 $ & $   0.4 $ & $  -0.1 $ & $   0.1 $  & $  -0.3 $ & $   0.2 $ & $  -0.7 $ & $  -0.6 $ & $  -1.1 $  & $  -2.9 $ \\ 
 $ c   29 $ & $   0.6 $ & $   0.6 $ & $   0.3 $ & $   0.1 $ & $  -0.9 $ & $   0.5 $ & $   0.0 $ & $   0.2 $  & $  -0.5 $ & $   0.3 $ & $  -1.6 $ & $  -1.0 $ & $   0.9 $  & $  -2.1 $ \\ 
\hline 
 $ b    1 $ & $   0.0 $ & $  -0.4 $ & $   0.6 $ & $  14.0 $ & $  -7.2 $ & $ -17.8 $ & $  -2.2 $ & $   9.0 $  & $  -2.2 $ & $   0.6 $ & $  -3.0 $ & $  -0.8 $ & $  -1.8 $  & $   9.2 $ \\ 
 $ b    2 $ & $  -0.2 $ & $  -1.1 $ & $  -0.4 $ & $  -1.3 $ & $  -7.3 $ & $ -17.3 $ & $  -3.8 $ & $  -1.0 $  & $  -1.2 $ & $   0.9 $ & $  -3.4 $ & $  -0.3 $ & $  -1.1 $  & $  11.4 $ \\ 
 $ b    3 $ & $  -0.3 $ & $  -2.6 $ & $   0.8 $ & $   4.7 $ & $ -13.7 $ & $ -14.5 $ & $   0.0 $ & $   8.2 $  & $  -3.9 $ & $   2.6 $ & $ -10.5 $ & $  -1.3 $ & $  -3.0 $  & $  12.9 $ \\ 
 $ b    4 $ & $  -0.2 $ & $  -0.5 $ & $  -0.1 $ & $  -1.0 $ & $  -3.8 $ & $ -15.2 $ & $  -1.8 $ & $  -1.4 $  & $  -0.9 $ & $   0.4 $ & $  -2.3 $ & $  -0.5 $ & $  -0.5 $  & $  10.7 $ \\ 
 $ b    5 $ & $  -0.6 $ & $  -0.3 $ & $   0.3 $ & $   7.1 $ & $  -8.1 $ & $ -11.9 $ & $   0.0 $ & $   7.7 $  & $  -3.4 $ & $   2.0 $ & $  -8.6 $ & $  -1.2 $ & $  -2.8 $  & $  13.3 $ \\ 
 $ b    6 $ & $  -0.3 $ & $  -0.8 $ & $  -0.4 $ & $  -2.9 $ & $  -2.8 $ & $ -10.2 $ & $  -1.4 $ & $  -0.6 $  & $  -1.4 $ & $   0.8 $ & $  -3.3 $ & $  -1.1 $ & $  -1.5 $  & $  12.1 $ \\ 
 $ b    7 $ & $  -0.4 $ & $  -0.5 $ & $   0.3 $ & $   7.2 $ & $  -3.0 $ & $  -7.4 $ & $   0.0 $ & $  10.1 $  & $  -2.9 $ & $   1.3 $ & $  -6.8 $ & $  -1.4 $ & $  -2.8 $  & $  14.8 $ \\ 
 $ b    8 $ & $  -0.4 $ & $  -1.8 $ & $  -0.6 $ & $  -2.1 $ & $  -1.3 $ & $  -8.7 $ & $   0.0 $ & $   0.8 $  & $  -0.9 $ & $   0.3 $ & $  -2.5 $ & $  -1.0 $ & $  -1.2 $  & $  12.6 $ \\ 
 $ b    9 $ & $  -0.3 $ & $  -2.4 $ & $   0.4 $ & $   4.7 $ & $  -1.4 $ & $  -5.1 $ & $   0.0 $ & $  10.3 $  & $  -2.0 $ & $   0.5 $ & $  -4.5 $ & $  -1.3 $ & $  -1.0 $  & $  15.1 $ \\ 
 $ b   10 $ & $  -0.8 $ & $  -1.5 $ & $  -1.0 $ & $  -2.7 $ & $  -1.0 $ & $  -5.4 $ & $  -1.9 $ & $   0.3 $  & $  -1.1 $ & $   0.3 $ & $  -4.0 $ & $  -2.2 $ & $  -2.8 $  & $  13.0 $ \\ 
 $ b   11 $ & $  -0.2 $ & $  -0.6 $ & $   0.1 $ & $   7.9 $ & $  -0.3 $ & $  -1.8 $ & $   0.0 $ & $  10.6 $  & $  -0.6 $ & $  -0.5 $ & $  -1.3 $ & $  -0.9 $ & $  -0.1 $  & $  13.0 $ \\ 
 $ b   12 $ & $  -0.9 $ & $  -0.2 $ & $   0.2 $ & $   0.8 $ & $   0.5 $ & $  -2.2 $ & $   2.4 $ & $   2.3 $  & $  -0.5 $ & $  -0.2 $ & $  -1.8 $ & $  -1.2 $ & $  -6.0 $  & $  11.8 $ \\ 
\hline 
\end{tabular} 

\end{center}
\normalsize
    \caption{ The correlated errors continued from
    table~\ref{tab:sig}. The first eight errors represent a $+ 1
    \sigma$ shift for the correlated systematic error contributions
    from: reweighting the $Q^2$ distribution, reweighting the $x$
    distribution, reweighting the jet transverse momentum distribution
    $P^{\rm jet}_T$, and reweighting the jet pseudorapidity $\eta^{\rm
    jet}$ distribution for $c$ and $b$ events. The remaining six
    columns show the contributions from: $c$ hadron branching
    fractions and multiplicities, and the $b$ quark multiplicity. Only
    those uncertainties where there is an effect of $>1\%$ for any
    $x$--$Q^2$ point are listed; the remaining uncertainties are
    included in the uncorrelated error.}
\label{tab:sysmod}
\end{sidewaystable}

\clearpage

\newpage

\clearpage

\end{document}